# Presolar Grains

Nan Liu

Institute for Astrophysical Research, Boston University, Boston, MA 02215, USA;
nanliu@bu.edu

Abstract

The chemical makeup of our solar system is a reflection of Galactic chemical evolution in the local interstellar medium (ISM) over the past ~9 Ga before the formation of solar system. Although the incorporated ISM dust was mostly destroyed during the solar system formation, a small fraction of the ISM dust, known as presolar grains, is preserved in pristine extraterrestrial materials and identified through their exotic isotopic compositions, pointing to their formation in gas outflows or explosions of ancient stars. Since their stellar birth at more than 4.6 Ga, presolar grains have borne witness to a diverse array of astrophysical and cosmochemical processes. In this chapter, I will review recent progress in utilizing the isotopic and structural compositions of presolar grains to constrain physical mixing processes and dust formation in stars, stellar nucleosynthesis, ISM processes, and the origin and evolution of the solar system.

**Key points**:

- Identification of presolar grains in primitive extraterrestrial materials
- Inventory of presolar grains in the solar system
- Stellar sources of presolar grains and challenges in identifying their parent stars
- Constraints on processes in stars, the ISM, and the solar system

# 1. INTRODUCTION

Since the first confirmation of exoplanet detection in 1992 (Wolszczan and Frail 1992), more than 5500 exoplanets in about 4000 planetary systems have been confirmed (03/2024; https://exoplanetarchive.ipac.caltech.edu/), and this number continues to grow rapidly. Among the 4000 exoplanetary systems, the type and variety of their architectures are highly diverse, ranging from systems containing hot Jupiters (e.g., Mayor and Queloz 1995) to super-Earth-sized water worlds (e.g., Cadieux et al. 2022). With the over 5500 confirmed exoplanets, the demographics of extrasolar systems begin to emerge, and the lack of detected exoplanetary systems like our own raises the question whether our solar system is unique or this simply reflects that finding an exoplanet system such as ours is beyond state-of-the-art technical capabilities. Are we unique and are we alone? While addressing this big science question requires studies in many research areas and at many levels, our knowledge of the origin and evolution of the solar system is at the heart of this challenge. To truly understand differences and similarities between extrasolar systems and the solar system, we need to unravel the processes that led the solar system to evolve from a cloud of gas and dust in the interstellar medium (ISM) to its current architecture over the past 4.6 Ga. This echoes the focus of cosmochemistry – "*the study of the chemical composition of matter in the Universe and the processes that led to those compositions* (McSween and Huss 2010) based on the chemical composition of extraterrestrial materials and other physical samples."

Cosmochemistry focuses primarily on objects in our own solar system as we have direct access to samples derived from asteroids (e.g., CI chondrites[1]), the Moon (e.g., Apollo samples), Mars (e.g., Martian meteorites), and comets (e.g., comet Wild 2 samples returned by Stardust mission). In addition, primitive extraterrestrial materials such as chondrites contain microscopic stardust grains, which originated in extrasolar systems that existed prior to the solar system formation. Given that such stardust grains are older than the solar system, they are commonly known as presolar grains. The presence of presolar grains in extraterrestrial materials provides cosmochemists direct access to bona fide extrasolar materials, enabling the investigation of a wide array of astrophysical processes in unprecedented detail.

[1] Chondrites are meteorites that are not modified by melting or differentiation of their parent bodies, and differentiation is the process in which different constituents of planetary materials are separated, resulting in the formation of distinct compositional layers.

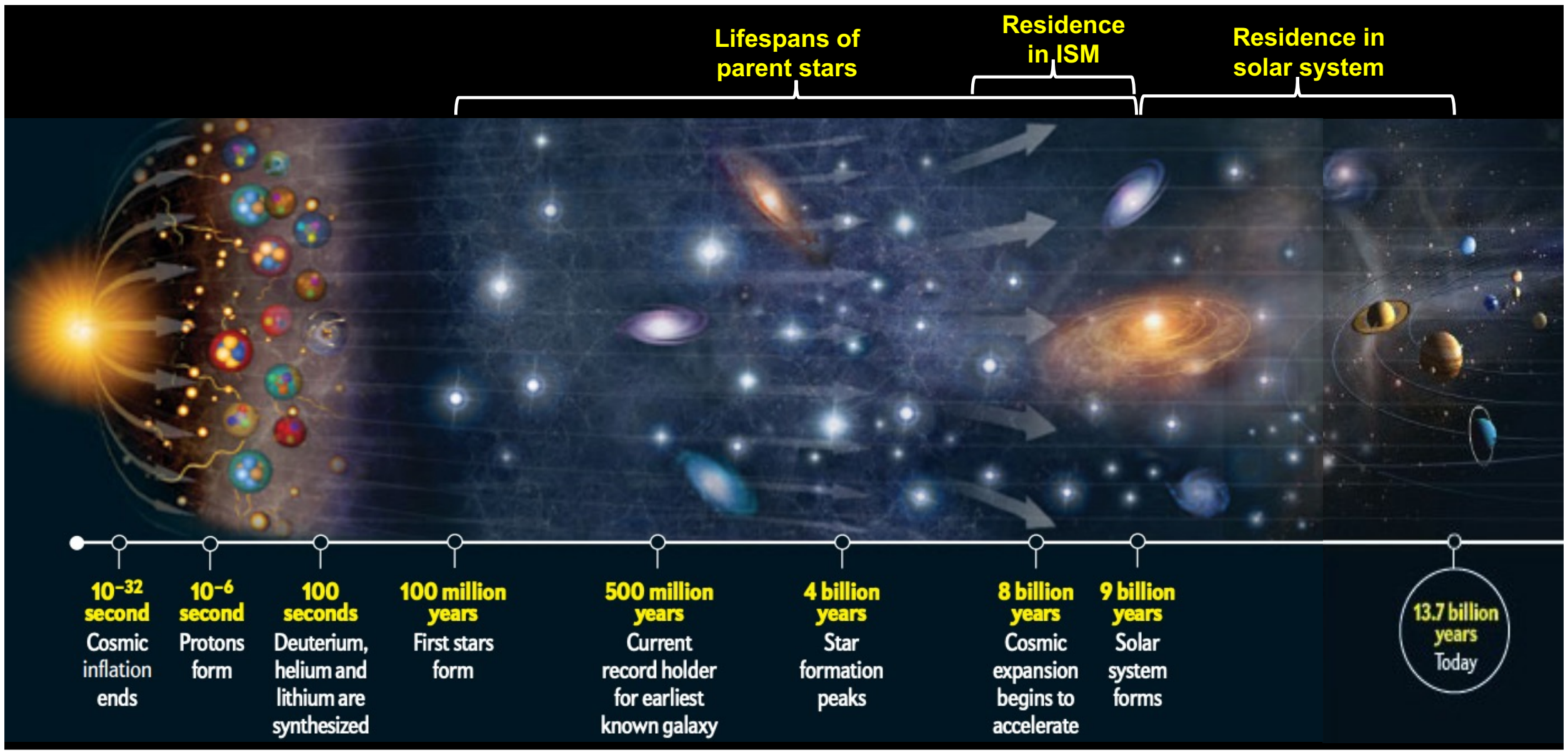

**Figure 1**. *The evolution of the Universe over 13.7 billion years (after Goldsmith 2012). The bottom timeline highlights milestones along the Universe evolution, and the curly brackets at the top mark three distinct timespans that are relevant to presolar grains.*

In the solar system, the number of rocky planets and their total mass, in general terms, are determined by the total mass of available rock-forming elements, which is directly related to the mass of ISM grains that were initially incorporated into the solar system. In other words, ISM dust grains are the building blocks of our planetary system, and presolar grains – extrasolar system fossils that once resided in the local ISM after their loss from the parent stars and before their incorporation into the solar system (Fig. 1) – are thus the leftover solar system building blocks, allowing us to directly probe the solar system birthplace. This initial condition for the solar system formation played a key role in determining the formation and evolution of planets in the solar system and, in turn, the habitability of Earth. For instance, radionuclides that had been produced by stellar nucleosynthesis – the formation of new atomic nuclei by nuclear reactions – in the parent stars of presolar grains, powered the thermal evolution of rocky planets in the solar system. The short-lived radionuclide $^{26}$Al ($t_{1/2}$ = 0.72 Ma) controlled the interior evolution of planetesimals, i.e., the building blocks of rocky planets, resulting in silicate melting and degassing of primordial water abundances (Wallis 1980; Merk et al. 2002; Šrámek et al. 2012; Lichtenberg et al. 2019). The long-term thermal evolution of rocky planets in the solar system is governed by several long-lived radionuclides (e.g., $^{40}$K, $^{232}$Th, $^{238}$U), which are responsible for the occurrence of planetary geological activities, e.g., plate tectonics (Frank et al. 2014; Unterborn et al. 2015; Nimmo et al. 2020). The initial abundances of $^{26}$Al, $^{232}$Th, and $^{238}$U are expected to vary by orders of magnitude between different planetary systems (Frank et al. 2014; Goriely and Martínez Pinedo 2015; Lichtenberg et al. 2016; Nimmo et al. 2020). Consequently, the formation and evolution of planetary systems could vary significantly, depending on the birthplace of their host stars (Lichtenberg et al. 2019; Nimmo et al. 2020). Laboratory analyses of presolar grains provide us a unique opportunity to directly probe the variety of stars that contributed their nucleosynthesis products to the solar system, which is a key piece of information for understanding whether the solar system is unique in comparison to extrasolar systems.

This review will be built upon previous overviews of presolar grain research (Davis 2011; Zinner 2014; Nittler and Ciesla 2016) with a focus on recent progress. The review will begin with an overview of the properties of presolar grains (Section 2) and delve into presolar grain studies over the past decade, which have led to new constraints on astrophysical processes in stars (Section 3), in the ISM (Section 4), and in the solar system (Section 5). The reader is referred to Zinner (2014) for a historical overview of presolar grain research in detail.

## 2. PRESOLAR GRAINS

### *2.1. Identification and Stellar Origins*

Presolar grains are identified in extraterrestrial materials by their exotic isotopic compositions that differ from those of solar system objects. The unique isotopic signatures of presolar grains, i.e., isotopic anomalies, cannot be explained by any known chemical or physical processes in the solar system and, instead, mainly reflect the compositions of their parent stars. Given its identification method, a presolar grain could become "destroyed" by isotopic equilibration with ISM and/or solar system materials (i.e., erasing its unique isotopic signatures) because of isotopic diffusion and exchange, in addition to physical destruction via evaporation and sputtering processes. While presolar grains were mostly destroyed during the melting and differentiation of planetesimals and planets, they are preserved in the primitive extraterrestrial materials that did not experience any significant heating, including chondrites (Floss and Haenecour 2016; Haenecour et al. 2018; Nittler et al. 2018a, 2021; Barosch et al. 2022b), chondritic micrometeorites (Yada et al. 2008), chondritic xenoliths in achondrites (Liu et al. 2020b), chondritic interplanetary dust particles (IDPs) (Nguyen et al. 2022), Wild 2 cometary sample (Floss et al. 2013), and asteroid samples returned by JAXA's Hayabusa 2 and NASA's OSIRIS-REx missions (Barosch et al. 2022a; Nguyen et al. 2023, 2024; Haenecour et al. 2024)[2]. Chondritic materials consist of three main components, including refractory inclusions [(Ca,Al-rich inclusions (CAIs) and amoeboid olivine aggregates (AOAs)], chondrules, and volatile-rich, fine-grained matrices (Scott and Krot 2014). While matrices are generally believed to have formed in the solar system at low temperatures (Buseck and Hua 1993), chondrules formed as molten, free-floating droplets (Scott and Krot 2014) and refractory inclusions formed in a very reducing environment at high temperatures (MacPherson 2014). The commonly held view is that presolar grains solely reside in primitive fine-grained matrices (Davidson et al. 2014; Floss and Haenecour 2016), because, in the hot nebular environments from which chondrules and refractory inclusions were derived, presolar grains likely became volatilized (Mendybaev et al. 2002) and/or isotopically equilibrated with solar nebular gas (Yamamoto et al. 2018). However, recently, Pravdivtseva et al. (2020) inferred the presence of presolar SiC grains in a fine-grained CAI based on noble gas isotopic signatures, challenging the conventional wisdom of matrices being the sole carrier of presolar grains. The presence of presolar SiC grains in refractory inclusions awaits confirmation by in situ isotopic studies. Nevertheless, refractory inclusions are unlikely to be a major host of presolar grains since (*i*) the inferred presolar SiC grain abundance in the fine-grained CAI (~7 ppm) is lower than those in primitive fine-grained matrices (~30 ppm) (Davidson et al. 2014), (*ii*) CAIs are always less abundant than matrices in chondrites (Scott and Krot 2014), and, most importantly, (*iii*) the dominant presolar mineral phase is silicates (up to several hundred ppm; Floss and Haenecour 2016), which were shown to exchange their O isotopes with water vapor rapidly at > 500 °C (Yamamoto et al. 2018) and should thus be absent in the CAI-forming region.

While nucleosynthesis in the parent star of a presolar grain largely determines the isotopic composition of the grain, its composition is also affected by the initial composition of its parent star, espeically if the isotopes of interest are not significantly produced in the star. Since the initial stellar composition reflects the composition of ISM gas in the solar neighborhood at different times (e.g., Lugaro et al. 1999), isotopic studies of presolar grains thus enable constraining the Galactic chemical evolution (GCE) – the process by which the elemental and isotopic composition of the

[2]This list of primitive extraterrestrial materials will be referred to as chondritic materials hereafter.

Galaxy varies in time and place because of stellar nucleosynthesis and material cycling between stars and the ISM (Nittler and Dauphas 2006) – in the solar neighborhood before the solar system formation. Earlier astronomical observations suggested that the ISM is quite chemically homogenous (within a factor of two variations) and has a solar-like composition on scales of hundreds of parsecs (for comparison, the Galaxy is 32,408 parsecs) based on elemental abundance measurements (e.g., Cartledge et al. 2006). However, De Cia et al. (2021) recently reported over a factor of 10 variations in metallicity in the vicinity of the Sun with the average metallicity being around 50% solar on scales of tens of parsecs. The finding of De Cia et al. (2021) is at odd with the presolar grain data that the majority of presolar SiC and silicate grains show higher-than-solar $^{29}Si/^{28}Si$ and $^{30}Si/^{28}Si$ ratios that are dominated by GCE (see discussions in detail in Section 3.1.3), suggesting that their parent stars, although being older, had higher metallicities than the Sun. This conundrum may be reconciled by the proposals that the dust production efficiency increases greatly with increasing stellar metallicity, resulting in preferential formation of dust in higher-than-solar metallicity stars (Lugaro et al. 2020) and that the lower-than-solar metallicity in the solar neighborhood is a consequence of pollution by infalling metal-poor gas from the Galactic halo over time (De Cia et al. 2021).

Extensive isotopic studies over the past three decades have largely confirmed that presolar grains were dominantly derived from stars reaching advanced evolutionary stages such as the asymptotic giant branch (AGB) phase and Type II core-collapse supernova (CCSN) explosions (Zinner 2014; Nittler and Ciesla 2016). The dominance of AGB and CCSN presolar grains in the solar system is in line with dust shells observed around AGB stars (e.g., Marengo et al. 1998; Zijlstra et al. 2006) and inferred dust formation in CCSN remnants (e.g., Gall et al. 2014; Indebetouw et al. 2014). This implies that AGB stars and CCSNe are the dominant contributors to dust reservoirs in the Universe, although it is still controversial whether the initially formed dust after a CCSN explosion could survive from destruction by the reserve supernova shock later (e.g., Micelotta et al. 2016). AGB stars and CCSNe are exceptional dust producers mainly because of their high mass loss rates and large mass ejections, respectively. Take our solar system for example, the current mass loss rate of the Sun is (2–3) × $10^{-14}$ $M_{\odot}$ per year (Cohen 2011) but will likely increase by several orders of magnitude to $10^{-8}$ – $10^{-5}$ $M_{\odot}$ per year (Höfner and Olofsson 2018) when it reaches the AGB phase. It is expected that 50 – 60 % of the solar mass will get lost to space during the AGB phase but eventually condense to form dust as it cools off. Other potential stellar sources that have been suggested for presolar grains include novae, J-type carbon stars, born-again AGB stars, the progenitor stars (Red supergiants) of CCSNe, and rare types of supernovae such as electron capture supernovae (ECSNe), but existing presolar grain isotope data, in many cases, are insufficient to distinguish between these different stellar sources. This topic will be discussed in detail in Section 3.

### *2.2. Chemical Composition, Abundance, and Grain Size*

**Table 1. Presolar Phases from Primitive Meteorites**

| Presolar Phase | Maximum Abundance | Main Stellar Source |
|---|---|---|
| SiC | ~30 ppm | AGB stars (>90%) |
| Graphite | ~10 ppm | AGB stars and CCSNe |
| $Si_3N_4$ | << 1 ppm | CCSNe (100%) |
| Silicates | ~200 ppm | AGB stars and CCSNe |
| Oxides | ~10 ppm | AGB stars and CCSNe |
| Nanodiamond | ~1400 ppm | Ambiguous |

Various presolar mineral grains have been identified (Table 1), including silicon carbide (SiC), graphite, and silicon nitride ($Si_3N_4$), silicates (e.g., olivine), oxides (e.g., spinel), and perhaps nanodiamond. This diverse array of mineral phases reflects vastly varied physicochemical conditions in the extrasolar systems from which presolar grains were originated. In the most primitive extraterrestrial materials, the matrix-normalized abundances of presolar silicates, oxides, SiC, graphite, and $Si_3N_4$ are up to several hundred ppm (e.g., Floss et al. 2013; Nittler et al. 2018a; Nguyen et al. 2022), ~10 ppm (Floss and Haenecour 2016), ~30 ppm (Huss et al. 2003; Davidson et al. 2014), ~10 ppm (Huss et al. 2003), and <<1 ppm (Nittler et al. 1995), respectively. In chondritic materials that experienced secondary processing (i.e., aqueous alteration and/or thermal metamorphism) on their parent bodies, presolar grains could have exchanged their isotopes with surrounding solar system materials, resulting in varying degrees of presolar grain destruction. Among different presolar phases, silicates are the most labile to secondary processing, resulting in their significantly lowered abundances (less than 10 ppm) in highly aqueously altered samples (Leitner et al. 2020; Liu et al. 2020b; Barosch et al. 2022a). The abundances of presolar SiC, graphite, and nanodiamond in chondrites seem not sensitive to aqueous alteration (e.g., Liu et al. 2020b; Barosch et al. 2022a; Nguyen et al. 2024) but significantly affected by thermal metamorphism; these presolar phases are inferred to be essentially absent in unequilibrated ordinary chondrites of petrologic type > 3.8 (Huss & Lewis 1995; Huss et al. 2003). The effect of thermal metamorphism on altering the presolar silicate abundance has not been systematically investigated and remains unknown. In achondritic materials, presolar grains are absent because they were either melted or became isotopically equilibrated during the partial melting of the parent bodies of achondritic materials.

Although nanodiamond is the most abundant among various presolar phases (Lewis et al. 1987), its origin remains enigmatic. Nanodiamond grains are typically small, averaging 2.6 nm in size. Their small sizes hinder isotope analyses of individual grains due to the limited number of atoms within each grain. Bulk noble gas analyses of nanodiamonds (Lewis et al. 1987; Richter et al. 1998) revealed large enrichments in both the heaviest (H) and lightest (L) isotopes of Xe and Te. This HL-enrichment pattern has led to the proposal of a CCSN origin for these nanodiamonds (Lewis et al. 1987; Richter et al. 1998). The proposed CCSN origin, however, is contradicted by the bulk C and N isotopic compositions of nanodiamonds being solar (Russell et al. 1991, 1996; Lewis et al. 2018), which are in great contrast to the wide ranges of C ($^{12}C/^{13}C \approx 5–10^{10}$) and N ($^{14}N/^{15}N \approx 0.05–10^{5}$) isotope ratios predicted for different shells across a CCSN (e.g., Xu et al. 2015). The transmission electron microscopic (TEM) study of nanodiamonds revealed that nanodiamond separates from the Allende and Murchison meteorites are actually mixtures of nanodiamonds and glassy carbon (Stroud et al. 2011). This observation led to the proposal that the HL-noble gases could have been implanted into organic materials such as glassy carbon in the ISM by a supernova shock, which also potentially transformed a fraction of the carbon into nanodiamonds. Due to their ambiguous presolar origin, nanodiamonds will not be further discussed below.

Presolar grains often contain subgrains that have more varied chemical compositions, ranging from common phases such as titanium carbide (TiC) in presolar SiC and graphite to rare phases such as oldhamite (CaS), aluminum nitride (AlN), titanium nitride (TiN), zirconium carbide (ZrC), and Fe,Ni-rich subgrains (Croat et al. 2003; Kodolányi et al. 2018; Singerling et al. 2021; Sanghani et al. 2022).

In terms of grain size, individual presolar grains range from nm to tens of μm. This upper limit reflects the limit of dust growth in stellar environments that is superimposed by the effect of dust destruction in the ISM and solar system. The lower limit, on the other hand, almost certainly reflects the detection limit of state-of-the-art analytical techniques. Most in situ searches of presolar grains were carried out by conducting NanoSIMS isotope analyses with a $Cs^+$ beam of ~100 nm (Fig. 2a-c), which is used for sputtering the sample surface and producing negatively charged secondary ions (Hoppe 2016). Hoppe et al. (2015a, 2017) repeated the search of presolar grains in three primitive chondrites by reducing the $Cs^+$ beam size to 50 nm (by reducing the beam current), and the reduced beam size enabled the identification of a higher abundance of smaller presolar grains. This observation highlights the importance of analytical spatial resolution in controlling the detection efficiency of presolar grains.

The limit of analytical spatial resolution for isotope analyses is less of a problem when searching for presolar grains in a meteoritic acid residue since the acid residue can be suspended in a water/isopropanol solution and dispersed onto a gold foil (Fig. 2d-f). In an ideal case, if a grain were located sufficiently far (≳10 μm) from all other grains, spatial resolution would no longer be a problem for isotope analyses, which instead would be limited by the number of atoms contained within the grain and the mass spectrometer useful yield, i.e., the ratio of detected ions to sputtered atoms. Take graphite for example and assume a C isotopic composition of $^{12}C/^{13}C = 500$,[3] a 1% useful yield for NanoSIMS C isotope analysis (Hillion et al. 1995) translates to 500 ± 456 (1σ error) in the measured $^{12}C/^{13}C$ ratio of a 10 nm graphite grain. The large statistical error results in the measured ratio overlapping with the terrestrial $^{12}C/^{13}C$ ratio at 89, thus making it extremely challenging to detect ≤ 10 nm presolar graphite grains with state-of-the-art analytical techniques. In practice, high analytical spatial resolution is always desired for suppressing terrestrial/asteroidal contamination from adjacent meteoritic grains and presolar grain rims (Fig. 2g-i) so that the intrinsic, true isotopic compositions of presolar grains can be obtained for investigating astrophysical processes without any ambiguity (Nittler et al. 2018b; Leitner and Hoppe 2019; Liu et al. 2021, 2022b, 2022c, 2023, 2024; Hoppe et al. 2023).

[3]Amari et al. (2014) showed that high-density presolar graphite grains exhibit a peak at 500 for their $^{12}C/^{13}C$ ratio distribution.

Oxygen- and C-rich presolar grains are expected to have condensed in different stellar environments. This is because, as a rule of thumb, carbon monoxide (CO) is one of the most stable gas molecules so that it consumes most of the C or most of the O in the envelopes of low-mass stars and in ejecta left by stellar explosions, depending on whether the gas is O- or C-rich. In turn, C-rich dust is predicted to condense around C-rich stars and ejecta (Lodders and Fegley 1995), and vice versa. This led to the common view that presolar SiC and graphite grains originated in C-rich stars and C-rich ejecta and that presolar oxides and silicates from oxygen stars and O-rich ejecta.

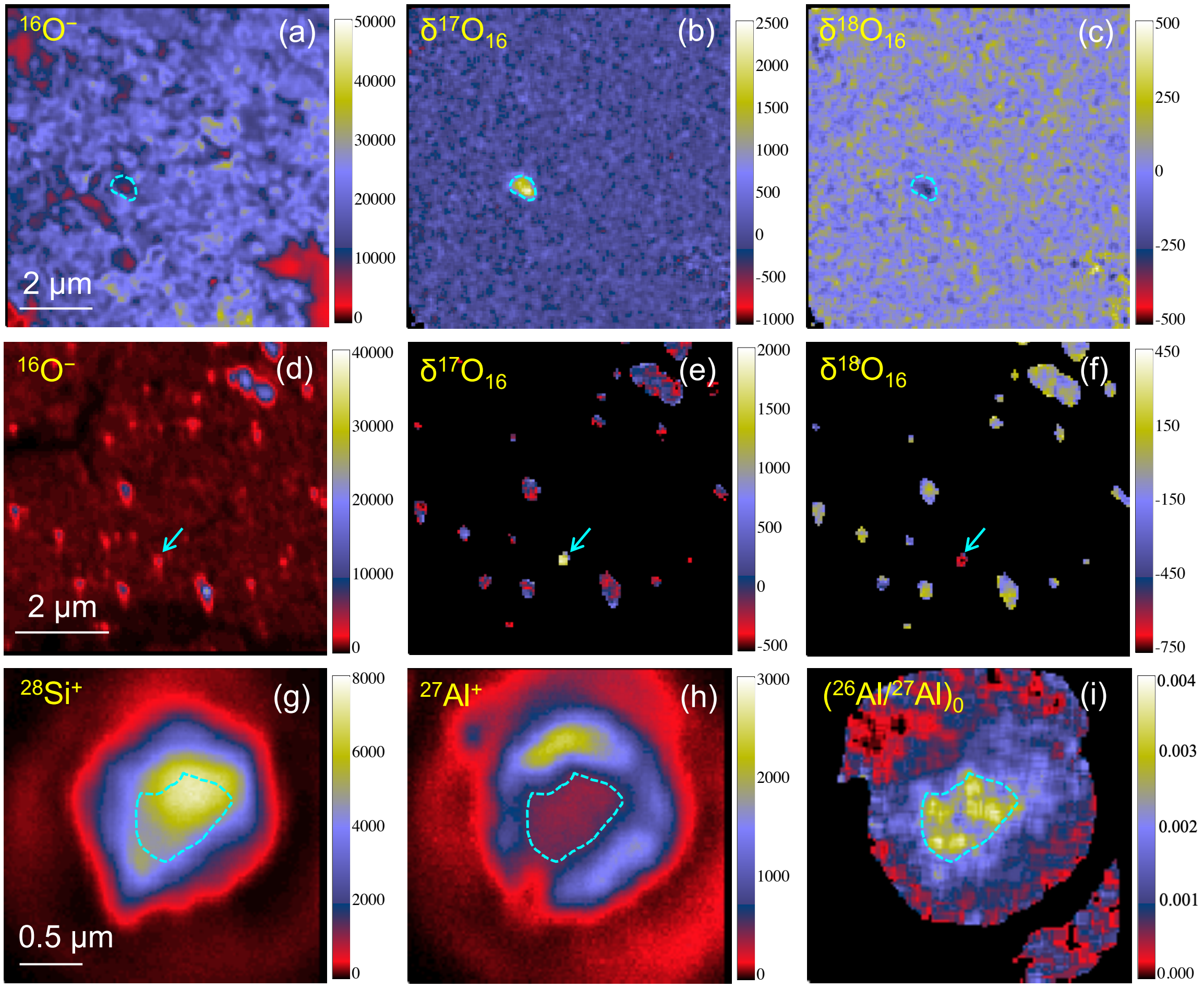

**Figure 2**. *NanoSIMS secondary ion and isotope ratio images of fine-grained matrices in the Antarctic CO3 chondrite Dominion Range 14359 studied by Liu & Ogliore (2021) (panels a-c), the spinel-rich acid residue of Orgueil CI chondrite (panels d-f), and a SiC grain from the SiC-rich acid residue of the CM2 chondrite Murchison (panels g-i). The scalebars for secondary ion images (panels a, d, g, h) and isotope ratio images (panels b, c, e, f) are in the units of counts and permil, respectively. Throughout this chapter,* $\delta^{i}A_{j}$ *(‰) is defined as* $[(^{i}A/^{j}A)_{grain}/(^{i}A/^{j}A)_{SS} - 1] \times$

*1000, in which $(^iA/^jA)_{grain}$ and $(^iA/^jA)_{SS}$ denote the isotope ratios for a grain and the solar system, respectively. The isotope ratio image in panel (i) represents the distribution of inferred initial $^{26}Al/^{27}Al$ ratios within the SiC grain. In panels (a)-(c), a $^{17}O$-rich, $^{18}O$-depleted presolar silicate grain is highlighted by cyan dashed contour lines. In panels (d)-(f), a presolar spinel grain with similar O isotopic composition is highlighted by cyan arrows (Liu et al. 2020b). In panels (g)-(i), the Al-poor core of the presolar SiC grain is highlighted by cyan dashed contour lines. The anticorrelation between the images in panels (h) and (i) points to asteroidal/terrestrial Al contamination in the Al-rich rim of the SiC grain (Liu et al. 2021).*

The fact that the solar system has a C/O ratio of 0.55 (Lodders 2021) eases the identification of presolar C-rich dust in extraterrestrial materials since dust formed in the solar system is predominantly O-rich (Grossman 1972). Almost all SiC grains in the SiC-rich acid residues of carbonaceous chondrites are isotopically anomalous in C, N, or Si, pointing to their circumstellar origins (Zinner 2014). The circumstellar origins of presolar SiC grains are further corroborated by their peculiar polytype structures. More than a hundred different SiC polytypes can be formed in the laboratory, and the formation of these polytypes depends strongly on growth conditions (e.g., temperature, pressure). Daulton et al. (2002, 2023) used TEM to study the microstructures of ~500 presolar SiC grains (with unknown isotopic compositions) in an acid residue of Murchison. They found that cubic SiC (β-SiC) is the dominant polytype (~80%) for presolar SiC grains that consist dominantly of SiC grains from ancient low-mass AGB stars (see discussion in Section 3.1). A few Murchison SiC grains were found to be isotopically normal but of 6H and 15R polytypes (Virag et al. 1992; Liu et al. 2017b). These two SiC polytypes are in contrast with the 2H and 3C polytypes that are commonly observed for isotopically anomalous SiC grains but are typical of SiC synthesized in the laboratory (Daulton et al. 2002, 2003), thus implying terrestrial contamination[4].

Compared to C-rich presolar phases, the detection efficiency of O-rich presolar grains depends more critically on the criteria adopted for identifying presolar O-rich grains, which are like most matrix minerals in chemical composition (e.g., olivine). Different studies adopted different criteria (3σ, 4σ, 5.3σ O isotope anomalies) for identifying presolar O-rich grains (e.g., Haenecour et al.

[4]Note that there are also isotopically anomalous SiC grains of higher order polytypes (e.g., 6H, 15R), which point to unique condensation environments in their parent stars (Liu et al. 2017b; Kodolányi et al. 2018; Singerling et al. 2021).

2018; Nittler et al. 2018a). The O isotope anomaly of a grain that meets the 3σ criterion may not satisfy the 5.3σ criterion, in which case the grain would not be identified as a presolar grain. This means that the adoption of different identification criteria must have contributed to the varying presolar O-rich grain abundances observed among chondritic materials, which thus cannot be used to directly probe the initial distribution of presolar grains across the solar nebula or the effect of secondary processing. Unified presolar grain identification criteria and analytical procedure are desired in future presolar grain studies to maximize the science return. On the other hand, the chosen criteria for identifying O-rich presolar grains could have caused missing presolar spinel grains with close-to-solar O isotopic compositions. How many presolar O-rich grains are missed owing to this isotopic-anomaly-based identification method or the analytical precision, is still an open question. Nittler (2009) argued based on O isotopes that presolar oxides all came from >1.15 $M_\odot$ O-rich stars, whose surface envelopes are predicted to exhibit $^{17}$O excesses with respect to the solar system. This argument stems from the fact that, given their long lifetimes, lower-mass stars (<1.15 $M_\odot$) would not end their lives before the formation of the solar system. Identifying presolar O-rich grains based on other isotope systematics, coupled with subsequent O isotope analyses, will provide more clues to test this proposal.

## 3. PRESOLAR GRAINS AND STELLAR PROCESSES

Presolar grains are often divided into different types based on their isotope ratios, and each type is believed to represent dust from a distinctive type of stars (Zinner 2014; Nittler and Ciesla 2016). The classification of these grains into different groups is often empirical and arises from observable clustering in multiple isotope ratio diagrams (e.g., Fig. 3c), which suggests common nucleosynthesis processes and, in turn, a common stellar origin for each cluster (group) of grains. Assigning a particular group of presolar grains to a specific stellar source entails a complex, iterative process that integrates both macroscopic telescope observations and stellar nucleosynthesis model calculations for many different isotope systematics (see discussions in detail in following sections). This attribution is especially challenging when there is no clear diagnostic isotopic signature that can unequivocally link the grains to particular types of stars. In many cases, the inferred sources remain ambiguous due to overlapping isotopic characteristics between different stellar environments, highlighting the intrinsic complexities and limitations of our current understanding of the stellar origins of presolar grains. Given the challenges associated with definitively linking presolar grains to their stellar sources, acquiring multielement isotopic data from each grain becomes essential. Such multielement isotope grain data provide multiple observables that help to constrain the large number of parameters in stellar nucleosynthesis models and, consequently, allow for a more robust cross-examination of their proposed stellar origins.

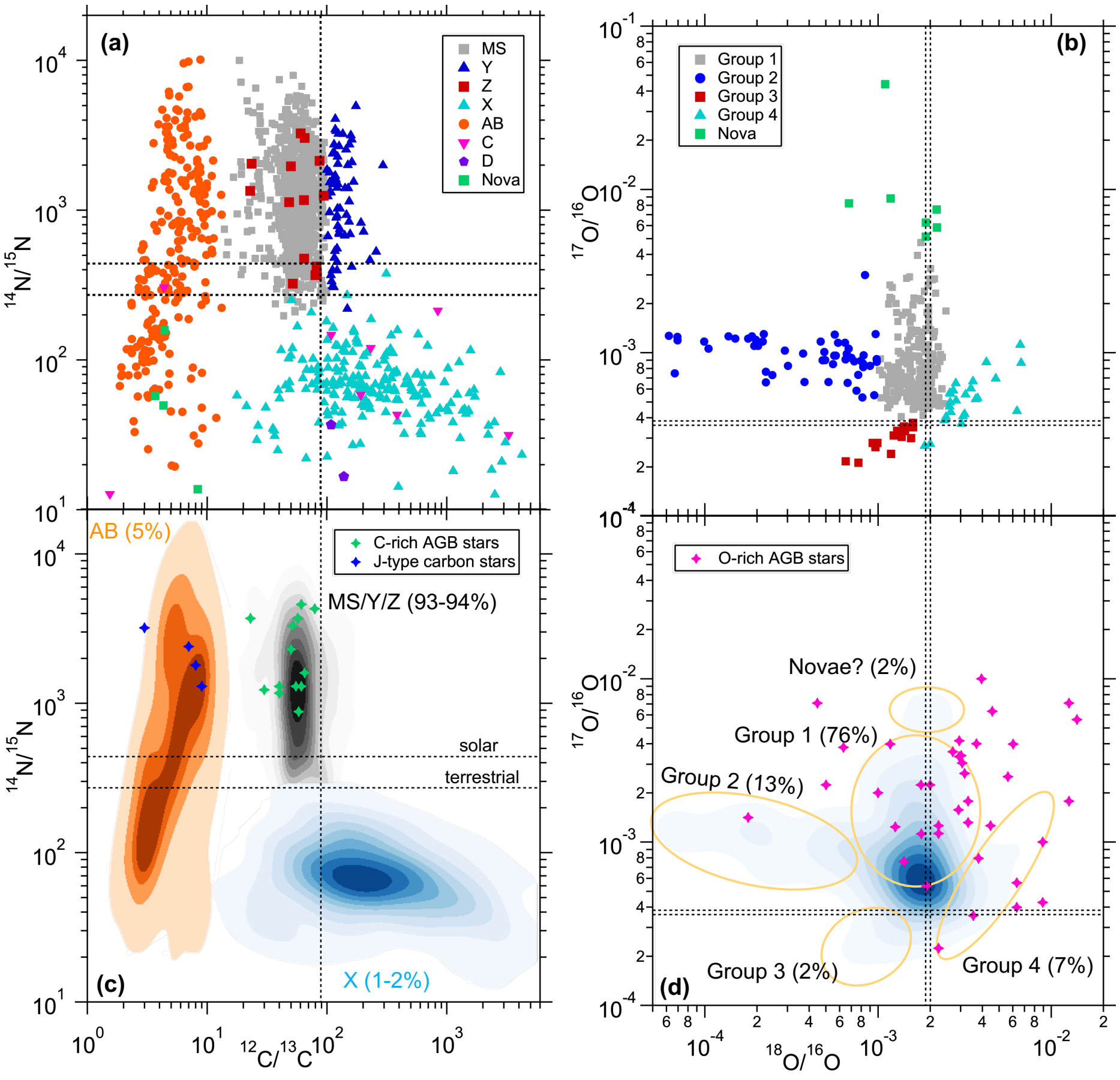

**Figure 3**. *Isotope plots of presolar SiC (panels a and c) and oxide (panels b and d) data. In panel (a), plotted are presolar SiC grain data compiled in the Presolar Grain Database (PGD) (10/30/2023 version; Stephan et al. 2024). Included are the PGD SiC grains with (i) complete C, N, and Si isotope data and (ii) the 1σ errors for Si isotope ratios ≤ 15 ‰ (≤ 100 ‰ for Type C grains). In panel (c), the color maps are the density distributions of AB, MS (and Y and Z), and X presolar SiC grains in panel (a), and the magenta stars denote the compositions of carbon stars reported by Hedrosa et al. (2013). Given their common origins in AGB stars (see discussions in Section 3.1.3), the MS, Y, and Z grains are combined in panel (c). In panel (b), plotted are the literature presolar oxide grains compiled in Liu et al. (2022b). In panel (d), the color map represents the density distribution of the presolar oxides in panel (b), and the magenta stars denote the compositions of O-rich AGB stars of up to 4.5 M☉ (mostly < 2 M☉) recently reported by Lebzelter et al. (2015) and Hinkle et al. (2016); also shown is the classification scheme for presolar O-rich grains given in Nittler et al. (2020). The terrestrial composition is shown as dashed lines in all panels, and the solar composition is also included when it differs from the terrestrial composition. The Sun is inferred*

*to have a 40% $^{15}N$ depletion (Marty et al. 2011) and 6% $^{16}O$ enrichment relative to Earth (McKeegan et al. 2011) as shown in panels (b) and (d). All the density maps (in linear scale) were generated using the seaborn kdeplot function (version 0.11.2) in Python (with default parameter values). The average 1$\sigma$ errors in stellar observations for $^{12}C/^{13}C$, $^{14}N/^{15}N$, $^{17}O/^{16}O$, and $^{18}O/^{16}O$ ratios are 30%, 66%, 32%, and 33%, respectively. The relative percentage of each type is given in the parenthesis after the type's name in panels (b) and (d).*

The isotope data for presolar SiC and oxide grains are shown in Fig. 3 for comparison with stellar observations since their isotope data are the least affected by isotopic dilution (see Section 2.2). In presolar SiC grains, Si and C are two major elements with the presence of up to several wt.% N (Huss et al. 1997). Thousands of presolar SiC grains have been measured for their C, N, and Si isotope ratios (Stephan et al. 2024), which vary by orders of magnitude and are used for grouping the grains. Presolar SiC grains have been divided into five main (≥ 1%) types, including mainstream (MS), Y, Z, AB, and X (Fig. 3a). Additional rare (<0.1%) types include putative nova, C, and D grains. Multielement isotope data for presolar SiC grains suggest that (*i*) MS, Y, and Z grains were all sourced from low-mass C-rich AGB stars but perhaps with different initial stellar metallicities, (*ii*) X, C, and D grains from CCSNe, and (*iii*) the stellar origins of AB and putative nova grains, both of which are extremely $^{13}C$-rich, remain ambiguous, and their proposed stellar origins include J-type carbon-stars, born-again AGB stars, CCSNe, and novae.

Thousands of presolar graphite grains have also been examined for isotopes of C, N, O, and Si, in addition to the major element C. The isotope ratios of trace elements (N in particular) in a significant fraction of presolar graphite, however, resemble solar/terrestrial values (Amari et al. 2014). This likely implies isotopic equilibration with ISM, asteroidal, and/or terrestrial materials in presolar graphite. To date, there is no established classification scheme based on isotope ratios for presolar graphite grains. These grains are typically categorized into high-density (>2.1 g/cm$^3$) and low-density (<2.1 g/cm$^3$) groups, a division determined through density separation during the acid extraction procedure for graphite (Amari et al. 1994). Multielement isotope data for presolar graphite grains have pointed to the general trends that high-density graphite grains are dominated by AGB dust and that the percentage of supernova graphite grains increases with decreasing density (Amari et al. 2014).

Presolar O-rich grains have been divided into five groups based on their O isotope ratios (Fig. 3b). Based largely on O isotopes, the conventional wisdom is that Group 1 and Group 2 grains primarily originated in low-mass (and intermediate-mass) RGB/AGB stars, Group 3 grains mostly in lower-metallicity RGB/AGB stars with a minor contribution from CCSNe, and Group 4 grains in CCSNe (Nittler et al. 2008; Lugaro et al. 2018; Palmerini et al. 2021b). Unlike presolar SiC grains and C-rich AGB stars (Fig. 3a), Fig. 3b reveals that the distribution of presolar oxides has a single compact core without clear separation between the defined groups, thus suggesting that the classification scheme for presolar O-rich grains based on O isotopes is not statistically significant. Indeed, Group 1 presolar silicates (Leitner and Hoppe 2019; Hoppe et al. 2021) and spinel grains (Zinner et al. 2005; Nittler et al. 2008; Gyngard et al. 2010) with similar O isotopic compositions can exhibit different Mg isotopic signatures, pointing to their different stellar origins. Furthermore, presolar oxides and silicates show both similarities and differences in the O isotope ratios (Floss and Haenecour 2016), and the differences are in part due to the fact that in situ isotope analyses of presolar silicates in fine-grained matrices are more likely to sample surrounding asteroidal materials than isotopic analyses of individual presolar oxides from meteoritic acid residues, leading to more diluted isotopic signatures of the former (Fig. 2).

In the following sections, the stellar origins of presolar grains will be discussed from the perspectives of stellar nucleosynthesis calculations and astronomical observations when available. Our following discussions will examine the shared stellar origins and diverse nucleosynthesis processes across presolar grains of various phases.

### *3.1. AGB Stars*

Type MS presolar SiC grains, the dominant type of presolar SiC grains ($\gtrapprox$ 85%), are inferred to have come from low-mass C-rich AGB stars. The supporting evidence includes the *s*-process isotopic signatures commonly found in MS grains (Savina et al. 2003; Barzyk et al. 2007; Liu et al. 2015) and the ubiquitous presence of SiC dust around such stars (Blanco et al. 1998; Speck et al. 2005, 2009). The AGB stellar origins of MS grains are further supported by the consistent C and N isotope ratios of MS grains and C-rich AGB stars (Liu et al. 2021). Although the MS grains in Figs. 3a,c have lower $^{14}N/^{15}N$ ratios on average than the C-rich AGB stars, this inconsistency was shown to have been caused by sampling asteroidal/terrestrial N contamination during grain analyses (Liu et al. 2021). Liu et al. (2021) further showed that after suppressing the N contamination the intrinsic N isotope data for MS grains are shifted to higher values and in line with the observations, thus further validating the AGB stellar origins of MS grains.

In addition to MS grains, Y and Z grains, two rare types of presolar SiC grains (about 1−5% each), are inferred to also have come from C-rich AGB stars but with lower initial stellar metallicities and/or higher initial masses based on light-element isotope data (Hoppe et al. 1997; Amari et al. 2001b; Zinner et al. 2007; Nguyen et al. 2018). The abundances of Y and Z grains were shown to increase with decreasing grain size (Hoppe et al. 1997; Zinner et al. 2007). Y grains are defined to have $^{12}C/^{13}C \geq 100$ (Fig. 3a), and Z grains, on average, have lower $^{29}Si/^{28}Si$ but higher $^{30}Si/^{28}Si$ ratios than MS and Y grains (Fig. 4a). Boujibar et al. (2021) and Hystad et al. (2022) recently applied natural clustering analysis techniques to examine the classification scheme for presolar SiC grains from a statistical perspective. These clustering analysis algorithms analyze a large set of data based on the statistics of data distribution without any prior knowledge of the related physics. The different algorithms employed in these studies generally differentiated MS, Y, and Z grains from the other types of presolar grains but failed to further distinguish between the three types. The results thus suggest that the current classification schemes for MS, Y, and Z grains are somewhat arbitrary and not statistically significant. The three groups of grains are thus combined in Fig. 3c to illustrate the C and N isotope distributions of dust from AGB stars.

In addition to MS, Y, and Z SiC grains, AGB stars are also believed to be the main stellar sources of high-density (>2.1 g/cm$^3$) graphite grains (e.g., Amari et al. 2014) and O-rich grains, including the majority of Group 1, Group 2, and Group 3 grains (e.g., Nittler et al. 2008).

The dominance and diversity of presolar grains identified as originating from AGB stars highlight the broader significance of these stars in shaping our understanding of the chemistry and evolution of our Galaxy. AGB stars play a pivotal role, not only as substantial contributors of dust and carbon to the Galaxy but also as essential sites for slow neutron-capture (*s*-process) nucleosynthesis that is responsible for producing half of the heavy elements beyond iron. Given the important role of AGB stars in stellar nucleosynthesis and their significant contributions to GCE, our discussion of the stellar origins of presolar grains will begin with those that originated from AGB stars.

### *3.1.1. Stellar Evolution and Nucleosynthesis*

Below, I will first introduce the terminology that is relevant to the evolution of low- to intermediate-mass stars to assist the discussion in this section. The reader is referred to Busso et al. (1999), Herwig (2005), and Straniero et al. (2006) for more in-depth descriptions of stellar evolution. Main sequence stars undergo H-burning to produce energy. In $\gtrapprox 1.3\ M_{\odot}$ stars[5], the CNO cycle occurs via a series of proton-capture reactions at the main sequence stage in the core to facilitate the transformation of H nuclei into He, producing isotopes like $^{13}C$, $^{14}N$, $^{17}O$ and depleting isotopes like $^{15}N$ and $^{18}O$.[6] When a main sequence star exhausts the H fuel in its core, the core starts to contract due to gravity. The core contraction releases potential energy that can heat up a shell around the core and eventually ignite H-burning in the shell, where the CNO cycle starts to operate. The energy produced by the H-burning causes the surface layers to expand outwards, resulting in lowered surface temperature (i.e., redden color) and enlarged stellar radius. This stage of stellar evolution is thus known as the red giant branch (RGB) for stars with initial masses of 0.5 – 8 $M_{\odot}$. Shortly after ascending the RGB, the star undergoes an episode of deep convection that is known as the first dredge-up (FDU), which changes the stellar surface composition by mixing partial H-burning products from deep layers of the star into the surface envelope. When the temperature in the core increases to approximately $3 \times 10^8\ K$ because of core contraction, the core begins He-burning, i.e., triple-alpha reaction. The He-burning eventually exhausts He and leaves behind C and O ashes in the core that is surrounded by a He shell. As the temperature is insufficient to fuse C and O in the core, the core contracts again. The core contraction heats up the outer He shell and eventually ignites He-burning in this shell, at which point the star evolves to the early AGB phase. After the He-burning exhausts the He fuel, the star reaches the thermally pulsing AGB phase.

A thermally pulsing AGB star consists of a partially degenerate CO core, a H shell and a He shell that burn alternatively, a thin region between the H and He shells (named He intershell), and a surface H-rich convective envelope. During the thermally pulsing AGB phase, H-burning is activated for most of the time in the H shell above the growing He intershell. The He-intershell is heated and compressed until the temperature and density at its bottom are high enough to ignite He-burning in the He shell, resulting in a thermonuclear runaway, i.e., a thermal pulse (TP). During a TP, He-burning products are mixed from the inner to outer layers of the He shell. At the quenching of a TP, the He shell becomes radiative, and $^{4}He$ is (almost) fully converted into $^{12}C$. After some time, if the H shell has been sufficiently lifted by the TP, the convective envelope penetrates the underlying region and brings newly synthesized materials to the surface. This process is named third dredge-up (TDU). Eventually, H-burning becomes reactivated in the H shell because of star contraction and the consequent temperature increase. This episode corresponds to the interpulse phase of an AGB star. An AGB star experiences strong stellar winds at the surface as discussed earlier, and the mass loss rate increases over time. When the mass of the H-rich envelope becomes very small, the star leaves the AGB phase toward the white dwarf cooling sequence.

[5]Throughout the chapter, $M_{\odot}$ denotes solar mass and $Z_{\odot}$ solar metallicity, and metallicity is defined as the mass fraction of elements heavier than H and He.

[6]During the hot CNO cycle that occurs at sufficiently high temperatures (e.g., in ONe novae), $^{15}N$, instead of being destroyed, is abundantly synthesized (José and Hernanz 1998).

In low-mass ($\lessapprox$ 3 M$_\odot$) AGB stars, during the interpulse phase the *s*-process occurs in the He intershell, which is powered by a main neutron source, the $^{13}C(\alpha,n)^{16}O$ reaction, at a low neutron density (~$10^{7}$–$10^{8}$ cm$^{-3}$). During the subsequent TP, *s*-process products can be further modified at an enhanced neutron density ($10^{9}$–$10^{10}$ cm$^{-3}$) once a minor neutron source, the $^{22}Ne(\alpha,n)^{25}Mg$ reaction, becomes partially activated in the He-intershell. This high-density neutron exposure controls the production of nuclides that are affected by *s*-process branch points, at which neutron capture rates are comparable to beta decay rates (Käppeler et al. 2011; Bisterzo et al. 2015). With repeated TDU episodes, C and *s*-process products that are produced in the He-intershell are brought up to the surface, thus increasing the surface C/O ratio and *s*-process isotope enrichments. The surface C/O ratio can increase to above unity for some low-mass (1.5 $M_\odot \lessapprox$ M $\lessapprox$ 3 $M_\odot$) AGB stars, enabling the formation of graphite and SiC dust grains in the cooling gas that is lost from the surface. Because of decreased TDU efficiency and/or increased dilution, lower-mass (0.5 M$_\odot$ $\lessapprox$ M $\lessapprox$ 1.5 $M_\odot$) and intermediate-mass AGB stars (3 $M_\odot \lessapprox$ M $\lessapprox$ 8 $M_\odot$) remain O-rich in the stellar envelope during the AGB phase, thus producing O-rich dust grains such as silicates and oxides. In intermediate-mass AGB stars, the $^{22}Ne(\alpha,n)^{25}Mg$ reaction plays a more dominant role in producing *s*-process nuclides because this reaction operates more efficiently at enhanced stellar temperature. Numerical modeling work suggested that the *s*-process in intermediate-mass AGB stars could have contributed to up to 15% of Sr, Y, and Zr in the solar system (Gallino et al. 2000) and that heavier *s*-process elements up to Bi in the solar system were dominantly contributed by the *s*-process in C-rich low-mass AGB stars (Arlandini et al. 1999; Bisterzo et al. 2014). In summary, AGB stars play a crucial role in governing the chemical (e.g., C, *s*-process isotopes) evolution of the Universe, including the chemical makeup of our solar system; they also serve as prime laboratories for investigating nucleosynthesis processes in which different physical and chemical processes work simultaneously on different timescales.

### *3.1.2. Scientific Challenges*

Despite the abovementioned consensus among AGB stellar models, the accuracy and precision of AGB nucleosynthesis calculations are directly controlled by uncertainties in the relevant nuclear reactions and involved stellar parameters (e.g., mass loss rate). The model calculations also suffer from "unknown" uncertainties as stellar models provide only a basic description and important physical mechanisms (e.g., convection) are sometimes oversimplified and others (e.g., rotation) are often neglected in the models. For instance, physical models based on different mechanisms have been adopted to (*i*) treat nonconvective mixing processes (e.g., cool bottom processing) through regions in which the energy transport is radiative in low-mass RGB/AGB stars and (*ii*) to introduce partial mixing of H from the convective envelope into the underlying radiative He-intershell for forming the major neutron source $^{13}$C via the reaction chain $^{12}C(p,\gamma)^{13}N(\beta^{+})^{13}C$. The former is related to the production of light elements (up to Al) in low-mass RGB/AGB stars, and the latter controls the production of heavy elements (mainly from Sr to Bi) via the *s*-process in low-mass AGB stars.

Cool bottom processing, sometimes also called extra mixing or deep mixing in the literature (deep mixing hereafter), refers to the circumstance where material is taken from the outer convective envelope of a low-mass star, transported to hot regions inside the star where it undergoes nuclear processing and then returned to the convective envelope (Wasserburg et al. 1995). The occurrence of deep mixing in a low-mass star can lead to enhanced enrichments in CNO burning products (e.g., $^{13}$C, $^{14}$N) and $^{26}$Al (if temperature is sufficiently high) at the stellar surface (Gilroy and Brown 1991; Nollett et al. 2003). Deep mixing has been proposed to occur because of thermohaline mixing in RGB stars (Eggleton et al. 2006). RGB stars indeed were observed to exhibit unusually low $^{12}C/^{13}C$ ratios, ascertaining the occurrence of the deep mixing process during the RGB phase (Gilroy and Brown 1991). Palmerini et al. (2017) proposed that the deep mixing process could be alternatively induced by magnetic buoyancy instabilities on the RGB/AGB, thus raising the possibility of producing larger isotopic shifts (at higher temperatures) by more efficient deep mixing occurring at the AGB stage. In comparison to deep mixing in low-mass stars, hot bottom burning (HBB) refers to the circumstance where the bottom of the convective envelope of an intermediate-mass star reaches temperatures that are sufficiently high for efficient proton capture reactions to take place, producing similar light-element isotopic signatures as deep mixing but to larger extents. The efficiency of HBB varies among state-of-the-art stellar models from different research groups mainly because the efficiency is affected by the physical inputs that affect surface properties such as the effective temperature, e.g., by the treatment of convection in stellar envelopes (Ventura and D'Antona 2005).

In addition to deep mixing and HBB, the formation of the major neutron source $^{13}$C in the He-intershell is a top fundamental unknown that is directly related to the *s*-process nucleosynthesis. The formation of $^{13}$C requires the $^{12}$C($p$,$\gamma$)$^{13}$N($\beta^{+}$)$^{13}$C reaction chain to take place in the He-intershell, thus pointing to mixing of H from the convective envelope border into the underlying He-intershell that is initially H free but contains abundant $^{12}$C produced by He-burning. Also, the amount of H that is mixed into the He-intershell needs to be limited (i.e., partial mixing of H) because, otherwise, $^{13}$C would undergo further proton capture via $^{13}$C($p$,$\gamma$)$^{14}$N and, consequently, reduce the effective neutron abundance in the He-intershell. Over the last several decades, multiple physical mechanisms have been proposed to account for the partial mixing of H into the underlying He-intershell, including overshooting (Freytag et al. 1996; Herwig et al. 1997; Straniero et al. 2006; Cristallo et al. 2009; Battino et al. 2016), gravity waves (Denissenkov and Tout 2003), rotation induced instabilities (Herwig et al. 2003; Piersanti et al. 2013), and magnetic buoyancy (Nucci and Busso 2014; Trippella et al. 2016). It is still highly controversial which is the mechanism primarily responsible for the partial mixing of H. Existing low-mass AGB nucleosynthesis models commonly include different physical prescriptions (based on different abovementioned physical mechanisms) to allow for the formation of $^{13}$C in the He-intershell. These models (hereafter standard AGB models), however, typically do not predict the occurrence of deep mixing in low-mass AGB stars (Cristallo et al. 2009, 2011; Karakas & Lugaro 2016; Battino et al. 2019), unless special treatment of this extra-mixing process is included (e.g., Palmerini et al. 2017). Below, I will highlight the unique role of presolar grains from AGB stars in constraining uncertainties in AGB nucleosynthesis model calculations and recent progress in this research area.

### *3.1.3. Constraints Derived from Presolar Grain Data*

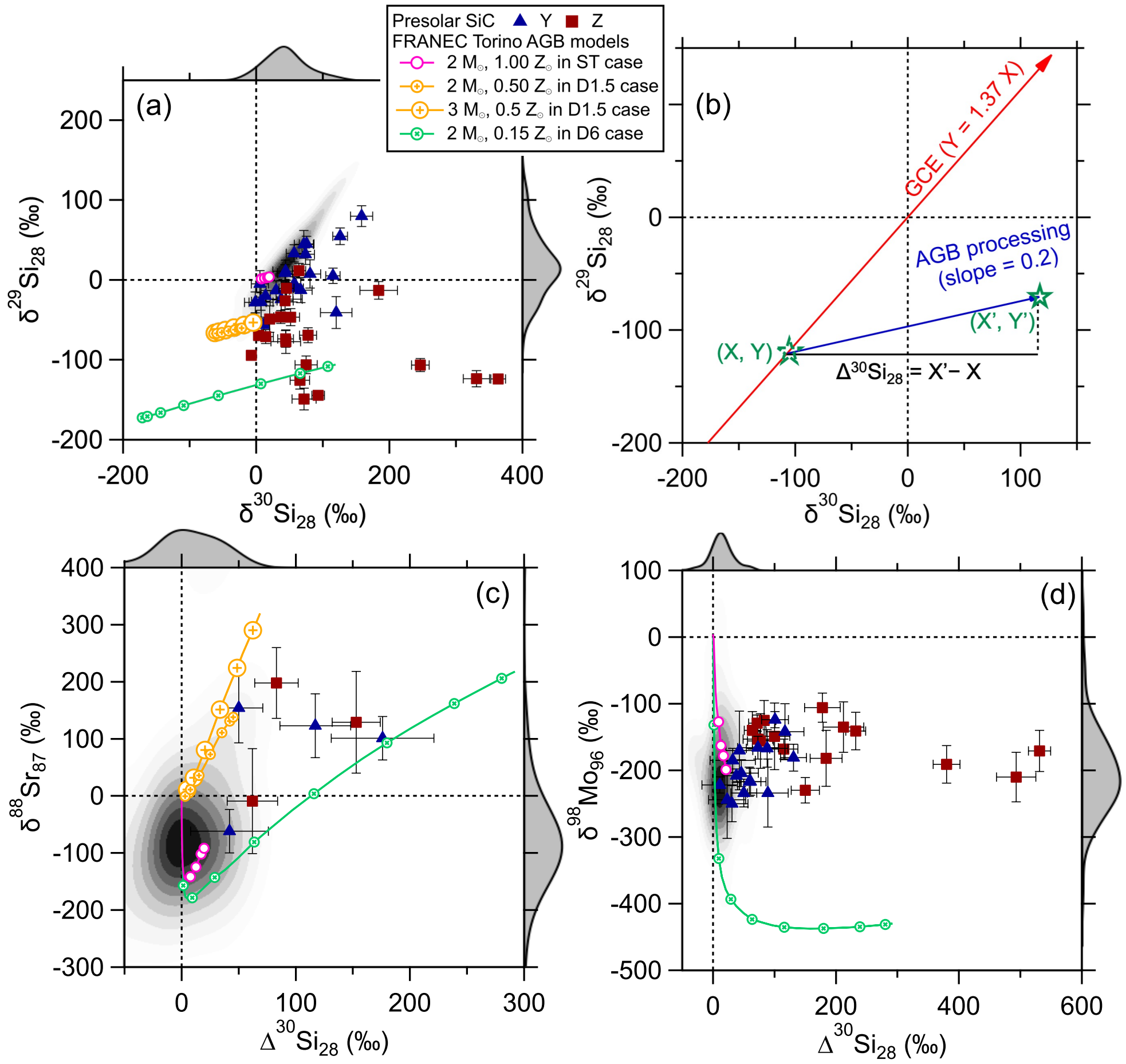

**Figure 4**. *Isotope plots comparing MS (density maps) with Y and Z grains for their light- and heavy-element isotopic compositions. The grey histograms in each plot denote the distribution of MS grain data from the PGD. The heavy-element isotope data for MS grains shown in panels (c) and (d) are from Barzyk et al. (2007), Liu et al. (2015, 2017c), and Stephan et al. (2018, 2019), and the data for Y and Z grains are from Liu et al. (2022e) with their corresponding Si isotope ratos plotted in panel (a). The grain data were chosen to have consistent ranges of $\delta^{84}Sr_{87}$ (<–500 ‰) and $\delta^{92}Mo_{96}$ (<–500 ‰) values to suppress the effect of contamination. Error bars are all 1σ. In panel (b), illustrated is the definition of $\Delta^{30}Si$, which represents the $^{30}Si$ excess produced by the s-process in AGB stars. Also plotted in panels (a)–(c) are postprocessing Torino AGB model calculations for the envelope compositions of AGB stars with varying initial stellar masses and metallicities. The AGB models treated the amount of H that is partially mixed into the He intershell as a free parameter, and the resulted $^{13}C$ densities in the D1.5 and D6 cases are divided by factors of 1.5 and 6, respectively, with respect to the standard (ST) case (Gallino et al.*

*1988). The O-rich phase is denoted by lines, and the C-rich phase by lines with symbols, each of which represents three TPs. The AGB stellar surface composition evolves from the adopted initial stellar composition toward C and s-process enrichments with repeated occurrence of TPs and TDUs.*

Modern isotope analyses of presolar grains can reach percent level precisions for many elements, ranging from light elements such as C (Hoppe 2006) to heavy elements such as Mo (Stephan et al. 2019). The achieved precision in the laboratory is more than an order of magnitude smaller than those of stellar spectroscopic observations (e.g., Hedrosa et al. 2013; Hinkle et al. 2016). Notably, isotope data for a wide range of elements, including C, N, Si, Mg-Al, S, Ti, Fe, Ni, Sr, Zr, Mo, Ru, and Ba, have been obtained in individual MS grains that range from submicron to tens of microns in size (see the PGD for details; Stephan et al. 2024). Compared to light elements ($A \leq 56$), the advantage of heavy elements ($A > 56$) is that some of their isotopes can be significantly produced by the *s*-process in AGB stars so that the related AGB model predictions are barely affected by the initial stellar composition. For instance, in a 2 $M_{\odot}$, 1 $Z_{\odot}$ AGB star, the *s*-process produces only up to 2% increase in the $^{30}$Si abundance at the stellar surface (Fig. 4a) but ~400% increase in the $^{96}$Mo (pure *s*-process isotope) abundance; in turn, compared to $\delta^{30}Si_{28}$, the predicted $\delta^{i}Mo_{96}$ values are much more dependent on the *s*-process production of Mo isotopes in AGB stars and less on the initial stellar composition (Liu et al. 2022a).

These high-precision multielement isotope data for presolar grains are thus unique and invaluable tools for scrutinizing stellar nucleosynthesis model calculations, based on which stringent constraints can be derived on relevant nuclear reaction rates and stellar parameters. However, while stellar observations have direct information on the properties (e.g., stellar mass) of targeted stellar objects, presolar grains came from ancient stars that died prior to the solar system formation and do not exist anymore. Thus, the properties of the parent stars of presolar grains are inferred, indirect information, and the associated uncertainties always need to be taken into consideration when using grain data to derive constraints on stellar processes.

#### 3.1.3.1. Type MS Presolar SiC Grains

Even for MS grains, the best-known type of AGB dust, the properties of their parent stars are highly controversial (Lugaro et al. 2018, 2020; Cristallo et al. 2020). MS grains are inferred to have come from AGB stars with initial masses of ~1.5 – 4 $M_{\odot}$, mainly because stellar models predict that such stars become C-rich in the envelope during the AGB phase so that SiC can form in their gas outflows (Lodders and Fegley 1995). However, the predicted initial stellar mass range for C-rich AGB stars is stellar-code dependent and varies among models developed by different research groups (Cristallo et al. 2011; Karakas and Lugaro 2016; Pignatari et al. 2016). For instance, these stellar models do not agree upon the upper limit for C-rich AGB stars at near solar metallicity.

The initial metallicities of the parent stars of MS grains are more ambiguous. Figure 4a shows that MS grains exhibit a wide range of $\delta^{29}Si_{28}$ values (Fig. 4a), in contrast to the limited increases in $\delta^{29}Si_{28}$ predicted for AGB stellar nucleosynthesis with respect to the adopted initial compositions. Thus, the range of $\delta^{29}Si_{28}$ values (~200‰) among MS grains was suggested to reflect variations in the initial compositions of their parent AGB stars as a result of GCE (Timmes and Clayton 1996). Since both the Si isotope ratios are expected to increase with increasing stellar metallicity (Fig. 4b), this implies that MS grains were mainly sourced from higher-than-solar metallicity AGB stars. Chemical heterogeneities in the Galaxy, however, could have blurred a simple, positive correlation between the Si isotope ratios and metallicity expected from homogenous GCE (Lugaro et al. 1999; Lewis et al. 2013), in which case the Si isotope ratios of MS grains cannot be used to directly infer the initial metallicities of their parent stars. Given the well-correlated Si and Ti isotopic ratios of MS grains, Nittler (2005) argued that the effect of heterogeneous GCE should be limited.

Recently, Lugaro et al. (2020) concluded that large MS grains (> 1 μm) came dominantly from higher-than-solar metallicity AGB stars by comparing the $^{88}Sr/^{86}Sr$ ratios of MS grains with the observed *s*-process enrichments of barium star[7] based on AGB stellar nucleosynthesis models. Cristallo et al. (2020) investigated the mass and metallicity distributions of the parent AGB stars of presolar SiC grains (~0.2 μm in size on average) in the solar system by coupling a Milky Way chemodynamical model with AGB stellar nucleosynthesis models and a SiC dust production model, based on which the authors concluded that presolar SiC grains predominantly originated in 2 $M_\odot$ AGB stars with slightly subsolar metallicities. Despite uncertainties in modeling parameters, the conclusion of Cristallo et al. (2020) is in general agreement with that of Gail et al. (2009), in which the authors investigated the same problem by coupling a different set of models. The tight mass and metallicity distributions inferred by the modeling work is in line with the finding that MS grains show a compact core in the distributions of their C and Si isotope ratios (Boujibar et al. 2021), which implies similarities in the properties (i.e., stellar mass and metallicity) of their parent stars. Given the different ranges of grain size investigated, the seemingly contradictory conclusions of Lugaro et al. (2020) and Cristallo et al. (2020) could be consistent if the initial stellar metallicity of the parent AGB stars of presolar SiC grains decreases with decreasing grain size. In summary, the current consensus is that MS grains originated in C-rich AGB stars with initial masses of ~1.5–4 $M_{\odot}$ and initial metallicities of ~1–2 $Z_{\odot}$ (the latter of which is potentially grain-size dependent).

[7]Barium stars are binary systems in which the *s*-process enrichments observed at the surface of a star are believed to have been transferred from the other star when it was at the C-rich AGB stage.

Regarding MS grains, the distributions of their intrinsic $^{12}C/^{13}C$, $^{14}N/^{15}N$, and inferred initial $^{26}Al/^{27}Al$ ratios[8] peak at ~50–60, ~2000, and ~0.001, respectively (Liu et al. 2021). The latter two distributions often have long tails toward lower values, reflecting the effects of asteroidal/terrestrial N and Al contamination. In addition to N and Al, isotope ratios of S, Fe, Sr, Mo, and Ba in presolar SiC grains are also susceptible to asteroidal/terrestrial contamination (Barzyk et al. 2007; Hoppe et al. 2015b; Trappitsch et al. 2018), whose effects need to be considered when conducting data-model comparison. Compared to MS grain data, standard AGB stellar nucleosynthesis models for low-mass C-rich AGB stars, which do not predict the occurrence of deep mixing, generally predict higher $^{12}C/^{13}C$ ratios at the stellar surface (e.g., Karakas & Lugaro 2016). Different studies showed that the MS grain data can be quantitatively explained by the AGB models if the initial $^{12}C/^{13}C$ ratio at the beginning of the AGB phase is greatly lowered, reflecting the common occurrence of deep mixing during the RGB phase (Zinner et al. 2006). Alternatively, the low $^{12}C/^{13}C$ ratios of MS grains could reflect that deep mixing occurred in their parent stars on the AGB, resulting from magnetic-buoyancy-induced instabilities (Palmerini et al. 2017).

[8] The inferred initial $^{26}Al/^{27}Al$ ratios reported in Liu et al. (2021) should all be increased by a factor of two, as suggested by Liu et al. (2024), in which the authors reported a factor of two decrease in the SIMS Al/Mg relative sensitivity factor (RSF) for SiC with respect to the value adopted by Liu et al. (2021). In turn, the new RSF value suggests that the peak of the initial $^{26}Al/^{27}Al$ ratios for MS grains should be revised to ~0.002.

A recent review on the heavy-element isotopic compositions of MS grains is given by Liu et al. (2022a) in detail, and this topic will be only briefly mentioned here. As discussed earlier, heavy-element isotopic compositions of MS grains can suppress the imprints of GCE, providing much more independent constraints on AGB stellar processes as compared to light-element isotopic compositions. Taking Mo isotopes for example, the Mo isotopic compositions of MS grains were shown to reflect dominantly the effect of AGB *s*-process nucleosynthesis and agree better with AGB models that adopt neutron-capture cross-sections recommended in the KADoNiS[9] v1.0 database with respect to those in KADoNiS v0.3 (Liu et al. 2019; Stephan et al. 2019). The differences in the neutron-capture cross-section values recommended in the two databases result from different normalization approaches based on the same set of experimental data (Liu et al. 2019). The presolar grain Mo isotope data have motivated new measurements of the neutron-capture cross-sections of Mo isotopes at the GELINA n_TOF facility in Belgium (Mucciola et al. 2023). The obtained preliminary data have been shown to allow an improved agreement with the presolar grain data by AGB model calculations. Furthermore, AGB model predictions for Sr, Zr, and Ba isotopes are sensitive to the two neutron sources for the *s*-process (Lugaro et al. 2003; Liu et al. 2014a, 2014b, 2015), and the MS grain data have been used to constrain the $^{13}C$ formation in the He-intershell (Liu et al. 2015, 2018a; Battino et al. 2016) and the efficiency of the $^{22}Ne(\alpha,n)^{25}Mg$ reaction in low-mass AGB stars (Liu et al. 2014b; Lugaro et al. 2014; Lugaro et al. 2018). Based on the same stellar code (known as FRUITY models), Vescovi et al. (2020) showed that AGB models that consider magnetic-buoyancy-induced $^{13}C$ formation provide significantly improved match to the heavy-element isotopic compositions of MS grains than the corresponding models that consider overshooting-induced $^{13}C$ formation. The underlying reason is that the former leads to deeper mixing of H into the He-intershell with more slowly decaying H abundance (which follows a power law) than the latter that introduces H mixing following an exponential decay (see Vescovi et al. 2020 for details). The magnetic AGB nucleosynthesis models also provide a more satisfying explanation to the F abundance observed for AGB stars of varying stellar metallicities (Vescovi et al. 2021). In comparison, Lugaro et al. (2018, 2020), based on a different stellar code

[9]KADoNiS stands for Karlsruhe Astrophysical Database of Nucleosynthesis in Stars. The version 0.3 database is available at https://www.kadonis.org/ (Dillmann et al. 2006), and the version 1.0 database at https://exp-astro.de/kadonis1.0/ (Dillmann 2014).

(known as Monash models), concluded that 2 $Z_\odot$ AGB stellar nucleosynthesis models that adopt the overshooting physical model can also explain the heavy element isotope data of MS grains. It remains to be seen whether adoption of the formula for magnetic buoyancy can further improve the data-model agreements for 2 $Z_\odot$ and lower-metallicity Monash models, given uncertainties in the initial metallicities of the parent stars of MS grains as discussed above. Magnetic buoyancy also needs to be systematically implemented in stellar codes to test its simultaneous effects on inducing the $^{13}$C formation in the He-intershell and deep mixing on the RGB/AGB. The reader is referred to Busso et al. (2021) and Palmerini et al. (2021a) for more discussions on the effects of magnetic buoyancy on AGB stellar evolution and nucleosynthesis. In conclusion, the correlated light- and heavy-element isotope data of MS grains have been invaluable in providing stringent constraints on relevant AGB stellar parameters.

#### 3.1.3.2. Types Y and Z Presolar SiC Grains

It is commonly believed that Y and Z grains originated from low-metallicity AGB stars based on their light-element isotope ratios, notably their large $^{30}$Si and $^{50}$Ti excesses (Hoppe et al. 1997; Amari et al. 2001b; Zinner et al. 2007; Nguyen et al. 2018). Heavy-element isotope data have been recently obtained for Y and Z grains and provided new insights into their stellar origins. Isotope ratios of Sr, Ba, and Mo (Fig. 4) were obtained for tens of Y and Z grains (Liu et al. 2019, 2022e) using the new generation of resonance ionization mass spectrometer (RIMS) instrument, the CHicago Instrument for Laser Ionization (CHILI) (Stephan et al. 2016). These isotope data are characterized by strong *s*-process isotope enrichments, thus corroborating the AGB stellar origins of Y and Z grains. However, it is challenging for AGB stellar nucleosynthesis model calculations to simultaneously explain the light- and heavy-element isotope data of Y and Z grains as illustrated in Fig. 4. Lower-than-solar metallicity AGB models generally predict larger $\Delta^{30}$Si values because of the increased efficiency of the $^{22}$Ne($\alpha$,$n$)$^{25}$Mg reaction at enhanced stellar temperatures (Zinner et al. 2007), thus explaining the increased $\Delta^{30}$Si values of Y and Z grains as compared to MS grains. These low-metallicity models also predict large increases in $\delta^{88}Sr_{87}$ and moderate decreases in $\delta^{98}Mo_{96}$ with decreasing metallicity, which, however, are not observed in Y and Z grains as compared to MS grains. The increases in $\delta^{88}Sr_{87}$ result from the increasing neutron/Fe ratio with decreasing metallicity (i.e., decreasing Fe abundance) if the amount of the major neutron source $^{13}$C is fixed (Liu et al. 2015, 2022e). The decreases in $\delta^{98}Mo_{96}$ are caused by changes in the relative

$^{98}$Mo and $^{96}$Mo production by the *s*-process at different stellar temperatures with varying stellar metallicity (Liu et al. 2019). Liu et al. (2022e) concluded that the $\delta^{88}Sr_{87}$ values of Y and Z grains could be explained if the amount of $^{13}$C were greatly lowered in the parent low-metallicity AGB stars of Y and Z grains than in those of MS grains (Fig. 4c). However, the comparable $\delta^{98}Mo_{96}$ values of MS, Y, and Z grains (Fig. 4d) cannot be explained by uncertainties in these stellar parameters since the changing $\delta^{98}Mo_{96}$ value with varying metallicity is mainly controlled by the nuclear properties of Mo isotopes, specifically changes in the $(n,\gamma)$ reaction rates in relevant AGB temperature regime (Liu et al. 2019). The new neutron-capture cross-sections of Mo isotope obtained at the GELINA n_TOF facility are currently under analysis (Mucciola et al. 2023) and will likely provide new insights into this puzzle.

In summary, the stellar origin of Y and Z grains currently remains an open question. While the low-metallicity AGB stellar origins of Y and Z grains are generally supported by their Si, Ti, Sr, and Ba isotopes, their Mo isotope data clearly argue against such an origin. New neutron-capture cross-section measurements of $^{95,96,97,98}$Mo using state-of-the-art facilities are needed to examine the energy dependence of these neutron-capture reactions across relevant AGB stellar temperatures. Another puzzle regarding the low-metallicity stellar origin of Y and Z grains is why their parent stars had much reduced amounts of $^{13}$C than low-metallicity AGB stars existing today (Busso et al. 2001). This could be a hint that the solar system sampled gas and dust that were lost and ejected from atypical stellar objects. In the future, the challenges will be to obtain correlated Sr, Zr, and Ba isotope data with better statistics and better precisions for Y and Z grains to constrain the mechanism that controls the $^{13}$C formation in their parent stars.

#### 3.1.3.3. High-Density Graphite Grains

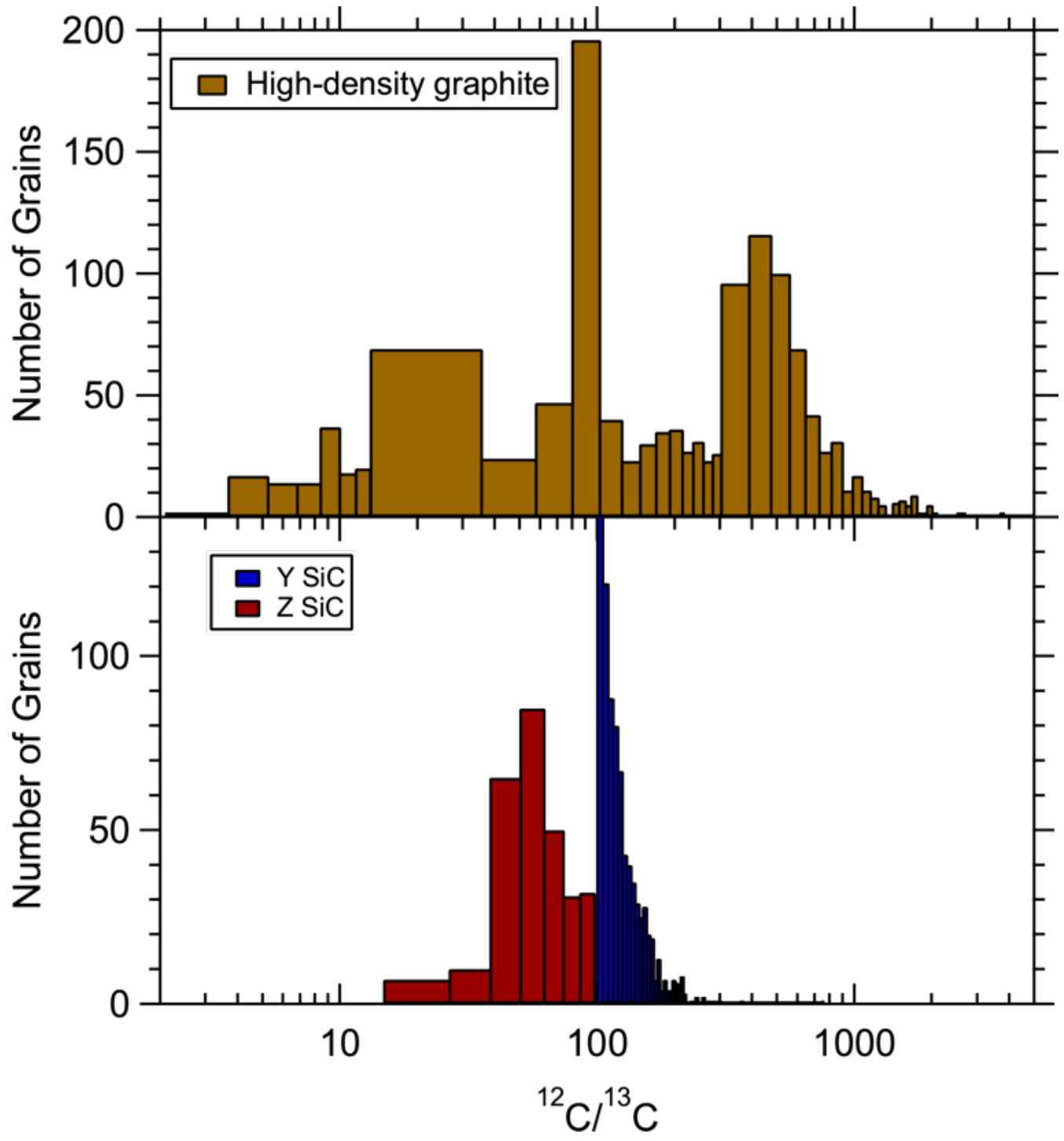

**Figure 5**. *Histograms comparing the $^{12}C/^{13}C$ ratios of high-density graphite, and Y and Z SiC grains, all of which were proposed to have originated in low-metallicity AGB stars. The Y and Z grain data are those complied in the SiC PGD (Stephan et al. 2024), and the high-density graphite data from the recently recompiled graphite PGD (Stephan and Trappitsch 2024). High-density graphite data exhibit a dominant peak at the solar $^{12}C/^{13}C$ ratio of 89, likely caused by asteroidal/terrestrial contamination (Amari et al. 2014).*

Bulk high-density (>2.1 g/cm$^3$) graphite grains exhibit *s*-process Kr isotopic signatures, suggesting that these graphite grains dominantly came from C-rich AGB stars (Amari et al. 1995). Based on individual grain data for the C isotope ratio, Amari et al. (2014) further constrained the mass (1–3 $M_\odot$) and metallicity ($\leq 0.30\ Z_\odot$) distributions of the parent stars for high-density graphite grains. Compared to Y and Z grains, a significantly larger fraction of high-density presolar graphite grains show high $^{12}C/^{13}C$ ratios with a dominant peak at ~500 (Fig. 5), contradicting the common view based on presolar SiC grain data that low-metallicity AGB stars have reduced $^{12}C/^{13}C$ ratios. Among the three types of presolar AGB SiC dust, Z grains have the largest $\Delta^{30}Si$ values (Fig. 4a), which are commonly believed to reflect that Z grains were sourced from AGB stars with the lowest metallicities. Z grains, however, do not show $^{12}C/^{13}C$ ratios higher than MS and Y grains, and their $^{12}C/^{13}C$ ratios are instead comparable to and lower than those of MS grains and Y grains, respectively (Fig. 3a). Standard AGB models (without deep mixing) generally predict increasing $^{12}C/^{13}C$ ratios with decreasing stellar metallicity, but AGB model predictions for $^{12}C/^{13}C$ are directly affected by the efficiency of the deep mixing process, whose dependence on stellar metallicity, however, is not well understood. Different studies showed that the $^{12}C/^{13}C$ ratios of Z grains could be quantitively explained by low-metallicity AGB models if the initial $^{12}C/^{13}C$ ratios at the beginning of the AGB phase adopted in the models were lower than those for MS grains (Zinner et al. 2006). This could imply that deep mixing occurs more efficiently in low-metallicity AGB stars than in near-solar-metallicity AGB stars. This implication is supported by the observation that very metal-poor stars always reach the equilibrium value of the CNO cycle ($^{12}C/^{13}C \approx 3.5$) on the RGB (Recio-Blanco and de Laverny 2007). It is plausible that, compared to Z grains, high-density graphite grains came from much lower-metallicity AGB stars so that their parent stars could have reached higher $^{12}C/^{13}C$ ratios at the stellar surface, as suggested by Amari et al. (2014). This scenario, however, seems not supported by the existing heavy-element isotope data. Nicolussi et al. (1998) and Pal et al. (2022) reported Mo and Zr isotope data for high-density graphite grains, and the grain data generally resemble the isotopic signatures of corresponding MS grain data. For instance, in addition to decreased $\delta^{98}Mo_{96}$ (Fig. 4d), enlarged $\delta^{100}Mo_{96}$ value is also expected for AGB stars with lower initial metallicity because the branch point at $^{99}Mo$ ($t_{1/2}$ = 66 h) can become more efficiently activated in low-metallicity AGB stars owing to higher neutron density exposure released by more efficient operation of the $^{22}Ne(\alpha,n)^{25}Mg$ reaction at enhanced stellar temperatures (Liu et al. 2019); however, the Mo isotope data for two dozens of high-density graphite grains from Nicolussi et al. (1998) do not show any decreases in $\delta^{98}Mo_{96}$ or increases in $\delta^{100}Mo_{96}$ as compared to MS SiC grains. In addition, low-metallicity AGB origins have been clearly ruled out for some high-density graphite grains based on (*i*) the inferred initial presence of $^{44}Ti$ ($t_{1/2}$ = 60 a) in a few of these grains (Jadhav et al. 2008, 2013a) and (*ii*) the finding of extremely anomalous Ca and Ti isotopic ratios in some of the grains with $^{12}C/^{13}C$ ratios < 20 (Jadhav et al. 2013b).

In summary, like Y and Z grains, the proposed low-metallicity stellar origin of high-density graphite grains remains ambiguous. In the future, the challenge will be to determine correlated light- and heavy-element isotope data with better statistics for Y, Z, and high-density graphite grains while suppressing the effect of contamination. This will enable scrutinizing their proposed low-metallicity stellar origins by directly comparing the isotopic composition of Y, Z, and high-density graphite grains.

#### 3.1.3.4. O-rich Grains of AGB Stellar Origin

For a long time, Group 1 and Group 2 O-rich grains (Fig. 3b) were thought to have primarily come from low-mass ($\lesssim$2.0 $M_\odot$) O-rich RGB/AGB stars (Nittler 2009). Since the mass loss rate increases significantly from the RGB to AGB phase, this implies that the two groups of grains mainly condensed out of the cooling gas that was lost during the AGB phase. Compared to O-rich AGB stars (Fig. 3b), Group 1 grains overall show lower $^{17}O/^{16}O$ and $^{18}O/^{16}O$ ratios. Hinkle et al. (2016) ascribed the higher $^{18}O/^{16}O$ ratios of present-day O-rich AGB stars to the result of GCE over the past ~4.6 Ga, suggesting that the parent stars of Group 1 O-rich grains had lower initial stellar metallicities than such stars present today. One, however, should be cautious about this conclusion, given that the stellar $^{18}O/^{16}O$ ratios reported in Hinkle et al. (2016) may suffer from larger uncertainties because the ratios were derived based on a single CO molecular line.

Different studies suggested that the large $^{18}O$ depletions of Group 2 grains resulted from the occurrence of deep mixing in the RGB phase (Wasserburg et al. 1995; Nollett et al. 2003; Palmerini et al. 2011; Lugaro et al. 2017). This is supported by the large $^{18}O$ depletion observed in a low-mass O-rich AGB star (albeit with large uncertainties) as shown in Fig. 3b. The origin of Group 2 grains, however, is equivocal. Lugaro et al. (2017) showed that, by adopting a recent $^{17}O(p,\alpha)^{14}N$ reaction rate, AGB stellar nucleosynthesis models for intermediate-mass stars (4−8 $M_{\odot}$) can also account for the O isotopic compositions of Group 2 grains. That Group 2 and possibly some Group 1 presolar O-rich grains came from intermediate-mass AGB stars, alleviates the conundrum that intermediate-mass AGB stars are expected to have contributed significantly to the presolar O-rich dust reservoir (Ferrarotti and Gail 2006; Zhukovska et al. 2008) but no presolar O-rich grains were inferred to originate in intermediate-mass AGB stars prior to the work of Lugaro et al. (2017). Although intermediate-mass stellar origins are plausible, recent studies (Palmerini et al. 2021b; Liu et al. 2022b) stressed that low-mass AGB stellar nucleosynthesis models, in which deep mixing takes place as a consequence of magnetic buoyancy, can equally account for the O and Mg-Al isotopic compositions of Group 1 and Group 2 oxides, especially those with low inferred initial $^{26}Al/^{27}Al$ ratios. Note that the predicted contribution of intermediate-mass AGB stars to the presolar O-rich dust reservoir suffers from uncertainties in two poorly constrained parameters in the models, including the mass loss rate of AGB stars and the metallicity and mass distributions of stars in the solar neighbor at ~4.6 Ga (Zhukovska and Henning 2013). Thus, it remains an open question what percentage of presolar O-rich grains truly came from intermediate-mass AGB stars.

Compared to Group 1 and Group 2 grains, the AGB stellar origin of Group 3 grains remains less certain. Their $^{16}O$ excesses are attributed to either GCE effects that are linked to the low metallicities of their parent stars, and/or $^{16}O$ overproduction in CCSNe. Although it is proposed that the majority of Group 3 grains originated in low-metallicity RGB/AGB stars primarily based on O isotopes (Nittler et al. 2008), the inferred initial presence of $^{44}$Ti in a Group 3 grain, along with its distinctive Ca isotopic pattern, unequivocally points to its CCSN origin (Nittler et al. 2011). Given the sparse multielement isotope data for Group 3 grains in the literature, more isotopic studies of Group 3 grains are needed to investigate what percentage of the grains came from each of the two stellar sources and whether the percentage depends on mineral phase, e.g., oxides versus silicates.

Magnesium-Al isotope data have been obtained for a number of Group 1 and Group 2 oxides and silicates, providing important clues to cross-examine their proposed stellar origins. Different studies concluded that the Mg isotopic compositions of Group 1 and Group 2 spinel grains are dominantly controlled by the GCE of Mg isotopes and the decay of $^{26}$Al (Zinner et al. 2005; Nittler et al. 2008; Gyngard et al. 2010; Liu et al. 2022b). These isotopic signatures are in general agreement with AGB stellar nucleosynthesis, thus corroborating their proposed AGB stellar origins (of low- to intermediate-mass) based on O isotopes. In contrast, large $^{25}$Mg excesses (~500−1000 ‰) were identified in ~10% of Group 1 silicate grains (Hoppe et al. 2021) and in a few Group 2 silicate grains (Leitner and Hoppe 2019; Verdier-Paoletti et al. 2019; Hoppe et al. 2021). This Mg isotopic signature (i.e., much larger enrichments in $^{25}$Mg than in $^{26}$Mg) points to proton capture onto $^{24}$Mg to form $^{25}$Mg via the $^{24}$Mg($p$,$\gamma$)$^{25}$Al($\beta^{+}$)$^{25}$Mg reaction chain at temperatures exceeding ~$9 \times 10^7$ $K$ (Doherty et al. 2014), thus excluding the low-mass AGB stellar origin due to their insufficient stellar temperatures. The reader is referred to a recent review on this topic by Hoppe et al. (2022). Briefly summarizing, grains with small $^{25}$Mg excesses ($\lesssim$500‰) could have originated in AGB stars with low-to-intermediate masses, while two scenarios have been proposed to explain these large $^{25}$Mg excesses ($\gtrsim$500‰), including the HBB process in more massive AGB stars (Nittler 2019) and explosive H burning in CCSNe (Leitner and Hoppe 2019). Explosive H burning in CCSNe has also been proposed to explain the large $^{13}$C, $^{15}$N, and inferred $^{26}$Al excesses observed in Type AB and putative nova SiC grains (Pignatari et al. 2015; Liu et al. 2016, 2017a; Hoppe et al. 2019), which will be discussed in detail in the next section.

In summary, Group 1, Group 2, and Group 3 O-rich grains were likely sourced from multiple stellar sources in addition to low-mass AGB stars, and the relative contribution of these stellar sources to the presolar O-rich dust reservoir probably depends on the mineral phase (e.g., silicates versus spinel). Since presolar O-rich grains with similar O isotopic compositions could have come from different types of stars, their stellar origins need to be scrutinized and validated based on more isotope systems, especially heavy-element isotopes in the future.

### *3.2. Type II Supernovae versus Novae*

The gamma-ray emissions of unstable isotope $^{44}$Ti have been unambiguously and consistently detected in the remnants of CCSNe Cassiopeia A (Iyudin et al. 1994) and SN 1987A (Grebenev et al. 2012). In comparison, the detection of $^{44}$Ti in Type Ia supernova[10] remnants, remains ambiguous and inconsistent (Borkowski et al. 2010; Troja et al. 2014; Lopez et al. 2015; Zoglauer et al. 2015), and the evidence for dust production in Type Ia supernova remnants is scarce (Borkowski et al. 2006). Thus, $^{44}$Ti is considered as the smoking gun to link presolar grains to a CCSN origin, which is also strongly supported by the other isotopic signatures of presolar SiC, $Si_3N_4$, and graphite grains with $^{44}$Ti excesses (Nittler et al. 1995; 1996). Namely, the initial presence of $^{44}$Ti, so far, has been inferred for Type X presolar SiC and $Si_3N_4$ grains with large $^{28}$Si excesses (Nittler et al. 1996; Lin et al. 2010), Type C presolar SiC grains with large $^{29,30}$Si excesses (Hoppe et al. 2012; Xu et al. 2015), a ungrouped[11] (reclassified as a D grain in the PGD) presolar SiC grain (Liu et al. 2018b), and two dozens of low-density and a few high-density presolar graphite grains (Nittler et al. 1996; Amari et al. 2014). These presolar grains are thus the most CCSN dust. Evidence for the initial presence of $^{44}$Ti in presolar oxides and silicates mostly remains elusive due to the small $^{44}$Ca excesses detected, which could equivocally point to GCE effects (Nittler et al. 2008, 2023; Leitner et al. 2022, 2024).

Since nova nucleosynthesis is dominated by explosive H-burning, it was believed that large enrichments in $^{13}$C, $^{15}$N, $^{17}$O, $^{25}$Mg, $^{33}$S, and/or high inferred initial $^{26}Al/^{27}Al$ ratios (≥ 0.01) are diagnostic of a nova origin (Amari et al. 2001a; José and Hernanz 2007; Parikh et al. 2014). Thus, presolar SiC and graphite grains with large $^{13}$C and $^{15}$N excesses and presolar O-rich grains with large $^{17}O$ excesses could have originated in novae and are thus named "nova" grains. This commonly held view, however, has been challenged by both the multielement isotopic compositions of presolar grains of potential nova origin (Nittler and Hoppe 2005; Liu et al. 2016) and CCSN model calculations that suggest explosive H-burning could also occur during CCSNe (Pignatari et al. 2015).

[10]Type Ia supernova is a type of supernovae that occurs in a binary system, which consists of a white dwarf and another star that can be anything from a giant star to a less massive white dwarf (Hillebrandt and Niemeyer 2000).

[11]Ungrouped grains in the literature include grains that do not fit into other types based on their C, N, or Si isotope ratios and grains for which no C, N, or Si isotope ratios were measured. The five D grains in Fig. 6 are U grains in the former case. Currently, all U grains in the PGD (Stephan et al. 2024) are those without any C, N, or Si isotope data.

Mounting observational evidence suggests that many supernova remnants are asymmetrical and chemically heterogeneous on large scales (e.g., Grefenstette et al. 2014; Boggs et al. 2015; Grefenstette et al. 2017). In this regard, presolar CCSN and nova grains should be considered as time capsules that captured the physicochemical condition and isotopic compositions of CCSN and nova ejecta at specific time steps in micron spatial resolution. It remains an open question how to quantitatively compare the compositions of these microscopic stellar fossils to nova and CCSN model calculations, given unknown uncertainties on micron scales in modeling mixing processes that take place during astronomical-scale explosions (Abarzhi et al. 2019). With these caveats in mind, it is noteworthy that presolar CCSN and nova grain data have been shown to act as powerful tools in highlighting the limitations of existing one-dimensional (1D) CCSN and nova nucleosynthesis models, and this will be discussed in detail.

### *3.2.1. Stellar Evolution and Nucleosynthesis*

CCSNe represent the final evolutionary stage of massive stars. A CCSN occurs when its Fe/Ni-rich core, which is left after successive burning stages, cannot burn further to sustain the pressure of the star and experiences neutronization. The star collapses under its own pressure, and the resulting rebound of outer shells onto the Fe/Ni-rich core leads to an outward expanding shockwave that heats the outer layers of the star (Woosley and Weaver 1995). Stellar models predict that the lower mass limit to produce a CCSN lies between 7 and 11 $M_{\odot}$ and that 0.5−1.0 $Z_{\odot}$ stars with initial masses above 25 $M_{\odot}$ may not be able to explode through the presumed core bounce and neutrino driven mechanism (Fryer 1999; Heger et al. 2003). In the latter case, massive stars above 25 $M_{\odot}$ collapse quietly to form black holes and either very faint CCSNe or none at all. In line with the model predictions, observations show that the progenitor mass of the most common type of CCSNe, Type II-P, lies between around 6 and 20 $M_{\odot}$ (Smartt 2009).

Novae are binary systems in which a white dwarf accretes H and He from a companion main sequence star, leading to rapid and episodic burning and ejection of the accreted material (Starrfield et al. 1972). The white dwarf is the leftover core of an AGB star and dominantly consists of either carbon/oxygen or oxygen/neon, depending on the initial stellar mass of the AGB star. The former are known as CO novae and the latter as ONe novae, and CO novae typically consist of lower-mass white dwarfs than ONe novae (José and Hernanz 1998). In the binary system, the white dwarf accretes H-rich material from the surface of the companion star. When the accreted material leads to sufficiently high temperatures at the base of the accreted envelope, it triggers a thermonuclear runaway at the surface of the white dwarf, resulting in nucleosynthesis that is dominated by hot CNO cycle, also known as explosive H burning, typically occurring at around $(1-3) \times 10^8$ $K$ (José et al. 2004; José and Hernanz 2007). After the fuel is exhausted, the white dwarf continues to accrete material from the companion star until another thermonuclear runway is triggered. Such nova outbursts can occur at regular intervals until the material accretion ceases. Given the dominance of explosive H burning, novae and presolar grains from novae are thus prime laboratories for studying proton-induced nuclear reactions in explosive conditions.

CCSN nucleosynthesis products are much more diverse than those of novae in chemical and isotopic compositions because a CCSN experiences nearly all types of stellar nucleosynthesis processes, ranging from fusion reactions to the *p*-process[12], which occur at different evolutionary stages under largely varying conditions. Although CCSNe occur much less frequently than novae, each CCSN ejects significantly a larger amount of processed materials. Gehrz et al. (1998) estimated that, at present, CCSNe contribute much more kinetic energy and processed stellar material to the ISM than novae, thus playing a more crucial role in governing the evolution of the Galaxy.

### *3.2.2. Scientific Challenges*

While novae and CCSNe are spectacular cosmic events, accurate modeling of these astrophysical explosions has been challenging, since it requires the coupling of hydrodynamics, gravity, thermonuclear reactions, and, in some cases, rotation and magnetic fields. These two types of astrophysical events are also characterized by a wide range of length and temporal scales, magnifying the demands on stellar codes to accurately simulate stellar evolution and associated stellar nucleosynthesis. 1D stellar models can simulate entire stellar evolution and investigate modeling uncertainties in a large parameter space but lack self-consistent physical descriptions of many stellar processes such as rotation (Herwig et al. 2003; Piersanti et al. 2013). In comparison, multidimensional stellar models are better at modeling physical processes in a self-consistent way, which, however, come at high computational cost on limited length and temporal scales (Woodward et al. 2019). In recent years, a practical approach is to couple 1D simulations of long, secular quiescent phases during the stellar evolution with multidimensional simulations of short, dynamic phases that could potentially open new nucleosynthesis pathways. This kind of approach has been adopted in recent studies of stellar nucleosynthesis, for example, in rapidly accreting white dwarfs (Stephens et al. 2021) and novae (José et al. 2020). Given the complexity of nova and CCSN nucleosynthesis and the challenges in properly simulating nova and CCSN stellar processes, it is challenging to determine what isotopic signatures can be used to confidently link presolar grains to nova and CCSN origins and what are the smoking guns to distinguish between presolar grains from these two stellar sources. Recent studies on this topic will be summarized and discussed in detail in this section.

[12]The *p*-process is an umbrella term and refers to any nucleosynthesis process that is supposed to be responsible for producing proton-rich nuclei (see Arnould and Goriely 2003 for a review).

Dust condensation in the ejecta of CCSNe and novae is another interesting, controversial topic. CCSNe are naturally thought to be important factories for producing dust observed in the ISM, given the relatively high occurrence frequency and the large mass of condensable elements in the ejecta. Infrared (IR) observations confirmed the formation of dust grains in a few supernova remnants, but, in most cases, it is challenging to place constraints on the dust chemical composition and the inferred amount of dust varies by orders of magnitude (e.g., Pozzo et al. 2004; Indebetouw et al. 2014; Andrews et al. 2016). Model calculations further suggest that dust that initially condensed in the remnants of CCSNe could be efficiently destroyed by the harsh million-kelvin gas of the debris through high-speed collisions with atoms and other grains (Silvia et al. 2010; Biscaro and Cherchneff 2016). The contribution of CCSNe to the ISM dust reservoir thus remains ambiguous. In addition, IR spectral observations showed that the temporal IR evolution of CO novae is controlled by the formation and evolution of dust in the outflow, but the IR spectra of ONe novae indicate little or no dust production during the evolution (Gehrz 1988,1999, 2015; Shore et al. 2018). The more efficient dust production in CO novae is likely a consequence of more mass ejected at lower velocities by CO novae, resulting in favorable conditions for dust condensation (Gehrz 1999). Although observations suggest preferential dust formation in CO nova ejecta, presolar grains of potential nova origin suggest otherwise, and this will be discussed in the next section. Finally, compared to the relatively static physical condition in the gas that is lost from the surface of AGB stars (Lodders and Fegley 1995), gas ejected from stellar explosions is expected to experience more rapid changes in pressure, temperature, and density, in which case dust of more diverse chemical and structural composition could form under kinetic instead of thermodynamic equilibrium conditions (Sarangi et al. 2018; Haenecour et al. 2019). If so, systematic differences are expected in the chemical and structural compositions between presolar AGB grains and those from novae and CCSNe. Recent results on this aspect will be summarized as follows.

*3.2.3. Constraints Derived from Presolar Grain Data*

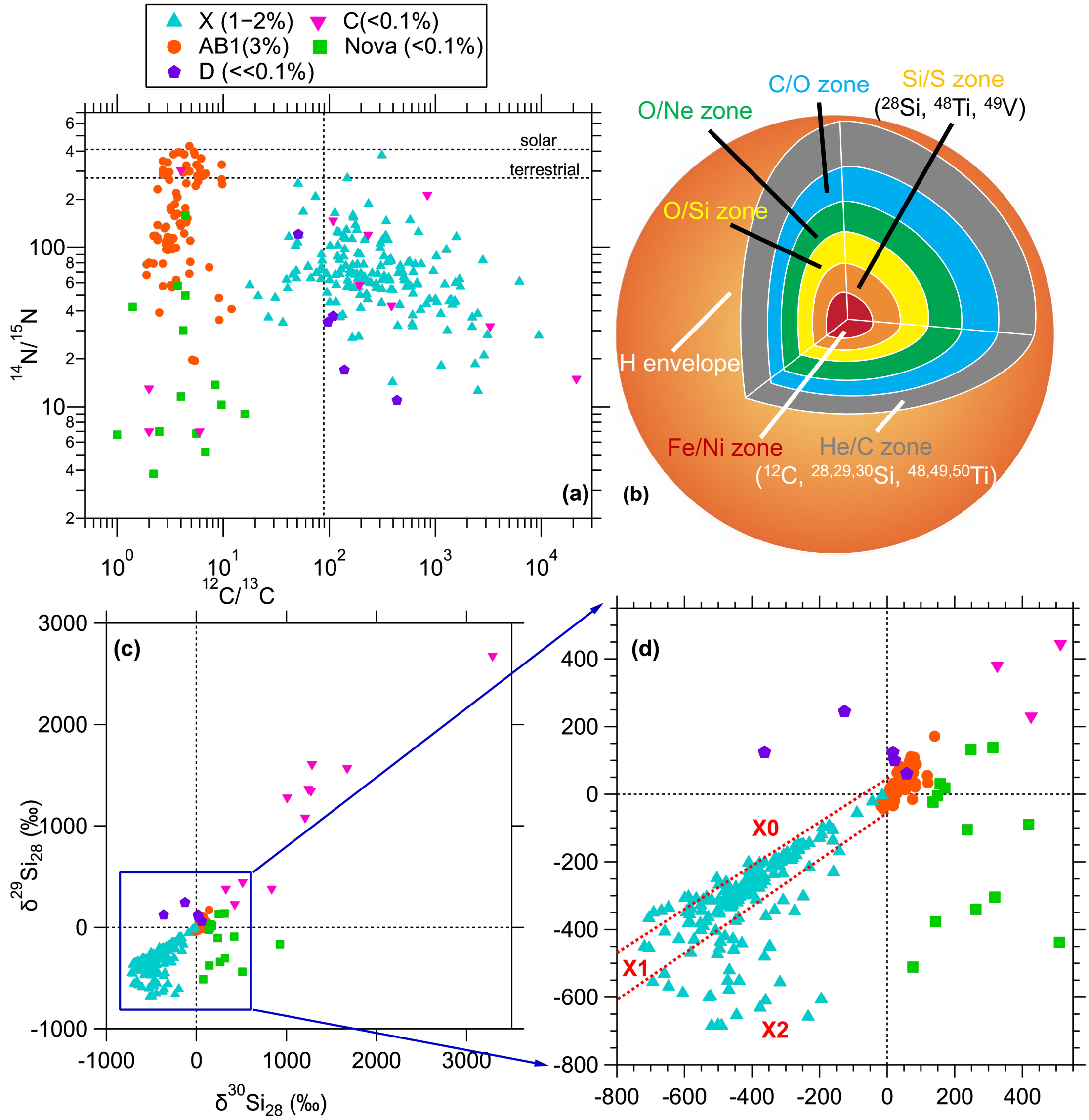

**Figure 6**. *Carbon, N, and Si isotopic compositions of the presolar SiC grains that could have originated in CCSNe and novae. The grain data are from the PGD (Stephan et al. 2024). In panel (b), given is a schematic diagram of the "onion-shell" internal structure of a presupernova massive star, in which zones are labeled by their most abundant elements (Meyer et al. 1995). Neutron capture taking place in the C-rich He/C zone converts* $^{28}Si$ *and* $^{48}Ti$ *to neutron-rich Si and Ti isotopes, respectively, whereas α-capture in the inner Si/S zone overproduces α-nuclides, including* $^{28}Si$ *and* $^{48}Ti$*. Abundant short-lived* $^{49}V$ *(*$t_{1/2}$ *= 330 d) is also made in the Si/S zone. The red dashed lines in panel (d) further divide X grains into three subtypes as proposed by Lin et al. (2010) and later modified by Stephan et al. (2024). In the legend, the number in the parenthesis represents the percentage of each type among presolar SiC grains.*

#### 3.2.3.1. Presolar SiC Grains from CCSNe

1D nucleosynthesis model calculations for CCSN ejecta are characterized by an onion shell structure as illustrated in Fig. 6b, resulting from varying stages of nuclear burning experienced during the presupernova evolution and explosive nucleosynthesis during the explosion (Meyer et al. 1995). In 1D models, this layered presupernova structure is maintained after the explosion because the chaotic mixing process occurring during the explosion, which is implied by the observed asymmetric structures of CCSN remnants, is omitted in the models. It thus requires randomly mixing ejecta from the different CCSN zones shown in Fig. 6b to reproduce the isotopic compositions of CCSN presolar grains based on 1D model calculations, and this method is often known as the *ad hoc* mixing approach (Nittler et al. 1995, 1996). Recently, Schulte et al. (2021) compared CCSN presolar grain data to 3D CCSN model calculations, which include the explosion-induced instabilities and physical mixing. The authors, however, did not investigate whether the multielement isotopic signatures of CCSN grains could be explained by the same CCSN ejecta. It thus remains to be seen whether the same localized ejecta yielded by the 3D models can provide simultaneous good matches to the multielement isotopic compositions of CCSN presolar grains and how CCSN presolar grain data can be used to constrain the 3D models. The following discussion will rely on 1D models to provide constraints on specific CCSN nucleosynthesis paths.

The dominant group of CCSN SiC grains, X grains, are characterized by large excesses in $^{28}$Si, and in some of the X grains also identified are the decay products of the short-lived nuclides $^{26}$Al, $^{41}$Ca ($t_{1/2}$ = 0.1 Ma), and $^{49}$V ($t_{1/2}$ = 330 d). X grains were inferred to have incorporated materials from at least the Si/S zone that produces pure $^{28}$Si and the outer C-rich He/C zone of a supernova that provides a C-rich condition for SiC to condense (Nittler et al. 1996). Additional pieces of evidence reported later (Lin et al. 2010; Liu et al. 2024) also point to the incorporate of Fe/Ni-rich materials. The selective mixing of materials from deep supernova regions with those from outer zones required in this scenario (Fig. 6b) could result from Rayleigh Taylor instability that is predicted to occur during the explosion (Abarzhi et al. 2019). In addition, recent 3D hydrodynamic model simulations suggest that the restricted mixing of Fe/Ni and Si/S materials with those from the outer zones is a natural consequence of shock deceleration and consequent pileup of metal-rich knots from the interior in the He-shell (Hammer et al. 2010; Wongwathanarat et al. 2017). In the literature, the Fe isotopic signatures of X grains were used to argue against the incorporation of Fe/Ni-rich materials (e.g., Marhas et al. 2008) because the CCSN models of Rauscher et al. (2002)

predict large $^{54}$Fe enrichments in the Fe/Ni zone, which have not observed in any of the X grains studied so far. Woosley and Hoffman (1992), however, showed that whether $\alpha$-rich freeze out[13] in the Fe/Ni zone predominantly produces $^{56}$Fe (the decay product of $^{56}$Ni) or $^{54}$Fe depends on the neutron richness there. Given the strong model dependence of $^{54}$Fe/$^{56}$Fe, detailed modeling investigation on the neutron-rich condition of the Fe/Ni zone during the $\alpha$-rich freeze out is needed to explore whether the Fe isotopic data could be explained.

On the other hand, Pignatari et al. (2013b) showed that if the temperature is sufficiently high at the base of the He/C zone, it results in a C/Si zone that is C-rich and contains $\alpha$-capture products such as $^{28}$Si and $^{44}$Ti, which could alternatively account for the unique isotopic signatures of X grains. In this case, no selective mixing is needed, and small-scale, local mixing across the outer zones from the He shell up to the surface is sufficient to explain the light- and heavy-element isotopic compositions of X grains (Pignatari et al. 2013b, 2018; Schofield et al. 2022). However, subsequent studies pointed out several problems in quantitively explaining the isotopic compositions of X grains in this latter scenario (Hoppe et al. 2018; Liu et al. 2018b, 2024), and the explosion energy required to achieve such high temperatures at the base of the He/C zone corresponds to hypernovae that are more massive and rarer than typical CCSNe. It remains an open question whether these problems are robust arguments against the existence of the Si/C zone in CCSNe or reflect uncertainties in current 1D simulations, e.g., asymmetric explosions that are not predicted by 1D models.

X grains, characterized by $^{28}$Si excesses to varying degrees, are divided into three subtypes, including X0, X1, and X2, based on their Si isotope ratios (Lin et al. 2010; Stephan et al. 2024).[14] Among the three subtypes, X1 grains are the most abundant (~70%) and form a tight, linear correlation with a slope of ~2/3 in the Si 3-isotope plot (Fig. 6d). No systematic differences in the isotopic compositions of light elements except for Si have been reported between the three subtypes. Pellin et al. (2000) found that X1 grains are characterized by unique Zr and Mo isotopic compositions, which point to a neutron density environment higher than that for the *s*-process but lower than that for the rapid neutron-capture process (*r*-process). Meyer et al. (2000) named this process "neutron burst", which could be powered by the $^{22}Ne(\alpha,n)^{25}Mg$ reaction during the CCSN explosion either in the He/C zone or in the C/O zone, depending on the explosion energy (Liu et al. 2018c). Alternatively, Bliss et al. (2018) proposed that the "neutron burst" isotopic signatures of X1 grains could be explained by neutron capture taking place in neutron-rich neutrino-driven winds associated with CCSNe. It, however, remains unclear whether the neutron-rich winds that can account for the X1 grain isotopic patterns are C-rich and contain sufficient Si and C so that SiC could condense. Stephan et al. (2018) reported the Sr and Ba isotopic compositions for two X2 grains, which differ substantially from those of X1 grains and point to lower neutron-density environments for X2 grains than for X1 grains. Furthermore, Liu et al. (2020a) showed that the Si isotopic compositions of X1 and X2 grains could be explained by varying explosion energies with lower explosion energies being favored by X2 grain data. Thus, if X2 grains came from CCSNe with lower explosion energies than X1 grains, this would provide a natural explanation to the systematic differences in the Si, Sr, and Ba isotopic compositions between X1 and X2 grains. However, more heavy-element isotopic data are needed for X grains to increase the statistics and consolidate the existence of systematic isotopic differences between the two subtypes. The heavy-element isotopic compositions of X0 grains remain unknown and need to be investigated in the future.

---

[13] In a CCSN, $\alpha$-rich freeze-out occurs when the core briefly reaches extremely high temperatures, causing nuclei to disassemble into alpha particles. As the core rapidly cools and expands, these alpha particles combine with other free nucleons and seed nuclei due to the high neutron and alpha particle densities, leading to the formation of heavier elements beyond iron (Jordan et al. 2003).

[14]Stephan et al. (2024) introduced a new classification scheme for dividing X grains into the subtypes X0, X1, and X2 based on Si isotopes (Fig. 6d), which differs from the scheme of Lin et al. (2010).

Contrary to the large $^{28}Si$ excesses identified in X grains, C grains are characterized by large excesses in the minor Si isotopes, $^{29}Si$ and $^{30}Si$ (Amari et al. 1999). Type C grains are inferred to have also originated in CCSNe based on the observations that some C grains are enriched in $^{32}S$ and exhibit high inferred initial $^{26}Al/^{27}Al$ and $^{44}Ti/^{48}Ti$ ratios (Hoppe et al. 2012; Xu et al. 2015). Type C grains were initially defined to have $^{12}C/^{13}C > 10$, but several grains with $^{12}C/^{13}C \leq 10$ and large $^{29,30}Si$ excesses have been also identified (Fig. 6a). Liu et al. (2016) suggested further dividing C grains into subtypes C1 ($^{12}C/^{13}C > 10$) and C2 ($^{12}C/^{13}C \leq 10$), given the clear gap between these two subpopulations as revealed in Fig. 6a. Different studies have shown that (*i*) the $^{29,30}Si$ excesses of C grains were produced by neutron burst in the He/C zone, (*ii*) the $^{32}S$ excesses of C1 grains resulted from the decay of the radiogenic nuclide $^{32}Si$ ($t_{1/2}$ = 153 year) that was also produced by neutron burst (Pignatari et al. 2013c; Xu et al. 2015) rather than pure $^{32}S$ that was produced by α-capture in the Si/S zone (Hoppe et al. 2012), and (*iii*) the inferred initial $^{44}$Ti in C1 grains could have been produced by α-capture in the Si/S zone (Nittler et al. 1996) and/or in the Si/C zone (Xu et al. 2015). In addition, Liu et al. (2016) showed that the large excess in $^{50}$Ti, the most neutron-rich Ti isotope, observed in a C2 grain is inconsistent with nova stellar nucleosynthesis during which no neutron source is predicted to exist (José et al. 2004); the authors concluded that the $^{50}$Ti excess points to $^{50}$Ti production by neutron burst in the He/C zone, thus implying the CCSN origin of the C2 grain. The different Si isotopic compositions between X and C grains could result from differences in the mixing process during the explosion (Xu et al. 2015) and/or in the properties of their parent CCSNe (Pignatari et al. 2013b).

In the recently recompiled PGD for presolar SiC grains, Stephan et al. (2024) introduced D grains (Fig. 6), which correspond to ungrouped and X0 grains in the literature. The five D grains shown in Fig. 6 were inferred to have been sourced from CCSNe based on their large $^{15}$N and inferred initial $^{26}$Al excesses (Nittler et al. 1996; Liu et al. 2018b) and/or their unique Ca and Ti isotopic compositions (Liu et al. 2018b).

In summary, X, C, and D grains shown in Fig. 6 are the most likely CCSN presolar SiC grains, which have widely varied C and Si isotope ratios by orders of magnitude but consistently lower-than-solar $^{14}$N/$^{15}$N ratios. The stellar production of abundant $^{15}$N requires explosive stellar nucleosynthesis at high temperatures, and CCSN models predict lower-than-solar $^{14}$N/$^{15}$N ratios for all zones beneath the He/N zone (Lin et al. 2010). The consistent $^{15}$N enrichments observed among the most probable CCSN presolar SiC grains are thus naturally expected. On the other hand, their wide ranges of C and Si isotope ratios could result from heterogenous mixing that occurred during the explosion of a CCSN or originating from multiple CCSNe.

CCSNe are believed to act as the major contributor to dust present in galaxies. The detection of 0.1 to 0.5 $M_{\odot}$ of dust in nearby CCSN remnants suggests efficient dust formation (Matsuura et al. 2011; Gomez et al. 2012; Indebetouw et al. 2014), but other observations reveal very little dust in CCSNe in the first few years after the explosions (Pozzo et al. 2004; Stritzinger et al. 2012; Andrews et al. 2016). This phenomenon has been interpreted as continuous buildup of dust over several years after supernova explosions (Gall et al. 2014; Wesson et al. 2015). A test of this idea requires long-term observations of in situ dust formation starting immediately after supernova explosions, which to date are available for only a few supernovae. For instance, it takes three decades to obtain a complete record of the dust formation process in the remnant of SN 1987A, the closest CCSNe observed in 400 years, which predicts that 99% of the dust should have formed after three years (Wesson et al. 2015). A test of whether the dust formation process in SN 1987A is representative of CCSNe in general by spectroscopy, clearly needs to await the explosion of another nearby supernova, which may not happen for hundreds of years. Alternatively, the SN 1987A observation can be immediately tested by presolar grain measurements to infer the timing of dust production.

Isotopic studies of X grains have provided constraints on the SiC dust formation timing after CCSN explosions. Given that SN 1987A observations predict that the majority of the dust should have formed after three years and reached maximum around 30 years (Wesson et al. 2015), the $^{49}V$–$^{49}Ti$ ($t_{1/2}$ = 330 d) and $^{137}Cs$–$^{137}Ba$ ($t_{1/2}$ = 30 a) chronometers can thus be adopted for dating CCSN dust formation. Earlier attempts to use the $^{49}V$–$^{49}Ti$ chronometer, however, produced controversial and conflicting results. The first systematic study of V-Ti isotopes in X grains observed a good correlation between their $\delta^{49}Ti$ and $^{51}V/^{48}Ti$ values, i.e., an isochron, suggesting that X grains condensed within several months of supernova explosions (Hoppe and Besmehn 2002). A subsequent study, however, did not reproduce the isochron, calling the inferred early formation timing into question (Lin et al. 2010). Liu et al. (2018b) observed that X grains show a positive correlation between their $^{49}Ti$ and $^{28}Si$ excesses, which is attributed to the radioactive decay of the short-lived $^{49}V$ to $^{49}Ti$ in the highly $^{28}Si$-rich Si/S zone (Fig. 6b). In addition, given that $^{50}Ti$ is only made in the He/C zone together with $^{49}Ti$ neutron burst, $^{50}Ti$ was used as a proxy to correct for $^{49}Ti$ incorporated into each X grain from the He/C zone (Fig. 6b). Based on this approach, Liu et al. (2018b) avoided conducting *ad hoc* mixing calculations and derived constraints on the condensation timing (> 2 years) of X grains by directly extrapolating the composition of the Si/S zone. Furthermore, based on Ba isotope data, Stephan et al. (2018) concluded that X1 grains condensed >10 years after the explosions of their parent CCSNe and X2 grains within a year. Later, Ott et al. (2019) reached a similar conclusion about the condensation timing (~20 years) of X1 grains by analyzing the literature Ba isotope data. While the Ba-data-derived timing constraint (>10 years) for X1 grains is consistent with the Ti-data-derived constraint (>2 years), the much earlier condensation of X2 grains suggested by Stephan et al. (2018) is, however, not supported by the Ti isotope data (Liu et al. 2018b). In summary, the recent new constraints on the condensation timing of X1 grains, the major subtype of X grains, is in line with recent dust formation model predictions (Sarangi and Cherchneff 2015) and the spectroscopic observations of dust formation in CCSN remnants (Wesson and Bevan 2021; Niculescu-Duvaz et al. 2022). Isotopic studies of more X grains based on more radiometric dating systems are needed to investigate the condensation timings of X1 versus X2 grains. Also, since spectroscopic observations suggest that glassy carbon is one of the dominant dust species in CCSN remnants (Gall et al. 2014; Wesson et al. 2015), dating CCSN presolar graphite grains is needed to compare with the dust condensation timing inferred from the SiC grain data.

In addition to the condensation timing constrained by their isotopic compositions, the structural compositions of X grains provide additional information on the physicochemical conditions for dust formation in CCSN remnants. Liu et al. (2017b, 2022d) performed coordinated Raman and isotopic measurements on ~450 presolar SiC grains, based on which the authors identified systematic differences in the Raman spectral features between MS and X grains. The broadened and lower-shifted Raman peaks observed in a significant fraction of X grains as compared to MS grains led to the suggestion that X grains often condensed more rapidly and at higher atmospheric densities and temperatures (Liu et al. 2017b, 2022d). Coordinated TEM and Raman studies (Liu et al. 2017b; Singerling et al. 2021), however, showed that although the Raman spectrum can generally be used to identify β-SiC, it sometimes misidentifies β-SiC as noncubic or highly disordered β-SiC for reasons that remain unclear. Continued efforts in correlating Raman and TEM results are desired to assess to what extent the Raman information is reliable in inferring SiC structural composition.

Finally, as witnesses to the birth of our solar system, the diversity and population of CCSN presolar grains provide a unique perspective on the formation mechanism for the solar system and hold great potential for addressing the longstanding cosmochemical question whether a shock from a supernova induced the solar system formation (Cameron and Truran 1977; Boss and Keiser 2012; Banerjee et al. 2016). With increasing explosion energy, CCSN model simulations predict that the Si isotopic compositions of X grains become increasingly enriched in $^{29}$Si (Liu et al. 2020a). The model calculations thus suggest that (*i*) X grains lying along the same line in the Si 3-isotope plot (Fig. 6d) must have condensed out of CCSN ejecta of similar explosion energy and (*ii*) X0, X1, and X2 grains, which lie along lines with different slopes in Fig. 6d, could have condensed from CCSN ejecta of different explosion energies. Thus, the fact that X1 grains, lying along the slope 2/3- line in the Si 3-isotope plot, are much more abundant than the other two subtypes, implies that a single CCSN dominantly contributed to the presolar CCSN dust reservoir in the solar system before or during its formation. This lends support to the proposal that a shock from a nearby CCSN induced the solar system formation. Alternatively, it has been proposed that a Wolf-Rayet (WR) star instead of a CCSN is responsible for contributing short-lived nuclides such as $^{26}$Al present in the early solar system (Young 2014). Since model simulations suggest that $^{26}$Al that was ejected into the solar system needs to be carried out by dust grains in the winds from the WR star to the solar system (Dwarkadas et al. 2017), a stringent test to this proposal is to identify presolar grains originating from WR stars in meteorites (Dwarkadas et al. 2020). The contribution of WR stars is likely minor to the presolar O-rich dust reservoir in the solar system. This is because (*i*) among C-, N-, and O-rich WR stars dust formation is observed to be limited around N-rich WR stars (e.g., Lau et al. 2020), (*ii*) O-rich WR stars are rare, and (*iii*) O-rich dust unlikely form efficiently, if not at all, around C-rich WR stars (Ebel 2006). Thus, future studies of C-rich presolar phases such as graphite and SiC are needed to identify isotopic signatures specific to C-rich WR stars. Such studies will offer direct evidence to evaluate the proposal of Young (2014).

#### 3.2.3.2. Presolar SiC Grains from Novae?

Different studies suggested that presolar SiC grains with low $^{12}$C/$^{13}$C ($\lessapprox$ 10) and $^{14}$N/$^{15}$N ($\lessapprox$ 200) ratios along with Z-grain-like Si isotopic signatures (Fig. 6) are potential nova grains (Amari et al. 2001a; José and Hernanz 2007). Earlier studies of putative nova grains pointed out several problems in using 1D nova models to quantitatively explain their isotopic compositions as summarized here. (1) Putative nova grains of both C- and O-rich mineral phases all have much less anomalous isotopic compositions compared to nova ejecta that thus need to be greatly diluted with isotopically normal material (>90%) to match the grain data (Amari et al. 2001a; Nittler and Hoppe 2005; Gyngard et al. 2010; Leitner et al. 2012; Nguyen and Messenger 2014; Liu et al. 2016). (2) Although thermodynamic equilibrium calculations predict that SiC can condense in the innermost ejected shell of ONe novae (José et al. 2004), mixing with >90% solar-like O-rich material would greatly lower the C/O ratio in nova ejecta and thus lower the probability for C-rich phases to condense during nova outbursts. (3) CO novae are more abundant (70–80%) and also more efficient in producing dust than ONe novae (Gehrz 1993), but most of the putative nova grains were inferred to have originated in ONe novae (Amari et al. 2001a; José and Hernanz 2007). Recent studies suggest that the three problems identified in these earlier studies could have been caused by nova stellar modeling uncertainties. Iliadis et al. (2018) investigated this problem by randomly sampling over realistic ranges of modeling parameters and found multiple solutions (i.e., different sets of stellar parameter values) to explain the multielement isotopic compositions of a number of putative nova grains by their predicted CO nova ejecta that require no dilution. Bose and Starrfield (2019) reached a similar conclusion by comparing the literature putative nova SiC grain data to a new set of CO nova models (Starrfield et al. 2020). Bose and Starrfield (2019) also argued that Types C2 and AB1 grains, both of which are characterized by large $^{13}$C and $^{15}$N excesses, could also have been sourced from CO novae. A caveat is that Iliadis et al. (2018) and Bose and Starrfield (2019) focused mainly on the C, N, and Si isotope data that are available for the majority of putative nova SiC grains; however, one of their most probable CO nova grains exhibits $^{44}$Ca and $^{49}$Ti excesses, and one of the C2 grains has a large $^{50}$Ti excess, all of which argue against a nova origin. Isotope data for S, Ca-Ti, and heavier elements are thus urgently needed to distinguish between putative nova grains from CO novae and CCSNe.

Since some, if not all, putative nova, C2 grains, and AB1 grains clearly came from CCSNe (Savina et al. 2007; Nittler and Hoppe 2005; Liu et al. 2016, 2023), it raises the question how CCSN ejecta could attain such low $^{12}$C/$^{13}$C ratios ($\lesssim$ 10) as observed in these grains. According to the 1D models of Woosley and Heger (2007), the CCSN shell that can reach $^{12}$C/$^{13}$C below 10 lies above the He/C zone. This shell is located within the so-called He/N zone in the 12 $M_{\odot}$ model and moves outward to the surface H envelope with increasing progenitor mass. It can attain such low $^{12}$C/$^{13}$C ratios because the abundant $^{12}$C made by He-burning reacts with H mixed inward from the surface envelope via the reaction chain $^{12}$C($p$,$\gamma$)$^{13}$N($\beta^{+}$)$^{13}$C during the presupernova evolution. Given the relatively low temperatures, $^{14}$N instead of $^{15}$N is made abundantly by CNO-cycle burning via the reaction $^{13}$C($p$,$\gamma$)$^{14}$N, resulting in $^{14}$N/$^{15}$N ratios of >10,000 in this shell. It was recognized in many studies (e.g., Nittler et al. 1995; Hoppe et al. 1996) that the admixture of material from this $^{13}$C-, $^{14}$N-rich shell in a way that preserves the condition C/O > 1 cannot yield $^{14}$N/$^{15}$N ratios $\lesssim$ 300, in contrast to the low $^{14}$N/$^{15}$N ratios ($\lesssim$ 300) of X, C, AB1, and D grains (Fig. 6a). Lin et al. (2010) overcame this problem by including a $^{15}$N-rich spike present in the He/N zone of the 25 $M_{\odot}$ CCSN model of Woosley and Heger (2007) to reproduce X grain data. This $^{15}$N-rich spike is likely caused by hot CNO-cycle burning when the CCSN shock hits the He/N shell with some H still present. Later, a new set of CCSN models by Pignatari et al. (2015) further suggests that the entire He-rich shell could have retained up to percent-level H mixed inward from the surface H-rich envelope until the explosion, during which hot CNO-cycle burning would occur and produce abundant $^{13}C$, $^{15}N$, and $^{26}Al$ that resemble nova nucleosynthesis. Motivated by the modeling results, Pignatari et al. (2015) investigated the effects of the amount of H present in the He shell and the explosion energy on CCSN nucleosynthesis model predictions, based on which the authors concluded that their predicted ejecta compositions can account for the isotopic signatures of $^{13}C$-, $^{15}N$-rich presolar SiC grains, including putative nova and AB1 grains.

The possibility of "nova-like" nucleosynthesis occurring during CCSN explosions magnifies the challenge in distinguishing between nova and CCSN presolar grains. For instance, Parikh et al. (2014) pointed out that classical nova and CCSN models predict $^{32}S/^{33}S$ ratios of 110−130 and 130−200, respectively, so that the $^{32}S/^{33}S$ ratio can be used as the smoking gun to distinguish between the two stellar sources. However, "nova-like" nucleosynthesis in CCSNe leads to an overlap in the $^{32}S/^{33}S$ ratios predicted by state-of-the-art nova and CCSN stellar nucleosynthesis models, and the $^{32}S/^{33}S$ ratio is thus not as diagnostic as previously thought. While neutron-capture isotopic signatures and inferred initial presence of $^{44}Ti$ can be used to exclude a nova origin (Liu et al. 2016), the lack of these signatures in $^{13}C$-, $^{15}N$-rich presolar grains does not guarantee a nova origin. Instead, compared to nova models (José et al. 2004; Denissenkov et al. 2014; Starrfield et al. 2020), one finds that the CCSN models of Pignatari et al. (2015) often provide more satisfying quantitative matches to the multielement isotopic compositions of $^{13}C$-, $^{15}N$-rich presolar grains, owing to the wide array of stellar nucleosynthesis processes occurring at different supernova evolutionary stages under largely varying conditions. For instance, although "nova-like" nucleosynthesis in CCSN deactivates the $^{22}Ne$ neutron source for the neutron-burst process because of the more efficient operation of the competing $^{22}Ne(p,\gamma)^{23}Na$ reaction, neutron burst could still occur in shells above and below the "nova-like-nucleosynthesis" shell so that mixing across these shells would yield both "nova-like" and neutron-capture isotopic signatures (Liu et al. 2018c). More nova stellar modeling efforts are needed to investigate whether the better agreement with CCSN models truly disputes nova origins or reflects limitations in current 1D nova models.

#### 3.2.3.3. Other Presolar Phases Potentially from CCSNe and Novae

In addition to SiC, other presolar phases from CCSNe and novae include silicates, oxides, and graphite grains. Furthermore, Nittler et al. (1995) and Hoppe et al. (1996) reported that $Si_3N_4$ grains in the acid residue of Murchison[15] all have X-grain-like isotopic compositions, pointing to their CCSN origin. Group 4 O-rich grains and a large fraction of low-density graphite grains likely came from CCSNe, and the $^{18}$O enrichments identified in many of these grains point to production via the $^{14}N(\alpha,\gamma)^{18}F(\beta^{+})^{18}O$ reaction chain in the He/C zone. Their proposed CCSNe origins are further supported by the inferred initial presence of $^{41}$Ca and $^{44}$Ti in some of the low-density graphite grains (Nittler et al. 1996; Amari et al. 1996) and large $^{26}$Mg enrichments (along with $^{25}$Mg depletions) in some of the Group 4 spinel and silicate grains (e.g., Nguyen & Messenger 2014). Different studies concluded that CCSN $Si_3N_4$, presolar graphite, and Group 4 grains sampled materials mainly in the outer CCSN zones from the O/C zone up to the surface (Nittler et al. 2008; Amari et al. 2014; Floss and Haenecour 2016), in general agreement with the conclusion drawn based on CCSN SiC grains discussed earlier. The lack of CCSN grains from deep CCSN zones could reflect unfavorable conditions for dust growth and/or more effective dust destruction due to gas-grain collisions (Amari et al. 2014). Recently, Leitner and Hoppe (2019) identified large $^{25}$Mg excesses in Group 1 silicate grains, which point to hot CNO-cycle burning via the $^{24}Mg(p,\gamma)^{25}Al(\beta^{+})^{25}Mg$ reaction chain at temperatures exceeding $\sim 9 \times 10^{7}$ $K$ (Doherty et al. 2014). What is puzzling about these $^{25}$Mg-rich Group 1 silicates is that $^{18}$O is expected to be effectively destroyed during the hot CNO-cycle burning, which, however, is not seen in any of the grains. Leitner and Hoppe (2019) were able to reproduce the grain data by invoking *ad hoc* mixing across different CCSN outer zones that include the "nova-like-nucleosynthesis" shell. Again, as in the case of putative nova SiC grains, although nova models cannot explain the lack of $^{18}$O depletions in these grains (Leitner and Hoppe 2019), it remains an open question whether the better agreement with CCSN models is a robust argument that disputes the nova origin or points to limitations in 1D nova models. Since multielement isotopic data for presolar silicate grains are, so far, limited to O, Mg, and Si, isotope analysis of heavier elements in the future will provide fresh insights into the dilemma.

Presolar graphite grains with isotopic signatures similar to putative nova SiC grains are inferred to have also originated in novae. In addition, presolar O-rich grains with large $^{17}$O excesses

[15]Different from the CM2 chondrite Murchison that formed in an oxidized condition, enstatite chondrites are among the most chemically reduced meteorites known and contain abundant $Si_3N_4$ grains that are inferred to have been formed in the solar system (Russell et al. 1995).

($^{17}O/^{16}O \gtrapprox 0.005$) are perhaps nova grains since nova models predict such large $^{17}O$ excesses and AGB stellar models do not. Isotopic investigations of a limited number of nova grains have shown that putative nova oxides are characterized by high inferred initial $^{26}Al/^{27}Al$ ratios (Gyngard et al. 2010; Liu et al. 2022b), and putative nova silicates by large excesses in $^{25}Mg$ and $^{26}Mg$ (the latter ascribed to $^{26}Al$ decay) (Nguyen and Messenger 2014). These isotopic signatures consistently point to hot CNO-cycle burning and agree with both "nova-like" nucleosynthesis in CCSNe and nova model calculations. Verifying the nova origin of putative graphite, oxides, and silicates is, however, faced with the same dilemma as in the case of putative SiC grains. The only difference is that the highest $^{17}O/^{16}O$ ratio predicted by the CCSN models of Pignatari et al. (2015) is only up to $2 \times 10^{-3}$, which is about one order of magnitude lower than the highest ratio observed in $^{17}O$-rich silicates and oxides (Liu et al. 2016). Thus, the O isotopic compositions of putative nova O-rich grains are in better agreement with a CO nova origin than those of putative nova SiC and graphite grains. Iliadis et al. (2018), however, found that the isotopic compositions of putative nova O-rich grains cannot all be reproduced by CO nova models even if uncertainties in stellar parameters are considered.

### 3.2.3.4. Other Potential Stellar Sources for Presolar Grains

In addition to AGB stars, novae, and CCSNe (and their progenitor stars), other stellar sources that have been proposed for presolar grains include J-type stars, born-again AGB stars, and ECSNe. Below, I highlight the controversies in assigning presolar grains to each of these stellar sources.

AB2 grains ($^{12}C/^{13}C \lessapprox 10$, $^{14}N/^{15}N \gtrapprox 410$) are inferred to have come from J-type stars[16], given their similar chemical and isotopic compositions, i.e., no *s*-process enrichments and similar $^{13}C$ and $^{14}N$ excesses (Liu et al. 2017c). Hoppe et al. (2019), however, argued that the isotopic signatures of AB2 grains could be alternatively explained by mixing more materials from the outer $^{14}N$-rich zones of CCSNe compared to AB1 grains, leaving the stellar origin of AB2 grains ambiguous. The natural clustering analysis results of Boujibar et al. (2021) and Hystad et al. (2022) suggest two clusters of AB grains that are in general agreement with the proposal of dividing AB grains into AB1 and AB2 grains based on the solar $^{14}N/^{15}N$ ratio (Liu et al. 2017a), although the exact dividing method varies with the chosen natural clustering analysis technique. The assessment of the confidence of the grain assignment by Hystad et al. (2022) further suggests that AB grains with $^{14}N/^{15}N \gtrapprox 1000$ and $^{14}N/^{15}N \lessapprox 270$ are two well separated populations and could represent presolar grains from two distinct stellar sources from the perspective of statistics. It remains questionable whether AB grains with $270 \lessapprox {}^{14}N/^{15}N \lessapprox 1000$ have multiple stellar sources or are more affected by terrestrial and/or asteroidal N contamination than the other AB grains, thus blurring the natural clustering analysis results.

[16] J-type stars are a class of red-giant C-rich stars that are characterized by low $^{12}C/^{13}C$ ratios and constitute a significant fraction (10–15%) of all carbon stars in the Galaxy (Abia and Isern 2000). The stellar evolution of J-type stars is poorly understood.

Schmidt et al. (2018) proposed that putative nova, C2, and AB1 grains could have come from J-type carbon stars based on their observation that a young, C-rich planetary nebula (K4-47) exhibits large $^{13}C$, $^{17}O$, and $^{15}N$ enrichments that resemble the isotopic signatures of these $^{13}C$-, $^{15}N$-rich presolar SiC grains. The authors proposed that the precursor of K4-47 was a J-type star that underwent a He-shell flash. However, the nature of K4-47 remains ambiguous as K4-47 could alternatively represent an extended nebula that was ejected by a pair of interacting binary stars during a nova-like explosion (Corradi et al. 2000). This suggestion is supported by the observation that high-velocity material is moving into the surrounding ISM (Gonçalves et al. 2004). Such a low $^{14}N/^{15}N$ ratio as observed by Schmidt et al. (2018), $13.6 \pm 6.5$, has not been observed for any J-type carbon stars yet, which, so far, have been determined to all exhibit $^{14}N/^{15}N \gtrapprox 1000$ (Fig. 3a).

Born-again AGB stars are H-deficient post AGB stars that experience another TP after they leave the AGB phase (Iben and MacDonald 1995). Born-again AGB stars have been proposed as the stellar source of $^{13}C$-rich high-density presolar graphite grains that exhibit exotic Ca and Ti isotopic anomalies (Jadhav et al. 2013b). The born-again AGB stellar models of Herwig et al. (2011) account for the neutron-rich Ca and Ti isotopic signatures found by Jadhav et al. (2013) since the $^{13}C(\alpha,n)^{16}O$ reaction is predicted to take place at the bottom of the He intershell, resulting in neutron densities that reach $\sim 10^{15}$ cm$^{-3}$. Neutron capture occurring in this neutron density regime was named the intermediate neutron capture process (*i*-process) by Cowan and Rose (1977), which occurs at higher neutron densities than the *s*-process but at lower densities than the neutron-burst process. Liu et al. (2014b) reported lower-than-solar $^{134}Ba/^{136}Ba$ ratios for two MS grains, which cannot be explained by AGB stellar nucleosynthesis and could point to the *i*-process in born-again

AGB stars. However, AGB model predictions for $^{134}Ba/^{136}Ba$ depend strongly on the activation strength at the $^{134}Cs$ ($t_{1/2}$ = 2.1 years) branch point along the *s*-process path and suffer from uncertainties in the $\beta^-$ decay rate of $^{134}Cs$ that has a strong temperature dependence (Bao and Käppeler 1987). Li et al. (2021) showed that the low $^{134}Ba/^{136}Ba$ ratios of the two MS grains can be explained by Monash AGB stellar models that adopt their newly calculated $^{134}Cs$ $\beta^-$ decay rate. Liu et al. (2022a), however, found that FRUITY AGB stellar models that adopt the new $^{134}Cs$ $\beta^-$ decay rate calculated by Taioli et al. (2022) cannot reach such low $^{134}Ba/^{136}Ba$ ratios. Given the existing uncertainties in the $^{134}Cs$ $\beta^-$ decay rate and in the maximum AGB stellar temperature, it remains ambiguous whether MS SiC grains with low $^{134}Ba/^{136}Ba$ ratios are sourced from AGB or born-again AGB stars. Amari et al. (2001c) proposed that AB grains could also have been sourced from born-again AGB stars given their similar $^{13}C$ enrichments. However, AB2 grains ($^{14}N$-rich), one of which had comparable *s*-process enrichment as MS grains (Stephan et al. 2019), generally lack *s*-process enrichments (Liu et al. 2017c), and AB1 grains ($^{15}N$-rich) show weaker *s*-process enrichments than MS grains (Liu et al. 2018c), both of which argue against the born-again stellar origin based on the models of Herwig et al. (2011). However, Herwig et al. (2011) developed the born-again AGB models with the goal of reproducing the peculiar chemical composition of Sakurai's Object. It remains unclear whether all born-again AGB stars experience the *i*-process and, if so, whether the *i*-process occurs in similar or vastly different conditions. All in all, the lack of unique Ca and Ti isotopic signatures in AB grains as observed in high-density graphite grains points to the varying relative contributions of distinctive stellar sources to the different presolar dust reservoirs in the solar system.

Rare type supernovae have been proposed as the stellar sources for the presolar Cr-rich nanospinel grains that carry large $^{54}Cr$ excesses (Nittler et al. 2018b), which have not been identified in any of the other presolar phases. These $^{54}Cr$-rich spinel grains were initially inferred to have come from CCSNe (Dauphas et al. 2010; Qin et al. 2011). Later, by suppressing asteroidal/terrestrial Cr contamination based on high resolution imaging, Nittler et al. (2018b) identified more extreme $^{54}Cr$ excesses (accompanied by $^{50}Ti$ excesses) in presolar nanospinel grains, based on which the authors argued that these nanospinel grains most likely originated in rare supernova events such as high-density Type Ia supernovae or ECSNe, the end evolution stage for 8−10 $M_{\odot}$ stars. ECSNe were first theoretically predicted by Nomoto (1987) and have been recently confirmed to exist in nature (Hiramatsu et al. 2021). den Hartogh et al. (2022), however, challenged the proposal of Nittler et al. (2018b) by showing that the $^{54}$Cr-rich presolar nanospinel grain data, in fact, can be well accounted for by CCSN nucleosynthesis according to three new sets of CCSN models, thus leaving their stellar origin even more ambiguous. Long-term spectroscopic observations of the ECSNe 2018zd (Hiramatsu et al. 2021) will provide us more clues about the dust production in this type of supernovae and assist our interpretation of the stellar origins of $^{54}$Cr-rich presolar nanospinel grains.

## 4. PRESOLAR GRAINS AND ISM PROCESSES

During their lifespans, presolar grains were initially born in cooling gas that was lost from the surface of a star or was ejected during stellar explosions and resided in the ISM for some time before their incorporation in the solar system (Fig. 1). Although presolar grains are all stardust and once ISM dust, ISM dust is not all stardust. In the ISM, dust could become shattered and vaporized, and new dust could form and grow (Jones 2004). These dust destruction processes in the ISM can be explored by studying the microstructures of surviving ISM dust – presolar grains. One such example was reported by Takigawa et al. (2018). Their high resolution TEM observations of a Group 1 corundum grain revealed that the grain experienced little modification by gas-grain and grain-grain collisions in the ISM and solar nebula; the high-Mg domains within this grain, however, imply that it underwent at least one very transient heating event – a single grain-grain collision in the ISM, which could also account for the one rough surface of and cavity within the grain.

ISM grains are inferred to mostly consist of silicates and carbonaceous materials (including diamond, graphite, amorphous carbon, hydrocarbons) based on the wavelength-dependent extinction of starlight and cosmic elemental abundances (Tielens and Allamandola 1986). Draine (2003) used Si as an example and estimated that the ISM residence time of stardust (~300 Ma) should be shorter than that of a Si atom (1000 Ma), indicating that Si would be in the gas phase if no dust formed in the ISM. However, observations suggest that more than 90% Si atoms are missing from the gas phase and should be locked in ISM dust (Field 1974; Jenkins 2009; De Cia et al. 2016; Roman-Duval et al. 2022), thus implying that ISM dust is dominated by dust that formed in situ in the ISM and that stardust is only a minor ISM dust component. This implication is in line with the observation that the presolar grain abundance is only up to percent levels in primitive cometary samples collected from comet Wild 2 and in primitive IDPs collected from the stratosphere (Floss et al. 2013; Nguyen et al. 2022), both of which are believed to largely consist of primordial ISM dust. Deuterium-rich and/or $^{15}N$-rich[17] organic matter (carbonaceous materials) and amorphous silicate-rich grains called GEMS (glass with embedded metal and sulfides) in chondritic materials are believed to represent the originally incorporated primordial ISM dust (Bradley et al. 1999; Busemann et al. 2006). These, however, are highly controversial topics, and solar origins have also been equally argued for the organic matter and GEMS (Keller and Messenger 2011; Alexander et al. 2017), pointing to challenges in unambiguously identifying ISM grains (except for presolar grains) in extraterrestrial materials. Thus, although presolar grains served only as a minor dust component in the ISM, they are the only ISM component that has been unequivocally identified in extraterrestrial materials so far, thus providing us direct access to probing both the contributors to the ISM dust reservoir and the ISM environment. The discussions in the previous section make it clear that AGB stars and CCSNe (and their progenitor massive stars) are two dominant contributors to the presolar dust reservoir in the ISM with additional contributions potentially from novae, J-type carbon stars, born-again AGB stars, high-density Type Ia supernovae, and ECSNe.

After its birth, a presolar grain could become isotopically modified (e.g., He, Li, Ne) due to spallation reactions with Galactic cosmic rays (GCR) that mostly consist of high energy protons and $\alpha$ particles (Heck et al. 2007; Gyngard et al. 2009). The spallation reactions produce specific nuclei whose abundances are proportional to the timespan during which the grain was irradiated (if the nuclei remain inside the grain), thus providing us a unique opportunity to investigate its ISM residence time and thus determine its "presolar age". Based on cosmogenic $^{21}Ne$, Heck et al. (2020) determined the GCR ages of 40 large presolar SiC grains, the majority of which have ISM lifetimes of <300 Ma with some ages reaching up to >1 Ga. It is noteworthy that the presolar SiC grains analyzed by Heck et al. (2020) are unusually large SiC grains (2–34 μm) that are atypical of meteoritic presolar SiC grains (a few hundred nm on average), which, in theory, should have spent less time in the ISM so that they could have survived and become incorporated into the solar system. While the GCR ages provide important time constraints on the residence time of presolar grains in the ISM, it still requires absolute ages to demonstrate that presolar grains indeed are older than the solar system. Such potential radioactive systems include the decay systems of $^{238}U$–$^{206}Pb$ ($t_{1/2}$ = 4.5 Ga) and $^{232}Th$–$^{208}Pb$ ($t_{1/2}$ = 14 Ga), and analyses of these isotopes in presolar grains could potentially be achieved by in situ measurements using the new generations of RIMS instruments (Stephan et al. 2016; Pal et al. 2022) with improved sensitivities in the future.

[17] The commonly held view is that large D and $^{15}N$ enrichments observed in organic matter result from low-temperature gas phase ion-molecule reactions (Terzieva and Herbst 2000; Goumans and Kästner 2011) instead of stellar nucleosynthesis, which would otherwise result in diverse isotopic signatures (e.g., $^{14}N$ and $^{15}N$ enrichments) through different nuclear reaction channels (e.g., neutron capture, proton capture).

## 5. PRESOLAR GRAINS AND SOLAR SYSTEM PROCESSES

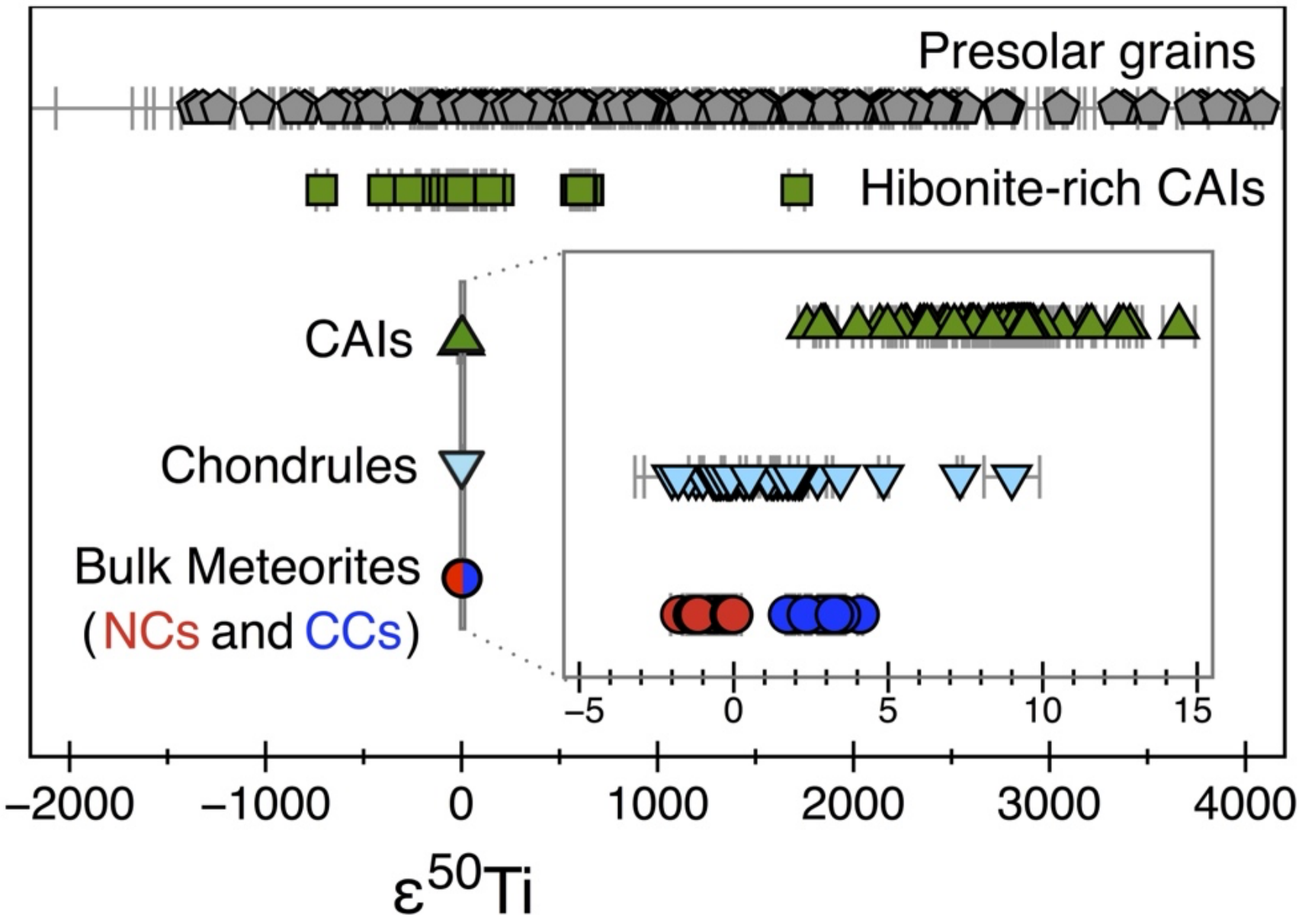

**Fig. 6**. *Titanium-50 variations in meteoritic components and bulk meteorites, expressed in ε notation ($\varepsilon^{50}Ti = \delta^{50}Ti / 10$) (from Bermingham et al. 2020).*

High-precision isotope analyses in the past decade have revealed nucleosynthetic isotopic variations[18] in various meteoritic components and bulk meteorites as well as systematic isotopic differences between non-carbonaceous and carbonaceous chondrite groups (Fig. 6; Trinquier et al. 2007, 2009; Leya et al. 2008; Warren 2011; Steele et al. 2012; Budde et al. 2016, 2019; Kruijer et al. 2017, 2020; Nanne et al. 2019; Yokoyama et al. 2019; Hopp et al. 2020; Spitzer et al. 2020; Nie et al. 2023; Martins et al. 2023). These nucleosynthetic isotopic variations and differences hold important clues about the earliest history of the solar system and the processes of planet formation. However, there is still a lack of a fundamental understanding of the process(es) that led to the presence of nucleosynthetic isotopic variations across the solar system (Bermingham et al. 2020). Multiple clues suggest that isotopic variations among non-carbonaceous and carbonaceous materials were caused by a heterogenous distribution of presolar grains in the protoplanetary disk (Qin and Carlson 2016; Hutchison et al. 2022). The gradually shrinking range of $^{50}$Ti anomalies observed from presolar grains to bulk meteorites shown in Fig. 6 reflects the homogenization of primordial dust and gas in the protoplanetary disk over time within the first few hundred million years after the solar system formation. Thus, presolar grains, the building blocks of our solar system, hold ground-truth information about the nucleosynthetic origin of isotopic variations across the solar system.

[18] Isotope variations in meteorites that cannot be related to the terrestrial composition by any known processes (e.g., radiogenic decay), are nucleosynthetic in origin, reflecting the fact that stellar nucleosynthesis products were not fully homogenized when the solar system formed (Dauphas and Schauble 2016).

Since the systematic isotopic differences between non-carbonaceous and carbonaceous chondrite groups have been observed for a number of refractory elements such as Ti and Mo, it is plausible that the isotopic anomalies were carried in the presolar grains that were not uniformly destroyed during the solar system formation and/or heterogeneously distributed across the protoplanetary disk. Indeed, presolar nanospinel grains that carry large $^{54}$Cr excesses were identified to be the carrier for observed bulk meteoritic Cr isotopic anomalies (Dauphas et al. 2010; Qin and Carlson 2016). As discussed earlier, these presolar nanospinel grains could have originated in high-density Type Ia supernovae, ECSNe, or CCSNe. Identification of carriers for bulk meteoritic isotopic anomalies of other elements will thus help further pinpoint the stellar source and answer the question whether the heterogenous distribution of identified carriers reflects the initial heterogenous distribution or the heterogenous destruction in the protoplanetary disk.

As presolar grains also spent billions of years residing inside the parent bodies of their host meteorites in the solar system, secondary processing like aqueous alteration and thermal metamorphism also affected the inventory of presolar phases in meteorites so that their abundances can be used to trace the evolution of small solar system bodies. Extensive in situ studies of primitive carbonaceous materials have shown that the abundance of presolar grains varies widely among different meteorite groups and petrologic types, largely reflecting the degree of secondary processing experienced by the host meteorite. Among different presolar phases, silicates are the most abundant type identified in primitive chondritic materials with a matrix-normalized abundance of up to ~200 ppm in unequilibrated chondrites (e.g., Nguyen & Zinner 2004; Nguyen et al. 2007; Haenecour et al. 2018; Nittler et al. 2018a; Barosch et al. 2022b). Higher abundances of presolar silicates were observed in anhydrous IDPs of possible cometary origin collected in the stratosphere (Messenger et al. 2003; Floss et al. 2006; Nguyen et al. 2022). Compared to anhydrous IDPs, the lowered presolar silicate abundances observed in the most primitive chondrites are interpreted as a consequence of slight parent body processing (Floss and Haenecour 2016). In addition, significantly lower abundances of presolar silicates have been observed in meteorites that experienced extensive aqueous alteration such as samples returned from asteroid Ryugu that is isotopically and petrologically similar to Ivuna-type chondrites (Barosch et al. 2022a). In this case, the significantly lowered abundance of presolar silicates likely resulted from altering labile presolar silicates or reequilibrating their oxygen isotopic compositions by exchange with a fluid. The reader is referred to Floss and Haenecour (2016) for a detailed overview of the subject.

In addition to presolar grains, we plausibly also have contemporaneous stardust grains coming from nearby supernovae as suggested by the finding of $^{60}$Fe and $^{244}$Pu present in deep-sea sediments (Knie et al. 2004; Wallner et al. 2015, 2016). Model simulations have shown that $^{60}$Fe and $^{244}$Pu atoms that are produced during supernova explosions could travel to the inner solar system only if they were carried in dust grains (Fry et al. 2016). Thus, it will be extremely exciting if such contemporaneous supernova stardust grains are identified in deep-sea sediments in the future so that they can be compared to presolar supernova dust grains for their chemical, isotopic, and structural compositions.

## 6. CONCLUSIONS

Since the discovery of presolar grains in meteorites in 1980s, a diverse array of information about processes in stars, ISM, and the solar system has been gleaned by studying them. Presolar grain studies rely on the synergy between astronomy, astrophysics, nuclear physics, and cosmochemistry. To understand the stellar sources of presolar grains, it is important to compare isotope data of presolar grains to astronomical observations for different types of stellar objects. When such astronomical observations are not currently available, stellar nucleosynthesis models must be relied upon, which require inputs of initial stellar composition estimated based on solar system nuclide abundances, stellar evolution models, and nuclear reaction rates determined by both theories and laboratory experiments. Uncertainties in these modelling inputs always need to be taken into consideration when inferring the stellar origins of presolar grains. Once the stellar source of a group of presolar grains is ascertained, isotope information extracted from the grains can then be used to constrain stellar mixing processes, nuclear reaction rates, the GCE, and the ISM residence times of the grains. Examples are Types MS and X SiC grains that are confirmed to have come from low-mass AGB stars and CCSNe, respectively, based on multielement isotope data, which have provided invaluable constraints on AGB and CCSN stellar nucleosynthesis and mixing processes and the timing of dust formation after supernova explosions. Continued efforts are needed to obtain multielement isotope data for more elements and more grains regarding presolar grains of ambiguous stellar origins so that similar stringent constraints could be obtained for other types of stars. In addition, crystal structures and chemical compositions of presolar grains can provide additional information to infer dust condensation conditions in their parent stars and ISM processes, while abundances of presolar grains in primitive chondrites can help constrain secondary processing experienced by the parent asteroids of their host chondrites.

Technological advances will almost certainly allow for the discovery of additional types of presolar grains and analysis of smaller, more typical presolar grains in the future. In the past decade, the invention of the Hyperion radio-frequency $O^-$ ion source enabled measuring isotopes of electropositive elements such as Mg and Ti at ~100 nm using NanoSIMS, and the significantly increased spatial resolution led to the discovery of $^{25}$Mg-rich Group 1 silicates and subgrains encapsuled within a presolar grain and our strengthened capability of obtaining intrinsic isotopic signatures of presolar grains in general. The development of the new generation of RIMS instruments also allowed measuring isotopes of up to three elements simultaneously with increased useful yields, thus greatly increasing the chance of obtaining multielement isotope data with better precisions from individual presolar grains. The increased signal-to-noise ratio and spatial resolution of Raman microscopes provided us a chance of screening for presolar grains with special structural compositions that can be studied in unprecedented detail under a TEM, and the obtained structural information allowed us to probe the wide range of physicochemical conditions in different types of stars. I am sure that these instrumental capabilities will continue to advance in the future, thus allowing us to peer into smaller grains with better accuracies and precisions.

**Acknowledgments**: I extend my heartfelt gratitude to my long-term collaborator and mentor, Dr. Larry Nittler, for his meticulous review and invaluable critiques of the manuscript, which have significantly enhanced the clarity and readability of this book chapter. I would like to express my sincere gratitude to the book editor Dr. Sujoy Mukhopadhyay for his diligent work in handling the manuscript, to the reviewer for their valuable and constructive comments, and to Dr. Katherine Bermingham for generously providing an original copy of Figure 6 from their paper. In addition, I am deeply grateful to my long-term collaborators Drs. Conel Alexander, Maurizio Busso, Sergio Cristallo, Bradley Meyer, Larry Nittler, and Sara Palmerini (in alphabetic order) for their insightful discussions on related topics in this book chapter. Their expertise and guidance have been invaluable in shaping the content of this work. This work was supported by NASA through grants 80NSSC23K1034 and 80NSSC24K0132 to N.L.

## REFERENCES

Abarzhi, S. I., Bhowmick, A. K., Naveh, A., Pandian, A., Swisher, N. C., Stellingwerf, R. F. & Arnett, W. D. 2019. Supernova, nuclear synthesis, fluid instabilities, and interfacial mixing. *Proceedings of the National Academy of Science,* 116**,** 18184-18192.

Abia, C. & Isern, J. 2000. The chemical composition of carbon stars. II. The J-type stars. *The Astrophysical Journal,* 536**,** 438-449.

Alexander, C. M. O' D., & Nittler, L. R. 1999. The Galactic evolution of Si, Ti, and O isotopic ratios. *The Astrophysical Journal,* 519**,** 222-235.

Alexander, C. M. O' D., Cody, G. D., De Gregorio, B. T., Nittler, L. R. & Stroud, R. M. 2017. The nature, origin and modification of insoluble organic matter in chondrites, the major source of Earth's C and N. *Chemie der Erde / Geochemistry,* 77**,** 227-256.

Amari, S., Lewis, R. S. & Anders, E. 1994. Interstellar grains in meteorites: I. Isolation of SiC, graphite and diamond; size distributions of SiC and graphite. *Geochimica et Cosmochimica Acta,* 58**,** 459-470.

Amari, S., Lewis, R. S. & Anders, E. 1995. Interstellar grains in meteorites: III. Graphite and its noble gases. *Geochimica et Cosmochimica Acta,* 59**,** 1411-1426.

Amari, S., Zinner, E. & Lewis, R. S. 1996. $^{41}$Ca in presolar graphite of supernova origin. *The Astrophysical Journal Letters*, 470, L101-L104.

Amari, S., Zinner, E. & Lewis, R. S. 1999. A singular presolar SiC grain with extreme $^{29}$Si and $^{30}$Si excesses. *The Astrophysical Journal Letters,* 517**,** L59-L62.

Amari, S., Gao, X., Nittler, L. R., Zinner, E., José, J., Hernanz, M. & Lewis, R. S. 2001a. Presolar grains from novae. *The Astrophysical Journal,* 551**,** 1065-1072.

Amari, S., Nittler, L. R., Zinner, E., Gallino, R., Lugaro, M. & Lewis, R. S. 2001b. Presolar SiC grains of type Y: Origin from low-metallicity asymptotic giant branch stars. *The Astrophysical Journal,* 546**,** 248-266.

Amari, S., Nittler, L. R., Zinner, E., Lodders, K. & Lewis, R. S. 2001c. Presolar SiC grains of type A and B: their isotopic compositions and stellar origins. *The Astrophysical Journal,* 559**,** 463-483.

Amari, S., Zinner, E. & Gallino, R. 2014. Presolar graphite from the Murchison meteorite: an isotopic study. *Geochimica et Cosmochimica Acta,* 133**,** 479-522.

Andrews, J. E., Krafton, K. M., Clayton, G. C., Montiel, E., Wesson, R., Sugerman, B. E. K., Barlow, M. J., Matsuura, M. & Drass, H. 2016. Early dust formation and a massive progenitor for SN 2011ja? *Monthly Notices of the Royal Astronomical Society,* 457**,** 3241-3253.

Arlandini, C., Käppeler, F., Wisshak, K., Gallino, R., Lugaro, M., Busso, M. & Straniero, O. 1999. Neutron capture in low-mass asymptotic giant branch stars: cross sections and abundance signatures. *The Astrophysical Journal,* 525**,** 886-900.

Arnould, M. & Goriely, S. 2003. The p-process of stellar nucleosynthesis: astrophysics and nuclear physics status. *Physics Reports,* 384**,** 1-84.

Banerjee, P., Qian, Y.-Z., Heger, A. & Haxton, W. C. 2016. Evidence from stable isotopes and $^{10}$Be for solar system formation triggered by a low-mass supernova. *Nature Communications,* 7**,** 13639.

Bao, Z. Y. & Käppeler, F. 1987. Neutron capture cross sections for *s*-process studies. *Atomic Data and Nuclear Data Tables,* 36**,** 411.

Barosch, J., Nittler, L. R., Wang, J., Alexander, C. M. O' D., De Gregorio, B. T., Engrand, C., Kebukawa, Y., Nagashima, K., Stroud, R. M., Yabuta, H., Abe, Y., Aléon, J., Amari, S., Amelin, Y., Bajo, K.-i., Bejach, L., Bizzarro, M., Bonal, L., Bouvier, A., Carlson, R. W., Chaussidon, M., Choi, B.-G., Cody, G. D., Dartois, E., Dauphas, N., Davis, A. M., Dazzi, A., Deniset-Besseau, A., Di Rocco, T., Duprat, J., Fujiya, W., Fukai, R., Gautam, I., Haba, M. K., Hashiguchi, M., Hibiya, Y., Hidaka, H., Homma, H., Hoppe, P., Huss, G. R., Ichida, K., Iizuka, T., Ireland, T. R., Ishikawa, A., Ito, M., Itoh, S., Kamide, K., Kawasaki, N., David Kilcoyne, A. L., Kita, N. T., Kitajima, K., Kleine, T., Komatani, S., Komatsu, M., Krot, A. N., Liu, M.-C., Martins, Z., Masuda, Y., Mathurin, J., McKeegan, K. D., Montagnac, G., Morita, M., Mostefaoui, S., Motomura, K., Moynier, F., Nakai, I., Nguyen, A. N., Ohigashi, T., Okumura, T., Onose, M., Pack, A., Park, C., Piani, L., Qin, L., Quirico, E., Remusat, L., Russell, S. S., Sakamoto, N., Sandford, S. A., Schönbächler, M., Shigenaka, M., Suga, H., Tafla, L., Takahashi, Y., Takeichi, Y., Tamenori, Y., Tang, H., Terada, K., Terada, Y., Usui, T., Verdier-Paoletti, M., Wada, S., Wadhwa, M., Wakabayashi, D., Walker, R. J., Yamashita, K., Yamashita, S., Yin, Q.-Z., Yokoyama, T., Yoneda, S., et al. 2022a. Presolar stardust in asteroid Ryugu. *The Astrophysical Journal Letters,* 935**,** L3.

Barosch, J., Nittler, L. R., Wang, J., Dobrică, E., Brearley, A. J., Hezel, D. C. & Alexander, C. M. O' D. 2022b. Presolar O- and C-anomalous grains in unequilibrated ordinary chondrite matrices. *Geochimica et Cosmochimica Acta,* 335**,** 169-182.

Barzyk, J. G., Savina, M. R., Davis, A. M., Gallino, R., Gyngard, F., Amari, S., Zinner, E., Pellin, M. J., Lewis, R. S. & Clayton, R. N. 2007. Constraining the $^{13}$C neutron source in AGB stars through isotopic analysis of trace elements in presolar SiC. *Meteoritics and Planetary Science,* 42**,** 1103-1119.

Battino, U., Pignatari, M., Ritter, C., Herwig, F., Denisenkov, P., Den Hartogh, J. W., Trappitsch, R., Hirschi, R., Freytag, B., Thielemann, F. & Paxton, B. 2016. Application of a theory and simulation-based convective boundary mixing model for AGB star evolution and nucleosynthesis. *The Astrophysical Journal,* 827**,** 30.

Bermingham, K. R., Füri, E., Lodders, K. & Marty, B. 2020. The NC-CC isotope dichotomy: implications for the chemical and isotopic evolution of the early Solar System. *Space Science Reviews,* 216**,** 133.

Biscaro, C. & Cherchneff, I. 2016. Molecules and dust in Cassiopeia A. II. Dust sputtering and diagnosis of supernova dust survival in remnants. *Astronomy and Astrophysics,* 589**,** A132.

Bisterzo, S., Travaglio, C., Gallino, R., Wiescher, M. & Käppeler, F. 2014. Galactic chemical evolution and solar s-process abundances: Dependence on the $^{13}$C-pocket structure. *The Astrophysical Journal,* 787**,** 10.

Bisterzo, S., Gallino, R., Käppeler, F., Wiescher, M., Imbriani, G., Straniero, O., Cristallo, S., Görres, J. & deBoer, R. J. 2015. The branchings of the main s-process: their sensitivity to α-induced reactions on $^{13}$C and $^{22}$Ne and to the uncertainties of the nuclear network. *Monthly Notices of the Royal Astronomical Society,* 449**,** 506-527.

Blanco, A., Borghesi, A., Fonti, S. & Orofino, V. 1998. Circumstellar emission from dust envelopes around carbon stars showing the silicon carbide feature. *Astronomy and Astrophysics,* 330**,** 505-514.

Bliss, J., Arcones, A. & Qian, Y.-Z. 2018. Production of Mo and Ru isotopes in neutrino-driven winds: implications for solar sbundances and presolar grains. *The Astrophysical Journal,* 866**,** 105.

Bloom, H. E., Stephan, T., Davis, A. M., Heck, P. R., Hoppe, P., Korsmeyer, J. M. & Amari, S. 2022. s-Process molybdenum, ruthenium, and barium in high-density presolar graphite. *in 53$^{rd}$ Lunar and Planetary Science Conference*, The Woodlands, TX., #2624 (abstract).

Boggs, S. E., Harrison, F. A., Miyasaka, H., Grefenstette, B. W., Zoglauer, A., Fryer, C. L., Reynolds, S. P., Alexander, D. M., An, H., Barret, D., Christensen, F. E., Craig, W. W., Forster, K., Giommi, P., Hailey, C. J., Hornstrup, A., Kitaguchi, T., Koglin, J. E., Madsen, K. K., Mao, P. H., Mori, K., Perri, M., Pivovaroff, M. J., Puccetti, S., Rana, V., Stern, D., Westergaard, N. J. & Zhang, W. W. 2015. $^{44}$Ti gamma-ray emission lines from SN1987A reveal an asymmetric explosion. *Science,* 348**,** 670-671.

Borkowski, K. J., Williams, B. J., Reynolds, S. P., Blair, W. P., Ghavamian, P., Sankrit, R., Hendrick, S. P., Long, K. S., Raymond, J. C., Smith, R. C., Points, S. & Winkler, P. F. 2006. Dust destruction in Type Ia supernova remnants in the Large Magellanic Cloud. *The Astrophysical Journal Letters,* 642**,** L141-L144.

Borkowski, K. J., Reynolds, S. P., Green, D. A., Hwang, U., Petre, R., Krishnamurthy, K. & Willett, R. 2010. Radioactive scandium in the youngest Galactic supernova remnant G1.9+0.3. *The Astrophysical Journal Letters,* 724**,** L161-L165.

Bose, M. & Starrfield, S. 2019. Condensation of SiC stardust in CO nova outbursts. *The Astrophysical Journal,* 873**,** 14.

Boss, A. P. & Keiser, S. A. 2012. Supernova-triggered molecular cloud core collapse and the Rayleigh-Taylor fingers that polluted the solar nebula. *The Astrophysical Journal Letters,* 756**,** L9.

Boujibar, A., Howell, S., Zhang, S., Hystad, G., Prabhu, A., Liu, N., Stephan, T., Narkar, S., Eleish, A., Morrison, S. M., Hazen, R. M. & Nittler, L. R. 2021. Cluster analysis of presolar silicon carbide grains: evaluation of their classification and astrophysical implications. *The Astrophysical Journal Letters,* 907**,** L39.

Bradley, J. P., Keller, L. P., Snow, T. P., Hanner, M. S., Flynn, G. J., Gezo, J. C., Clemett, S. J., Brownlee, D. E. & Bowey, J. E. 1999. An infrared spectral match between GEMS and interstellar grains. *Science,* 285**,** 1716-1718.

Budde, G., Burkhardt, C., Brennecka, G. A., Fischer-Gödde, M., Kruijer, T. S. & Kleine, T. 2016. Molybdenum isotopic evidence for the origin of chondrules and a distinct genetic heritage of carbonaceous and non-carbonaceous meteorites. *Earth and Planetary Science Letters,* 454**,** 293-303.

Budde, G., Burkhardt, C. & Kleine, T. 2019. Molybdenum isotopic evidence for the late accretion of outer solar system material to Earth. *Nature Astronomy,* 3**,** 736-741.

Buseck, P. R. & Hua, X. 1993. Matrices of carbonaceous chondrite meteorites. *Annual Review of Earth and Planetary Sciences,* 21**,** 255-305.

Busemann, H., Young, A. F., Alexander, C. M. O' D., Hoppe, P., Mukhopadhyay, S. & Nittler, L. R. 2006. Interstellar chemistry recorded in organic matter from primitive meteorites. *Science,* 312**,** 727-730.

Busso, M., Gallino, R. & Wasserburg, G. J. 1999. Nucleosynthesis in asymptotic giant branch stars: relevance for Galactic enrichment and solar system formation. *Annual Review of Astronomy and Astrophysics,* 37**,** 239-309.

Busso, M., Gallino, R., Lambert, D. L., Travaglio, C. & Smith, V. V. 2001. Nucleosynthesis and mixing on the asymptotic giant branch. III. predicted and observed *s*-process abundances. *The Astrophysical Journal,* 557**,** 802-821.

Busso, M., Vescovi, D., Palmerini, S., Cristallo, S. & Antonuccio-Delogu, V. 2021. s-processing in AGB stars revisited. III. Neutron captures from MHD mixing at different metallicities and observational constraints. *The Astrophysical Journal,* 908**,** 55.

Cadieux, C., Doyon, R., Plotnykov, M., Hébrard, G., Jahandar, F., Artigau, É., Valencia, D., Cook, N. J., Martioli, E., Vandal, T., Donati, J.-F., Cloutier, R., Narita, N., Fukui, A., Hirano, T., Bouchy, F., Cowan, N. B., Gonzales, E. J., Ciardi, D. R., Stassun, K. G., Arnold, L., Benneke, B., Boisse, I., Bonfils, X., Carmona, A., Cortés-Zuleta, P., Delfosse, X., Forveille, T., Fouqué, P., Gomes da Silva, J., Jenkins, J. M., Kiefer, F., Kóspál, Á., Lafrenière, D., Martins, J. H. C., Moutou, C., do Nascimento, J.-D., Ould-Elhkim, M., Pelletier, S., Twicken, J. D., Bouma, L. G., Cartwright, S., Darveau-Bernier, A., Grankin, K., Ikoma, M., Kagetani, T., Kawauchi, K., Kodama, T., Kotani, T., Latham, D. W., Menou, K., Ricker, G., Seager, S., Tamura, M., Vanderspek, R. & Watanabe, N. 2022. TOI-1452 b: SPIRou and TESS reveal a super-Earth in a temperate orbit transiting an M4 dwarf. *The Astronomical Journal,* 164**,** 96.

Cameron, A. G. W. & Truran, J. W. 1977. The supernova trigger for formation of the solar system. *Icarus,* 30**,** 447-461.

Cartledge, S. I. B., Lauroesch, J. T., Meyer, D. M. & Sofia, U. J. 2006. The homogeneity of interstellar elemental abundances in the Galactic disk. *The Astrophysical Journal,* 641**,** 327-346.

Cohen, O. 2011. The independency of stellar mass-loss rates on stellar X-ray luminosity and activity level based on solar X-ray flux and solar wind observations. *Monthly Notices of the Royal Astronomical Society,* 417**,** 2592-2600.

Corradi, R. L. M., Gonçalves, D. R., Villaver, E., Mampaso, A., Perinotto, M., Schwarz, H. E. & Zanin, C. 2000. High-velocity collimated outflows in planetary nebulae: NGC 6337, HE 2-186, and K4-47. *The Astrophysical Journal,* 535**,** 823-832.

Cowan, J. J. & Rose, W. K. 1977. Production of $^{14}C$ and neutrons in red giants. *The Astrophysical Journal,* 212**,** 149-158.

Cowan, J. J. & Thielemann, F.-K. 2004. R-process nucleosynthesis in supernovae. *Physics Today,* 57**,** 10.47.

Cristallo, S., Straniero, O., Gallino, R., Piersanti, L., Domínguez, I. & Lederer, M. T. 2009. Evolution, nucleosynthesis, and yields of low-mass asymptotic giant branch stars at different metallicities. *The Astrophysical Journal,* 696**,** 797-820.

Cristallo, S., Piersanti, L., Straniero, O., Gallino, R., Domínguez, I., Abia, C., Di Rico, G., Quintini, M. & Bisterzo, S. 2011. Evolution, nucleosynthesis, and yields of low-mass asymptotic giant branch stars at different metallicities. II. The FRUITY database. *The Astrophysical Journal Supplement Series,* 197**,** 17.

Cristallo, S., Nanni, A., Cescutti, G., Minchev, I., Liu, N., Vescovi, D., Gobrecht, D. & Piersanti, L. 2020. Mass and metallicity distribution of parent AGB stars of presolar SiC. *Astronomy and Astrophysics,* 644**,** A8.

Croat, T. K., Bernatowicz, T., Amari, S., Messenger, S. & Stadermann, F. J. 2003. Structural, chemical, and isotopic microanalytical investigations of graphite from supernovae. *Geochimica et Cosmochimica Acta*, 67, 4705-4725.

Daulton, T. L., Bernatowicz, T. J., Lewis, R. S., Messenger, S., Stadermann, F. J. & Amari, S. 2002. Polytype distribution in circumstellar silicon carbide. *Science,* 296**,** 1852-1855.

Daulton, T. L., Bernatowicz, T. J., Messenger, S., Stadermann, F. J., and Amari, S. 2003. Polytype distribution of circumstellar silicon carbide: Microstructural characterization by transmission electron microscopy. *Geochimica et Cosmochimica Acta*, 67, 4743-4767.

Dauphas, N., Remusat, L., Chen, J. H., Roskosz, M., Papanastassiou, D. A., Stodolna, J., Guan, Y., Ma, C. & Eiler, J. M. 2010. Neutron-rich chromium isotope anomalies in supernova nanoparticles. *The Astrophysical Journal,* 720**,** 1577-1591.

Dauphas, N. & Schauble, E. A. 2016. Mass fractionation laws, mass-independent effects, and isotopic anomalies. *Annual Review of Earth and Planetary Sciences,* 44**,** 709-783.

Davidson, J., Busemann, H., Nittler, L. R., Alexander, C. M. O' D., Orthous-Daunay, F.-R., Franchi, I. A. & Hoppe, P. 2014. Abundances of presolar silicon carbide grains in primitive meteorites determined by NanoSIMS. *Geochimica et Cosmochimica Acta,* 139**,** 248-266.

Davis, A. M. 2011. Cosmochemistry special feature: stardust in meteorites. *Proceedings of the National Academy of Science,* 108**,** 19142-19146.

De Cia, A., Ledoux, C., Mattsson, L., Petitjean, P., Srianand, R., Gavignaud, I. & Jenkins, E. B. 2016. Dust-depletion sequences in damped Lyman-α absorbers. A unified picture from low-metallicity systems to the Galaxy. *Astronomy and Astrophysics,* 596**,** A97.

De Cia, A., Jenkins, E. B., Fox, A. J., Ledoux, C., Ramburuth-Hurt, T., Konstantopoulou, C., Petitjean, P. & Krogager, J.-K. 2021. Large metallicity variations in the Galactic interstellar medium. *Nature,* 597**,** 206-208.

den Hartogh, J., Petö, M. K., Lawson, T., Sieverding, A., Brinkman, H., Pignatari, M. & Lugaro, M. 2022. Comparison between core-collapse supernova nucleosynthesis and meteoric stardust grains: Investigating magnesium, aluminium, and chromium. *The Astrophysical Journal,* 927**,** 220.

Denissenkov, P. A. & Tout, C. A. 2003. Partial mixing and formation of the $^{13}C$ pocket by internal gravity waves in asymptotic giant branch stars. *Monthly Notices of the Royal Astronomical Society,* 340**,** 722-732.

Denissenkov, P., Herwig, F., Pignatari, M. & Truran, J. W. 2014. MESA and NuGrid simulations of classical nova outbursts and nucleosynthesis. *In:* WOUDT, P. A. & RIBEIRO, V. A. R. M., eds. Stellar Novae: Past and Future Decades, December 01, 303.

Dillmann, I., Heil, M., Käppeler, F., Plag, R., Rauscher, T. & Thielemann, F.-K. 2006. KADoNiS - The Karlsruhe Astrophysical Database of Nucleosynthesis in Stars. *In:* WOEHR, A. & APRAHAMIAN, A., eds. Capture Gamma-Ray Spectroscopy and Related Topics, March 01, 123-127.

Dillmann, I. 2014. The new KADoNiS v1.0 and its influence on the s-process. XIII Nuclei in the Cosmos (NIC XIII), January 01, 57.

Doherty, C. L., Gil-Pons, P., Lau, H. H. B., Lattanzio, J. C. & Siess, L. 2014. Super and massive AGB stars - II. Nucleosynthesis and yields - Z = 0.02, 0.008 and 0.004. *Monthly Notices of the Royal Astronomical Society,* 437**,** 195-214.

Draine, B. T. 2003. Interstellar dust grains. *Annual Review of Astronomy and Astrophysics,* 41**,** 241-289.

Dwarkadas, V. V., Dauphas, N., Meyer, B., Boyajian, P. & Bojazi, M. 2017. Triggered star formation inside the shell of a Wolf-Rayet bubble as the origin of the solar system. *The Astrophysical Journal,* 851**,** 147.

Dwarkadas, V. V., Dilmohamed, S., Ekstrom, S., Meynet, G., Liu, N., Meyer, B. & Dauphas, N. 2020. Searching for the signatures of presolar grains in massive stars. *in 51st Lunar and Planetary Science Conference*, virtual, #2968.

Ebel, D. S. 2006. Condensation of rocky material in astrophysical environments. In *Meteorites and the Early Solar System II,* edited by Ante S. Lauretta and Harry Y. McSween, 253.

Eggleton, P. P., Dearborn, D. S. P. & Lattanzio, J. C. 2006. Deep mixing of $^{3}He$: reconciling Big Bang and stellar nucleosynthesis. *Science,* 314**,** 1580.

Ferrarotti, A. S. & Gail, H.-P. 2006. Composition and quantities of dust produced by AGB-stars and returned to the interstellar medium. *Astronomy and Astrophysics,* 447**,** 553-576.

Field, G. B. 1974. Interstellar abundances: gas and dust. *The Astrophysical Journal,* 187**,** 453-459.

Floss, C., Stadermann, F. J., Bradley, J. P., Dai, Z. R., Bajt, S., Graham, G. & Lea, A. S. 2006. Identification of isotopically primitive interplanetary dust particles: A NanoSIMS isotopic imaging study. *Geochimica et Cosmochimica Acta,* 70**,** 2371-2399.

Floss, C., Stadermann, F. J., Kearsley, A. T., Burchell, M. J. & Ong, W. J. 2013. The abundance of presolar grains in Comet 81P/Wild 2. *The Astrophysical Journal,* 763**,** 140.

Floss, C. & Haenecour, P. 2016. Presolar silicate grains: abundances, isotopic and elemental compositions, and the effects of secondary processing. *Geochemical Journal,* 50**,** 3-25.

Frank, E. A., Meyer, B. S. & Mojzsis, S. J. 2014. A radiogenic heating evolution model for cosmochemically Earth-like exoplanets. *Icarus,* 243**,** 274-286.

Freytag, B., Ludwig, H.-G. & Steffen, M. 1996. Hydrodynamical models of stellar convection. The role of overshoot in DA white dwarfs, A-type stars, and the Sun. *Astronomy and Astrophysics,* 313**,** 497-516.

Fry, B. J., Fields, B. D. & Ellis, J. R. 2016. Radioactive iron rain: Transporting $^{60}$Fe in supernova dust to the ocean floor. *The Astrophysical Journal,* 827**,** 48.

Fryer, C. L. 1999. Mass limits for black hole formation. *The Astrophysical Journal,* 522, 413-418.

Gail, H.-P., Zhukovska, S. V., Hoppe, P. & Trieloff, M. 2009. Stardust from asymptotic giant branch stars. *The Astrophysical Journal,* 698**,** 1136-1154.

Gall, C., Hjorth, J., Watson, D., Dwek, E., Maund, J. R., Fox, O., Leloudas, G., Malesani, D. & Day-Jones, A. C. 2014. Rapid formation of large dust grains in the luminous supernova 2010jl. *Nature,* 511**,** 326-329.

Gallino, R., Busso, M., Picchio, G., Raiteri, C. M. & Renzini, A. 1988. On the role of low-Mass asymptotic giant branch stars in producing a solar system distribution of *s*-process isotopes. *The Astrophysical Journal,* 334**,** L45.

Gallino, R., Busso, M., Lugaro, M., Travaglio, C. & Straniero, O. 2000. s-Process nucleosynthesis in intermediate mass AGB stars and its metallicity dependence. *In:* NOELS, A., MAGAIN, P., CARO, D., JEHIN, E., PARMENTIER, G. & THOUL, A. A., eds. Liege International Astrophysical Colloquia, January 01, 81.

Gehrz, R. D. 1988. The infrared temporal development of classical novae. *Annual Review of Astronomy and Astrophysics,* 26**,** 377-412.

Gehrz, R. D. 1993. Recent infrared observations of novae in outburst. *In:* REGEV, O. & SHAVIV, G., eds. Cataclysmic Variables and Related Physics, January 01, 100.

Gehrz, R. D. 1999. Infrared studies of classical novae and their contributions to the ISM. *Physics Reports,* 311**,** 405-418.

Gehrz, R. D., Truran, J. W., Williams, R. E. & Starrfield, S. 1998. Nucleosynthesis in classical novae and its contribution to the interstellar medium. *Publications of the Astronomical Society of the Pacific,* 110**,** 3-26.

Gehrz, R. D., Evans, A., Helton, L. A., Shenoy, D. P., Banerjee, D. P. K., Woodward, C. E., Vacca, W. D., Dykhoff, D. A., Ashok, N. M., Cass, A. C., Carlon, R. L., Corgan, D. T., Eyres, S. P. S., Joshi, V., Keller, L. D., Krautter, J., Liimets, T., Rushton, M. & Starrfield, S. 2015. The early infrared temporal development of nova Delphini 2013 (V339 DEL) observed with the stratospheric observatory for infrared astronomy (SOFIA) and from the ground. *The Astrophysical Journal,* 812**,** 132.

Gilroy, K. K. & Brown, J. A. 1991. Carbon isotope ratios along the giant branch of M67. *The Astrophysical Journal,* 371**,** 578.

Goldsmith, D. 2012. The far, far future of stars. *Scientific American,* 306**,** 32-39.

Gomez, H. L., Krause, O., Barlow, M. J., Swinyard, B. M., Owen, P. J., Clark, C. J. R., Matsuura, M., Gomez, E. L., Rho, J., Besel, M.-A., Bouwman, J., Gear, W. K., Henning, T., Ivison, R. J., Polehampton, E. T. & Sibthorpe, B. 2012. A cool dust factory in the Crab Nebula: a Herschel study of the filaments. *The Astrophysical Journal,* 760**,** 96.

Gonçalves, D. R., Mampaso, A., Corradi, R. L. M., Perinotto, M., Riera, A. & López-Martín, L. 2004. K 4-47: a planetary nebula excited by photons and shocks. *Monthly Notices of the Royal Astronomical Society,* 355**,** 37-43.

Goriely, S. & Martínez Pinedo, G. 2015. The production of transuranium elements by the r-process nucleosynthesis. *Nuclear Physics A,* 944**,** 158-176.

Goumans, T. P. M. & Kästner, J. 2011. Deuterium enrichment of interstellar methanol explained by atom tunneling. *Journal of Physical Chemistry A,* 115**,** 10767-10774.

Grebenev, S. A., Lutovinov, A. A., Tsygankov, S. S. & Winkler, C. 2012. Hard-X-ray emission lines from the decay of $^{44}$Ti in the remnant of supernova 1987A. *Nature,* 490**,** 373-375.

Grefenstette, B. W., Harrison, F. A., Boggs, S. E., Reynolds, S. P., Fryer, C. L., Madsen, K. K., Wik, D. R., Zoglauer, A., Ellinger, C. I., Alexander, D. M., An, H., Barret, D., Christensen, F. E., Craig, W. W., Forster, K., Giommi, P., Hailey, C. J., Hornstrup, A., Kaspi, V. M., Kitaguchi, T., Koglin, J. E., Mao, P. H., Miyasaka, H., Mori, K., Perri, M., Pivovaroff, M. J., Puccetti, S., Rana, V., Stern, D., Westergaard, N. J. & Zhang, W. W. 2014. Asymmetries in core-collapse supernovae from maps of radioactive $^{44}$Ti in Cassiopeia A. *Nature,* 506**,** 339-342.

Grefenstette, B. W., Fryer, C. L., Harrison, F. A., Boggs, S. E., DeLaney, T., Laming, J. M., Reynolds, S. P., Alexander, D. M., Barret, D., Christensen, F. E., Craig, W. W., Forster, K., Giommi, P., Hailey, C. J., Hornstrup, A., Kitaguchi, T., Koglin, J. E., Lopez, L., Mao, P. H., Madsen, K. K., Miyasaka, H., Mori, K., Perri, M., Pivovaroff, M. J., Puccetti, S., Rana, V., Stern, D., Westergaard, N. J., Wik, D. R., Zhang, W. W. & Zoglauer, A. 2017. The distribution of radioactive $^{44}$Ti in Cassiopeia A. *The Astrophysical Journal,* 834**,** 19.

Grossman, L. 1972. Condensation in the primitive solar nebula. *Geochimica et Cosmochimica Acta,* 36**,** 597-619.

Gyngard, F., Amari, S., Zinner, E. & Ott, U. 2009. Interstellar exposure ages of large presolar SiC grains from the Murchison meteorite. *The Astrophysical Journal,* 694**,** 359-366.

Gyngard, F., Zinner, E., Nittler, L. R., Morgand, A., Stadermann, F. J. & Mairin Hynes, K. 2010. Automated NanoSIMS measurements of spinel stardust from the Murray meteorite. *The Astrophysical Journal,* 717**,** 107-120.

Gyngard, F., Amari, S., Zinner, E., and Marhas, K. K. 2018. Correlated silicon and titanium isotopic compositions of presolar SiC grains from the Murchison CM2 chondrite. *Geochimica et Cosmochimica Acta*, 221, 145-161.

Haenecour, P., Floss, C., Zega, T. J., Croat, T. K., Wang, A., Jolliff, B. L. & Carpenter, P. 2018. Presolar silicates in the matrix and fine-grained rims around chondrules in primitive CO3.0 chondrites: Evidence for pre-accretionary aqueous alteration of the rims in the solar nebula. *Geochimica et Cosmochimica Acta,* 221**,** 379-405.

Haenecour, P., How, J. Y., Zega, T. J., Amari, S., Lodders, K., José, J., Kaji, K., Sunaoshi, T., and Muto, A. 2019. Laboratory evidence for co-codnesed oxygen- and carbon-rich meteoritic stardust from nova outbursts. *Nature Astronomy*, 3, 626-630.

Haenecour, P., Barnes, J. J., Smith, L. R., Hills, D., Bloch, E., Zega, T. J., McCoy, T. J., Thompson, M. S., Keller, L. P., King, A. J., Glavin, D. P., Dowrkin, J. P., Nguyen, A. N., Connolly Jr., H. C., and Lauretta, D. S. 2024. Search for presolar materials and isotopically anomalous diffuse insoluble organic matter in samples from asteroid 101955 Bennu. *in 55$^{th}$ Lunar and Planetary Science Conference*, The Woodlands, TX, #1134.

Hammer, N. J., Janka, H.-Th., and Müller, E. 2010. Three-dimensional simulations of mixing instabilities in supernova explosions. *The Astrophysical Journal*, 714, 1371.

Heck, P. R., Marhas, K. K., Hoppe, P., Gallino, R., Baur, H. & Wieler, R. 2007. Presolar He and Ne isotopes in single circumstellar SiC grains. *The Astrophysical Journal,* 656**,** 1208-1222.

Heck, P. R., Greer, J., Kööp, L., Trappitsch, R., Gyngard, F., Busemann, H., Maden, C., Ávila, J. N., Davis, A. M. & Wieler, R. 2020. Lifetimes of interstellar dust from cosmic ray exposure ages of presolar silicon carbide. *Proceedings of the National Academy of Science,* 117**,** 1884-1889.

Hedrosa, R. P., Abia, C., Busso, M., Cristallo, S., Domínguez, I., Palmerini, S., Plez, B. & Straniero, O. 2013. Nitrogen isotopes in asymptotic giant branch carbon stars and presolar SiC grains: a challenge for stellar nucleosynthesis. *The Astrophysical Journal Letters,* 768**,** L11.

Heger, A., Fryer, C. L., Woosley, S. E., Langer, N. & Hartmann, D. H. 2003. How massive single stars end their life. *The Astrophysical Journal,* 591**,** 288-300.

Herwig, F. 2005. Evolution of asymptotic giant branch stars. *Annual Review of Astronomy and Astrophysics,* 43**,** 435-479.

Herwig, F., Bloecker, T., Schoenberner, D. & El Eid, M. 1997. Stellar evolution of low and intermediate-mass stars. IV. Hydrodynamically-based overshoot and nucleosynthesis in AGB stars. *Astronomy and Astrophysics,* 324**,** L81-L84.

Herwig, F., Langer, N. & Lugaro, M. 2003. The s-process in rotating asymptotic giant branch stars. *The Astrophysical Journal,* 593**,** 1056-1073.

Herwig, F., Pignatari, M., Woodward, P. R., Porter, D. H., Rockefeller, G., Fryer, C. L., Bennett, M. & Hirschi, R. 2011. Convective-reactive Proton-$^{12}$C Combustion in Sakurai's Object (V4334 Sagittarii) and Implications for the Evolution and Yields from the First Generations of Stars. *The Astrophysical Journal,* 727**,** 89.

Hillebrandt, W. & Niemeyer, J. C. 2000. Type IA supernova explosion models. *Annual Review of Astronomy and Astrophysics,* 38**,** 191-230.

Hillion, F., Daigne, B., Girard, F., and Slodzian, G. 1995. The CAMECA "NanoSIMS 50" experimental results. In: A. Benninghoven, B. Hagenhoff, and H. W. Werner (eds), *Secondary Ion Mass Spectrometery SIMS* X. Wiley (Chichester), 979-982.

Hinkle, K. H., Lebzelter, T. & Straniero, O. 2016. Carbon and oxygen isotopic ratios for nearby Miras. *The Astrophysical Journal,* 825**,** 38.

Hiramatsu, D., Howell, D. A., Van Dyk, S. D., Goldberg, J. A., Maeda, K., Moriya, T. J., Tominaga, N., Nomoto, K. i., Hosseinzadeh, G., Arcavi, I., McCully, C., Burke, J., Bostroem, K. A., Valenti, S., Dong, Y., Brown, P. J., Andrews, J. E., Bilinski, C., Williams, G. G., Smith, P. S., Smith, N., Sand, D. J., Anand, G. S., Xu, C., Filippenko, A. V., Bersten, M. C., Folatelli, G., Kelly, P. L., Noguchi, T. & Itagaki, K. 2021. The electron-capture origin of supernova 2018zd. *Nature Astronomy,* 5**,** 903-910.

Höfner, S. & Olofsson, H. 2018. Mass loss of stars on the asymptotic giant branch. Mechanisms, models and measurements. *Astronomy and Astrophysics Review,* 26**,** 1.

Hopp, T., Budde, G. & Kleine, T. 2020. Heterogeneous accretion of Earth inferred from Mo-Ru isotope systematics. *Earth and Planetary Science Letters,* 534**,** 116065.

Hoppe, P. 2006. NanoSIMS: a new tool in cosmochemistry. *Applied Surface Science,* 252**,** 7102-7106.

Hoppe, P. 2016. NanoSIMS and more: new tools in nuclear astrophysics. *Journal of Physics Conference Series*, January 01, 012075.

Hoppe, P., Strebel, R., Eberhardt, P., Amari, S. & Lewis, R. S. 1996. Small SiC grains and a nitride grain of circumstellar origin from the Murchison meteorite: Implications for stellar evolution and nucleosynthesis. *Geochimica et Cosmochimica Acta,* 60**,** 883-907.

Hoppe, P., Annen, P., Strebel, R., Eberhardt, P., Gallino, R., Lugaro, M., Amari, S. & Lewis, R. S. 1997. Meteoritic silicon carbide grains with unusual Si isotopic compositions: evidence for an origin in low-mass, low-metallicity asymptotic giant branch stars. *The Astrophysical Journal,* 487**,** L101-L104.

Hoppe, P. & Besmehn, A. 2002. Evidence for extinct vanadium-49 in presolar silicon carbide grains from supernovae. *The Astrophysical Journal Letters,* 576**,** L69-L72.

Hoppe, P., Leitner, J., Gröner, E., Marhas, K. K., Meyer, B. S. & Amari, S. 2010. NanoSIMS studies of small presolar SiC grains: new insights into supernova nucleosynthesis, chemistry, and dust formation. *The Astrophysical Journal,* 719**,** 1370-1384.

Hoppe, P., Fujiya, W. & Zinner, E. 2012. Sulfur molecule chemistry in supernova ejecta recorded by silicon carbide stardust. *The Astrophysical Journal Letters,* 745**,** L26.

Hoppe, P., Leitner, J. & Kodolányi, J. 2015a. New constraints on the abundances of silicate and oxide stardust from supernovae in the Acfer 094 meteorite. *The Astrophysical Journal Letters,* 808**,** L9.

Hoppe, P., Lodders, K. & Fujiya, W. 2015b. Sulfur in presolar silicon carbide grains from asymptotic giant branch stars. *Meteoritics and Planetary Science,* 50**,** 1122-1138.

Hoppe, P., Leitner, J. & Kodolányi, J. 2017. The stardust abundance in the local interstellar cloud at the birth of the Solar System. *Nature Astronomy,* 1**,** 617-620.

Hoppe, P., Pignatari, M., Kodolányi, J., Gröner, E. & Amari, S. 2018. NanoSIMS isotope studies of rare types of presolar silicon carbide grains from the Murchison meteorite: Implications for supernova models and the role of $^{14}C$. *Geochimica et Cosmochimica Acta,* 221**,** 182-199.

Hoppe, P., Stancliffe, R. J., Pignatari, M. & Amari, S. 2019. Isotopic signatures of supernova nucleosynthesis in presolar silicon carbide grains of type AB with supersolar $^{14}N/^{15}N$ Ratios. *The Astrophysical Journal,* 887**,** 8.

Hoppe, P., Leitner, J., Kodolányi, J. & Vollmer, C. 2021. Isotope systematics of presolar silicate grains: new insights from magnesium and silicon. *The Astrophysical Journal,* 913**,** 10.

Hoppe, P., Leitner, J., Kodolányi, J., Borrmann, S. & Jones, A. P. 2022. Dust from supernovae and their progenitors in the solar nebula. *Nature Astronomy,* 6**,** 1027-1034.

Hoppe, P., Leitner, J., Pignatari, M., and Amari, S. 2023. New constraints for supernova models from presolar silicon carbide X grains with very high $^{26}Al/^{27}Al$ ratios. *The Astrophysical Journal Letters*, 943, L22.

Huss, G. R. & Lewis, R. S. 1995. Presolar diamond, SiC, and graphite in primitive chondrites: Abundances as a function of meteorite class and petrologic type. *Geochimica et Cosmochimica Acta*, 59, 115-160.

Huss, G. R., Hutcheon, I. D. & Wasserburg, G. J. 1997. Isotopic systematics of presolar silicon carbide from the Orgueil (CI) chondrite: Implications for Solar System formation and stellar nucleosynthesis. *Geochimica et Cosmochimica Acta,* 61**,** 5117-5148.

Huss, G. R., Meshik, A. P., Smith, J. B. & Hohenberg, C. M. 2003. Presolar diamond, silicon carbide, and graphite in carbonaceous chondrites: implications for thermal processing in the solar nebula. *Geochimica et Cosmochimica Acta,* 67**,** 4823-4848.

Hutchison, M. A., Bodénan, J.-D., Mayer, L., and Schönbächler, M. 2022. Presolar grain dynamics: Creating nucleosynthetic variations through a combination of draf and viscous evolution. *Monthly Notices of the Royal Astronomical Society*, 512, 5874-5894.

Hystad, G., Boujibar, A., Liu, N., Nittler, L. R. & Hazen, R. M. 2022. Evaluation of the classification of pre-solar silicon carbide grains using consensus clustering with resampling methods: An assessment of the confidence of grain assignments. *Monthly Notices of the Royal Astronomical Society,* 510**,** 334-350.

Iben, I., Jr. & MacDonald, J. 1995. The born again AGB phenomenon. *In:* KOESTER, D. & WERNER, K. (eds.) *White Dwarfs.*

Iliadis, C., Downen, L. N., José, J., Nittler, L. R. & Starrfield, S. 2018. On presolar stardust grains from CO classical novae. *The Astrophysical Journal,* 855**,** 76.

Indebetouw, R., Matsuura, M., Dwek, E., Zanardo, G., Barlow, M. J., Baes, M., Bouchet, P., Burrows, D. N., Chevalier, R., Clayton, G. C., Fransson, C., Gaensler, B., Kirshner, R., Lakićević, M., Long, K. S., Lundqvist, P., Martí-Vidal, I., Marcaide, J., McCray, R., Meixner, M., Ng, C.-Y., Park, S., Sonneborn, G., Staveley-Smith, L., Vlahakis, C. & van Loon, J. 2014. Dust production and particle acceleration in supernova 1987A revealed with ALMA. *The Astrophysical Journal Letters,* 782**,** L2.

Iyudin, A. F., Diehl, R., Bloemen, H., Hermsen, W., Lichti, G. G., Morris, D., Ryan, J., Schoenfelder, V., Steinle, H., Varendorff, M., de Vries, C. & Winkler, C. 1994. COMPTEL observations of $^{44}Ti$ gamma-ray line emission form CAS A. *Astronomy and Astrophysics,* 284**,** L1-L4.

Jadhav, M., Amari, S., Marhas, K. K., Zinner, E., Maruoka, T. &Gallino, R. 2008. New stellar sources for high-density presolar graphite grains. *The Astrophysical Journal,* 682, 1479-1485.

Jadhav, M., Zinner, E., Amari, S., Maruoka, T., Marhas, K. K. & Gallino, R. Multi-element isotopic analyses of presolar graphite grains from Orgueil. *Geochimica et Cosmochimica Acta*, 113, 193-224.

Jadhav, M., Pignatari, M., Herwig, F., Zinner, E., Gallino, R. & Huss, G. R. 2013b. Relics of ancient post-AGB stars in a primitive meteorite. *The Astrophysical Journal Letters,* 777**,** L27.

Jenkins, E. B. 2009. A unified representation of gas-phase element depletions in the interstellar medium. *The Astrophysical Journal,* 700**,** 1299-1348.

Jones, A. P. 2004. Dust destruction processes. *In:* WITT, A. N., CLAYTON, G. C. & DRAINE, B. T., eds. Astrophysics of Dust, May 01, 347.

Jordan, G. C., Gupta, S. S. & Myer, B. S. Nuclear reactions important in α-rich freeze-outs. *Physical Review C*, 68, 065801.

José, J. & Hernanz, M. 1998. Nucleosynthesis in classical novae: CO versus ONe white dwarfs. *The Astrophysical Journal,* 494**,** 680-690.

José, J., Hernanz, M., Amari, S., Lodders, K. & Zinner, E. 2004. The imprint of nova nucleosynthesis in presolar grains. *The Astrophysical Journal,* 612**,** 414-428.

José, J. & Hernanz, M. 2007. The origin of presolar nova grains. *Meteoritics and Planetary Science,* 42**,** 1135-1143.

José, J., Shore, S. N. & Casanova, J. 2020. 123-321 models of classical novae. *Astronomy and Astrophysics,* 634**,** A5.

Käppeler, F., Gallino, R., Bisterzo, S. & Aoki, W. 2011. The *s* process: Nuclear physics, stellar models, and observations. *Reviews of Modern Physics,* 83**,** 157-194.

Karakas, A. I. & Lugaro, M. 2016. Stellar yields from metal-rich asymptotic giant branch models. *The Astrophysical Journal,* 825**,** 26.

Keller, L. P. & Messenger, S. 2011. On the origins of GEMS grains. *Geochimica et Cosmochimica Acta,* 75**,** 5336-5365.

Knie, K., Korschinek, G., Faestermann, T., Dorfi, E. A., Rugel, G. & Wallner, A. 2004. $^{60}$Fe Anomaly in a deep-sea manganese crust and implications for a nearby supernova source. *Physical Review Letters,* 93**,** 171103.

Kobayashi, C., Karakas, A. I. & Lugaro, M. 2020. The origin of elements from carbon to uranium. *The Astrophysical Journal,* 900**,** 179.

Kodolányi, J., Vollmer, C., Hoppe, P., Müller, M. 2018. Structural investigation of silicon carbide X grains: Constraints on condensation in supernova ejecta. *The Astrophysical Journal*, 868, 34.

Kruijer, T. S., Burkhardt, C., Budde, G. & Kleine, T. 2017. Age of Jupiter inferred from the distinct genetics and formation times of meteorites. *Proceedings of the National Academy of Science,* 114**,** 6712-6716.

Kruijer, T. S., Kleine, T. & Borg, L. E. 2020. The great isotopic dichotomy of the early Solar System. *Nature Astronomy,* 4**,** 32-40.

Lau, R. M., Eldridge, J. J., Hankins, M. J., Lamberts, A., Sakon, I., and Williams, P. M. 2020. Revisting the impact of dust production from carbon-rich Wolf-Rayet binaries. *The Astrophysical Journal*, 898, 74.

Lebzelter, T., Straniero, O., Hinkle, K. H., Nowotny, W. & Aringer, B. 2015. Oxygen isotopic ratios in intermediate-mass red giants. *Astronomy and Astrophysics,* 578**,** A33.

Leitner, J., Kodolányi, J., Hoppe, P. & Floss, C. 2012. Laboratory analysis of presolar silicate stardust from a nova. *The Astrophysical Journal Letters,* 754**,** L41.

Leitner, J. & Hoppe, P. 2019. A new population of dust from stellar explosions among meteoritic stardust. *Nature Astronomy,* 3**,** 725-729.

Leitner, J., Metzler, K., Vollmer, C., Floss, C., Haenecour, P., Kodolányi, J., Harries, D. & Hoppe, P. 2020. The presolar grain inventory of fine-grained chondrule rims in the Mighei-type (CM) chondrites. *Meteoritics and Planetary Science,* 55**,** 1176-1206.

Leitner, J. & Hoppe, P. 2022. Ca-isotope investigation of silicate stardust. *53rd Lunar and Planetary Science Conference*, #1949 (abstract).

Leitner, J., Hoppe, P. & Trieloff, M. 2024. Calcium and potassium isotope studies of presolar silicates. *55th Lunar and Planetary Science Conference*, #1443 (abstract).

Lewis, R. S., Ming, T., Wacker, J. F., Anders, E. & Steel, E. 1987. Interstellar diamonds in meteorites. *Nature*, 326, 160-162.

Lewis, K. M., Lugaro, M., Gibson, B. K. & Pilkington, K. 2013. Decoding the message from meteoritic stardust silicon carbide grains. *The Astrophysical Journal Letters Letters,* 768**,** L19.

Lewis, J. B., Floss, C. & Gyngard, F. 2018. Origin of nanodiamonds from Allende constrained by statistical analysis of C isotopes from small clusters of acid residue by NanoSIMS. *Geochimica et Cosmochimica Acta*, 21, 237-254.

Leya, I., Schönbächler, M., Wiechert, U., Krähenbühl, U. & Halliday, A. N. 2008. Titanium isotopes and the radial heterogeneity of the solar system. *Earth and Planetary Science Letters,* 266**,** 233-244.

Li, K.-A., Qi, C., Lugaro, M., Yagüe López, A., Karakas, A. I., den Hartogh, J., Gao, B.-S. & Tang, X.-D. 2021. The stellar β-decay rate of $^{134}$Cs and its impact on the barium nucleosynthesis in the s-process. *The Astrophysical Journal Letters,* 919**,** L19.

Lichtenberg, T., Parker, R. J. & Meyer, M. R. 2016. Isotopic enrichment of forming planetary systems from supernova pollution. *Monthly Notices of the Royal Astronomical Society,* 462**,** 3979-3992.

Lichtenberg, T., Golabek, G. J., Burn, R., Meyer, M. R., Alibert, Y., Gerya, T. V. & Mordasini, C. 2019. A water budget dichotomy of rocky protoplanets from $^{26}$Al-heating. *Nature Astronomy,* 3**,** 307-313.

Lin, Y., Gyngard, F. & Zinner, E. 2010. Isotopic analysis of supernova SiC and $Si_3N_4$ Grains from the Qingzhen (EH3) chondrite. *The Astrophysical Journal,* 709**,** 1157-1173.

Liu, N., Gallino, R., Bisterzo, S., Davis, A. M., Savina, M. R. & Pellin, M. J. 2014a. The $^{13}$C-pocket structure in AGB models: constraints from zirconium isotope abundances in single mainstream SiC grains. *The Astrophysical Journal,* 788**,** 163.

Liu, N., Savina, M. R., Davis, A. M., Gallino, R., Straniero, O., Gyngard, F., Pellin, M. J., Willingham, D. G., Dauphas, N., Pignatari, M., Bisterzo, S., Cristallo, S. & Herwig, F. 2014b. Barium isotopic composition of mainstream silicon carbides from Murchison: constraints for *s*-process nucleosynthesis in asymptotic giant branch stars. *The Astrophysical Journal,* 786**,** 66.

Liu, N., Savina, M. R., Gallino, R., Davis, A. M., Bisterzo, S., Gyngard, F., Käppeler, F., Cristallo, S., Dauphas, N., Pellin, M. J. & Dillmann, I. 2015. Correlated strontium and barium isotopic compositions of acid-cleaned single mainstream silicon carbides from Murchison. *The Astrophysical Journal,* 803**,** 12.

Liu, N., Nittler, L. R., Alexander, C. M. O' D., Wang, J., Pignatari, M., José, J. & Nguyen, A. 2016. Stellar origins of extremely $^{13}$C- and $^{15}$N-enriched presolar SiC grains: novae or supernovae? *The Astrophysical Journal,* 820**,** 140.

Liu, N., Nittler, L. R., Pignatari, M., Alexander, C. M. O' D. & Wang, J. 2017a. Stellar origin of $^{15}$N-rich presolar SiC grains of type AB: supernovae with explosive hydrogen burning. *The Astrophysical Journal Letters,* 842**,** L1.

Liu, N., Steele, A., Nittler, L. R., Stroud, R. M., De Gregorio, B. T., Alexander, C. M. O' D. & Wang, J. 2017b. Coordinated EDX and micro-Raman analysis of presolar silicon carbide: A novel, nondestructive method to identify rare subgroup SiC. *Meteoritics and Planetary Science,* 52**,** 2550-2569.

Liu, N., Stephan, T., Boehnke, P., Nittler, L. R., Alexander, C. M. O' D., Wang, J., Davis, A. M., Trappitsch, R. & Pellin, M. J. 2017c. J-type carbon stars: a dominant source of $^{14}$N-rich presolar SiC grains of type AB. *The Astrophysical Journal Letters,* 844**,** L12.

Liu, N., Gallino, R., Cristallo, S., Bisterzo, S., Davis, A. M., Trappitsch, R. & Nittler, L. R. 2018a. New constraints on the major neutron source in low-mass AGB stars. *The Astrophysical Journal,* 865**,** 112.

Liu, N., Nittler, L. R., Alexander, C. M. O' D. & Wang, J. 2018b. Late formation of silicon carbide in type II supernovae. *Science Advances,* 4**,** eaao1054.

Liu, N., Stephan, T., Boehnke, P., Nittler, L. R., Meyer, B. S., Alexander, C. M. O' D., Davis, A. M., Trappitsch, R. & Pellin, M. J. 2018c. Common occurrence of explosive hydrogen burning in Type II supernovae. *The Astrophysical Journal,* 855**,** 144.

Liu, N., Stephan, T., Cristallo, S., Gallino, R., Boehnke, P., Nittler, L. R., Alexander, C. M. O' D., Davis, A. M., Trappitsch, R., Pellin, M. J. & Dillmann, I. 2019. Presolar silicon carbide grains of types Y and Z: their molybdenum isotopic compositions and stellar origins. *The Astrophysical Journal,* 881**,** 28.

Liu, N., Meyer, B. S., Nittler, L. R. & Alexander, C. M. O' D. 2020a. A new method for constraining explosive environments in Type II supernovae using presolar silicon carbide X grain isotopic data. *in 51$^{st}$ Lunar and Planetary Science Conference*, The Woodlands, TX., #2379 (abstract).

Liu, N., Ogliore, R. C. & Vacher, L. G. 2020b. NanoSIMS isotopic investigation of xenolithic carbonaceous clasts from the Kapoeta howardite. *Geochimica et Cosmochimica Acta,* 283**,** 243-264.

Liu, N., Barosch, J., Nittler, L. R., Alexander, C. M. O' D., Wang, J., Cristallo, S., Busso, M. & Palmerini, S. 2021. New multielement isotopic compositions of presolar SiC grains: implications for their stellar origins. *The Astrophysical Journal Letters,* 920**,** L26.

Liu, N. & Ogliore, R. C. 2021. NanoSIMS isotopic investigation of the CO2 chondrite Dominion Range 14359. *in 84th Annual Meeting of the Meteoritical Society, Chicago*, #6069 (abstract).

Liu, N., Cristallo, S. & Vescovi, D. 2022a. Slow Neutron-Capture Process: Low-Mass Asymptotic Giant Branch Stars and Presolar Silicon Carbide Grains. *Universe,* 8**,** 362.

Liu, N., Dauphas, N., Cristallo, S., Palmerini, S. & Busso, M. 2022b. Oxygen and aluminum-magnesium isotopic systematics of presolar nanospinel grains from CI chondrite Orgueil. *Geochimica et Cosmochimica Acta,* 319**,** 296-317.

Liu, N., Alexander, C. M. O' D. & Nittler, L. R. 2022c. Intrinsic nitrogen isotope ratios of presolar silicon carbide grains. 85th Annual Meeting of The Meteoritical Society, Glascow, U.K., 6384.

Liu, N., Steele, A., Alexander, C. M. O' D., Nittler, L. R. & Barosch, J. 2022d. Distributions of chemical, isotopic, and structural compositions of presolar silicon carbide grains. *in 53rd Lunar and Planetary Science Conference,* The Woodlands, TX., #1434 (abstract).

Liu, N., Stephan, T., Cristallo, S., Vescovi, D., Gallino, R., Nittler, L. R., Alexander, C. M. O' D. & Davis, A. M. 2022e. Presolar silicon carbide grains of types Y and Z: their strontium and barium isotopic compositions and stellar origins. *The European Physical Journal A* 58, 216.

Liu, N., Alexander, C. M. O' D., Wang, J., Nittler, L. R., and Meyer, B. S. 2023. Radioactive $^{44}$Ti found in $^{15}$N-rich AB presolar SiC grains. *in 86th Annual Meeting of the Meteoritical Society, Los Angeles*, #6230 (abstract).

Liu, N., Alexander C. M. O' D., Meyer, B. S., Nittler, L. R., Wang, J., and Stroud, R. M. 2024. Explosive nucleosynthesis in core-collapse Type II supernovae: Insights from new C, N, Si, and Al-Mg isotopic compositions of presolar grains. *The Astrophysical Journal Letters*, 961, L22 (14 pp).

Lodders, K. 2021. Relative atomic solar system abundances, mass fractions, and atomic masses of the elements and their isotopes, composition of the solar photosphere, and compositions of the major chondritic meteorite groups. *Space Science Reviews,* 217**,** 44.

Lodders, K. & Fegley, B., Jr. 1995. The origin of circumstellar silicon carbide grains found in meteorites. *Meteoritics,* 30**,** 661.

Lopez, L. A., Grefenstette, B. W., Reynolds, S. P., An, H., Boggs, S. E., Christensen, F. E., Craig, W. W., Eriksen, K. A., Fryer, C. L., Hailey, C. J., Harrison, F. A., Madsen, K. K., Stern, D. K., Zhang, W. W. & Zoglauer, A. 2015. A spatially resolved study of the synchrotron emission and titanium in Tycho's supernova remnant using NuSTAR. *The Astrophysical Journal,* 814**,** 132.

Lugaro, M., Zinner, E., Gallino, R. & Amari, S. 1999. Si isotopic ratios in mainstream presolar SiC grains revisited. *The Astrophysical Journal*, 527, 369.

Lugaro, M., Davis, A. M., Gallino, R., Pellin, M. J., Straniero, O. & Käppeler, F. 2003. Isotopic compositions of strontium, zirconium, molybdenum, and barium in single presolar SiC grains and asymptotic giant branch stars. *The Astrophysical Journal,* 593**,** 486-508.

Lugaro, M., Tagliente, G., Karakas, A. I., Milazzo, P. M., Käppeler, F., Davis, A. M. & Savina, M. R. 2014. The impact of updated Zr neutron-capture cross sections and new asymptotic giant branch models on our understanding of the s process and the origin of stardust. *The Astrophysical Journal,* 780**,** 95.

Lugaro, M., Karakas, A. I., Bruno, C. G., Aliotta, M., Nittler, L. R., Bemmerer, D., Best, A., Boeltzig, A., Broggini, C., Caciolli, A., Cavanna, F., Ciani, G. F., Corvisiero, P., Davinson, T., Depalo, R., di Leva, A., Elekes, Z., Ferraro, F., Formicola, A., Fülöp, Z., Gervino, G., Guglielmetti, A., Gustavino, C., Gyürky, G., Imbriani, G., Junker, M., Menegazzo, R., Mossa, V., Pantaleo, F. R., Piatti, D., Prati, P., Scott, D. A., Straniero, O., Strieder, F., Szücs, T., Takács, M. P. & Trezzi, D. 2017. Origin of meteoritic stardust unveiled by a revised proton-capture rate of $^{17}$O. *Nature Astronomy,* 1**,** 0027.

Lugaro, M., Karakas, A. I., Pető, M. & Plachy, E. 2018. Do meteoritic silicon carbide grains originate from asymptotic giant branch stars of super-solar metallicity? *Geochimica et Cosmochimica Acta,* 221**,** 6-20.

Lugaro, M., Cseh, B., Világos, B., Karakas, A. I., Ventura, P., Dell'Agli, F., Trappitsch, R., Hampel, M., D'Orazi, V., Pereira, C. B., Tagliente, G., Szabó, G. M., Pignatari, M., Battino, U., Tattersall, A., Ek, M., Schönbächler, M., Hron, J. & Nittler, L. R. 2020. Origin of large meteoritic SiC stardust grains in metal-rich AGB stars. *The Astrophysical Journal,* 898**,** 96.

MacPherson, G. J. 2014. Calcium-aluminum-rich inclusions in chondritic meteorites. *In:* DAVIS, A. M. (ed.) *Meteorites and Cosmochemical Processes.* Second ed.: Elsevier.

Marengo, M., Busso, M., Silvestro, G. & Fazio, G. 1998. AGB circumstellar envelope dust mineralogy from ground based mid-IR imaging and photometry. *IAU Symposium,* 191**,** P310.

Marhas, K. K., Amari, S., Gyngard, F., Zinner, E., and Gallino, R. 2008. Iron and nickel isotopic ratios in presolar SiC grains. *The Astrophysical Journal*, 689, 622-645.

Martins, R., Kuthning, S., Coles, B. J., Kreissig, K., and Rehkämper, M. 2023. Nucleosynthetic isotope anomalies of zinc in meteorites constrain the origin of Earth's volatiles. *Science*, 379, 369.

Matsuura, M., Dwek, E., Meixner, M., Otsuka, M., Babler, B., Barlow, M. J., Roman-Duval, J., Engelbracht, C., Sandstrom, K., Lakićević, M., van Loon, J. T., Sonneborn, G., Clayton, G. C., Long, K. S., Lundqvist, P., Nozawa, T., Gordon, K. D., Hony, S., Panuzzo, P., Okumura, K., Misselt, K. A., Montiel, E. & Sauvage, M. 2011. Herschel detects a massive dust reservoir in supernova 1987A. *Science,* 333**,** 1258.

Marty, B., Chaussidon, M., Wiens, R. C., Jurewicz, A. J. G. & Burnett, D. S. 2011. A $^{15}$N-poor isotopic composition for the solar system as shown by Genesis solar wind samples. *Science*, 332, 1533-1536.

Mayor, M., & Queloz, D. 1995. A Jupiter-mass companion to a solar-type star. *Nature*, 378, 355-359.

McKeegan, K. D., Kallio, A. P. A., Heber, V. S., Jarzebinski, G., Mao, P. H., Coath, C. D., Kunihiro, T., Wiens, R. C., Nordholt, J. E., Moses, R. W., Reisenfeld, D. B., Jurewicz, A. J. G. & Burnett, D. S. 2011. The oxygen isotopic composition of the Sun inferred from captured solar wind. *Science,* 332**,** 1528.

McSween, J., H. & Huss, G. R. 2010. Introduction to cosmochemistry. *In:* MCSWEEN, J., H. & HUSS, G. R. (eds.) *Cosmochemistry.* Cambridge, UK: Cambridge University Press.

Mendybaev, R. A., Beckett, J. R., Grossman, L., Stolper, E., Cooper, R. F. & Bradley, J. P. 2002. Volatilization kinetics of silicon carbide in reducing gases: an experimental study with applications to the survival of presolar grains in the solar nebula. *Geochimica et Cosmochimica Acta,* 66**,** 661-682.

Merk, R., Breuer, D. & Spohn, T. 2002. Numerical modeling of $^{26}$Al-induced radioactive melting of asteroids considering accretion. *Icarus,* 159**,** 183-191.

Messenger, S., Keller, L. P., Stadermann, F. J., Walker, R. M. & Zinner, E. 2003. Samples of stars beyond the solar system: Silicate grains in interplanetary dust. *Science,* 300**,** 105-108.

Meyer, B. S., Weaver, T. A. & Woosley, S. E. 1995. Isotope source table for a 25 Msun supemova. *Meteoritics,* 30**,** 325.

Meyer, B. S., Clayton, D. D. & The, L.-S. 2000. Molybdenum and zirconium isotopes from a supernova neutron burst. *The Astrophysical Journal Letters,* 540**,** L49-L52.

Micelotta, E. R., Dwek, E. & Slavin, J. D. 2016. Dust destruction by the reverse shock in the Cassiopeia A supernova remnant. *Astronomy and Astrophysics,* 590**,** A65.

Mucciola, R., Paradela, C., Manna, A., Alaerts, G., Massimi, C., Kopecky, S., Mengoni, A., Moens, A., Schillebeeckx, P., Wynants, R., Aberle, O., Alcayne, V., Altieri, S., Amaducci, S., Amar Es-Sghir, H., Andrzejewski, J., Babiano-Suarez, V., Bacak, M., Balibrea, J., Bennett, S., Bernardes, A.P., Berthoumieux, E., Bosnar, D., Busso, M., Caamaño, M., Calviño, F., Calviani, M., Cano-Ott, D., Casanovas, A., Castelluccio, D.M., Cerutti, F., Cescutti, G., Chasapoglou, S., Chiaveri, E., Colombetti, P., Colonna, N., Console Camprini, P.C., Cortés, G., Cortés-Giraldo, M.A., Cosentino, L., Cristallo, S., Di Castro, M., Diacono, D., Diakaki, M., Dietz, M., Domingo-Pardo, C., Dressler, R., Dupont, E., Durán, I., Eleme, Z., Fargier, S., Fernández-Domínguez, B., Finocchiaro, P., Fiore, S., Furman, V., García-Infantes, F., Gawlik-Ramięga, A., Gervino, G., Gilardoni, S., González-Romero, E., Guerrero, C., Gunsing, F., Gustavino, C., Heyse, J., Jenkins, D.G., Jericha, E., Junghans, A., Kadi, Y., Katabuchi, T., Knapová, I., Kokkoris, M., Kopatch, Y., Krtička, M., Kurtulgil, D., Ladarescu, I., Lederer-Woods, C., Lerendegui-Marco, J., Lerner, G., Martínez, T., Masi, A., Mastinu, P., Mastromarco, M., Matteucci, F., Maugeri, E.A., Mazzone, A., Mendoza, E., Michalopoulou, V., Milazzo, P.M., Murtas, F., Musacchio-Gonzalez, E., Musumarra, A., Negret, A., Oprea, A., Pérez-Maroto, P., Patronis, N., Pavón-Rodríguez, J.A., Pellegriti, M.G., Perkowski, J., Petrone, C., Piersanti, L., Pirovano, E., Pomp, S., Porras, I., Praena, J., Protti, N., Quesada, J.M., Rauscher, T., Reifarth, R., Rochman, D., Romanets, Y., Romano, F., Rubbia, C., Sánchez, A., Sabaté-Gilarte, M., Schumann, D., Sekhar, A., Smith, A.G., Sosnin, N.V., Spelta, M., Stamati, M.E., Tagliente, G., Tarifeño-Saldivia, A., Tarrío, D., Terranova, N., Torres-Sánchez, P., Urlass, S., Valenta, S., Variale, V., Vaz, P., Vescovi, D., Vlachoudis, V., Vlastou, R., Wallner, A., Woods, P.J., Wright, T. & Žugec, P. 2023. Neutron capture and total cross-section measurements on $^{94,95,96}$Mo at n_TOF and GELINA. *European Physical Journal Web of Conferences*, 284, 01031.

Nanne, J. A. M., Nimmo, F., Cuzzi, J. N. & Kleine, T. 2019. Origin of the non-carbonaceous-carbonaceous meteorite dichotomy. *Earth and Planetary Science Letters,* 511**,** 44-54.

Nguyen, A. N. & Zinner, E. 2004. Discovery of ancient silicate stardust in a meteorite. *Science,* 303**,** 1496-1499.

Nguyen, A. N., Stadermann, F. J., Zinner, E., Stroud, R. M., Alexander, C. M. O' D. & Nittler, L. R. 2007. Characterization of presolar silicate and oxide grains in primitive carbonaceous chondrites. *The Astrophysical Journal,* 656**,** 1223-1240.

Nguyen, A. N. & Messenger, S. 2014. Resolving the stellar sources of isotopically rare presolar silicate grains through Mg and Fe isotopic analyses. *The Astrophysical Journal,* 784**,** 149.

Nguyen, A. N., Nittler, L. R., Alexander, C. M. O' D., and Hoppe, P. 2018. Titanium isotopic compositions of rare presolar SiC grain types form the Murchison meteorite. *Geochimica et Cosmochimica Acta*, 221, 162-181.

Nguyen, A. N., Nakamura-Messenger, K., Keller, L. P. & Messenger, S. 2022. Diverse sssemblage of presolar and solar system materials in anhydrous interplanetary dust particles: coordinated NanoSIMS and TEM analyses. *Geochimica et Cosmochimica Acta*, 336, 131-149.

Nguyen, A.N., Mane, P., Keller, L.P., Piani, L., Abe, Y., Aléon, J., Alexander, C. M. O' D., Amari, S., Amelin, Y., Bajo, K.-i., Bizzarro, M., Bouvier, A., Carlson, R.W., Chaussidon, M., Choi, B.-G., Dauphas, N., Davis, A.M., Di Rocco, T., Fujiya, W., Fukai, R., Gautam, I., Haba, M.K., Hibiya, Y., Hidaka, H., Homma, H., Hoppe, P., Huss, G.R., Ichida, K., Iizuka, T., Ireland, T.R., Ishikawa, A., Itoh, S., Kawasaki, N., Kita, N.T., Kitajima, K., Kleine, T., Komatani, S., Krot, A.N., Liu, M.-C., Masuda, Y., McKeegan, K.D., Morita, M., Motomura, K., Moynier, F., Nakai, I., Nagashima, K., Nesvorný, D., Nittler, L., Onose, M., Pack, A., Park, C., Qin, L., Russell, S.S., Sakamoto, N., Schönbächler, M., Tafla, L., Tang, H., Terada, K., Terada, Y., Usui, T., Wada, S., Wadhwa, M., Walker, R.J., Yamashita, K., Yin, Q.-Z., Yokoyama, T., Yoneda, S., Young, E.D., Yui, H., Zhang, A.-C., Nakamura, T., Naraoka, H., Noguchi, T., Okazaki, R., Sakamoto, K., Yabuta, H., Abe, M., Miyazaki, A., Nakato, A., Nishimura, M., Okada, T., Yada, T., Yogata, K., Nakazawa, S., Saiki, T., Tanaka, S., Terui, F., Tsuda, Y., Watanabe, S.-i., Yoshikawa, M., Tachibana, S., and Yurimoto, H. (2023) Abundant presolar grains and primordial organics preserved in carbon-rich exogenous clasts in asteroid Ryugu. *Science Advances,* 9, eadh1003.

Nguyen, A. N., Keller, L. P., Clemett, S. J., Barnes, J. J., Zega, T. J., McCoy, T. J., Thompson, M. S., Haenecour, P., King, A. J., Connolly Jr., H. C., and Lauretta, D. S. 2024. Identification of presolar grains in quick-look samples of asteroid Bennu returned by the OSIRIS-REx Mission. *in 55$^{th}$ Lunar and Planetary Science Conference*, The Woodlands, TX., #2580 (abstract).

Nicolussi, G. K., Pellin, M. J., Lewis, R. S., Davis, A. M., Clayton, R. N. & Amari, S. 1998. Zirconium and molybdenum in individual circumstellar graphite grains: new isotopic data on the nucleosynthesis of heavy elements. *The Astrophysical Journal,* 504**,** 492-499.

Niculescu-Duvaz, M., Barlow, M. J., Bevan, A., Wesson, R., Milisavljevic, D., De Looze, I., Clayton, G. C., Krafton, K., Matsuura, M. & Brady, R. 2022. Dust masses for a large sample of core-collapse supernovae from optical emission line asymmetries: dust formation on 30-year time-scales. *Monthly Notices of the Royal Astronomical Society,* 515**,** 4302-4343.

Nie, N. X., Wang, D., Torrano, Z. A., Carlson, R. W., Alexander, C. M. O' D., and Shahar, A. 2023. Meteorites have interited nucleosynthetic anomalies of potassium-40 produced in supernovae. *Science*, 379, 372.

Nimmo, F., Primack, J., Faber, S. M., Ramirez-Ruiz, E. & Safarzadeh, M. 2020. Radiogenic heating and its influence on rocky planet dynamos and habitability. *The Astrophysical Journal Letters,* 903**,** L37.

Nittler, L. R. 2005. Constraints on heterogeneous Galactic chemical evolution from meteoritic stardust. *The Astrophysical Journal,* 618**,** 281-296.

Nittler, L. R. 2009. On the mass and metallicity distributions of the parent AGB stars of O-rich presolar stardust grains. *Publications of the Astronomical Society of Australia,* 26**,** 271-277.

Nittler, L. R. 2019. Isotopic imprints of super-AGB stars and their supernovae in the solar system. *82nd Annual Meeting of The Meteoritical Society*, July 01, #6424 (abstract).

Nittler, L. R., Hoppe, P., Alexander, C. M. O' D., Amari, S., Eberhardt, P., Gao, X., Lewis, R. S., Strebel, R., Walker, R. M. & Zinner, E. 1995. Silicon nitride from supernovae. *The Astrophysical Journal,* 453**,** L25.

Nittler, L. R., Amari, S., Zinner, E., Woosley, S. E. & Lewis, R. S. 1996. Extinct $^{44}$Ti in presolar graphite and SiC: proof of a supernova origin. *The Astrophysical Journal,* 462**,** L31.

Nittler, L. R. & Alexander, C. M. O' D. 2003. Automated isotopic measurements of micron-sized dust: application to meteoritic presolar silicon carbide. *Geochimica et Cosmochimica Acta,* 67**,** 4961-4980.

Nittler, L. R. & Hoppe, P. 2005. Are presolar silicon carbide grains from novae actually from supernovae? *The Astrophysical Journal Letters,* 631**,** L89-L92.

Nittler, L. R. & Dauphas, N. 2006. Meteorites and the Chemical Evolution of the Milky Way. *In:* LAURETTA, D. S. & MCSWEEN, H. Y. (eds.) *Meteorites and the Early Solar System II.*

Nittler, L. R., Alexander, C. M. O' D., Gallino, R., Hoppe, P., Nguyen, A. N., Stadermann, F. J. & Zinner, E. K. 2008. Aluminum-, calcium- and titanium-rich oxide stardust in ordinary chondrite meteorites. *The Astrophysical Journal,* 682**,** 1450-1478.

Nittler, L.R., Gyngard, F., Zinner, E.& Stroud, R.M. 2011. Mg and Ca isotopic anomalies in presolar oxices: large anoamlies in a Group 3 hibonite grain. in *42nd Lunar and Planetary Science Conferenc*e, The Woodlands, TX, #1872 (abstract).

Nittler, L. R. & Ciesla, F. 2016. Astrophysics with extraterrestrial materials. *Annual Review of Astronomy and Astrophysics,* 54**,** 53-93.

Nittler, L. R., Alexander, C. M. O' D., Davidson, J., Riebe, M. E. I., Stroud, R. M. & Wang, J. 2018a. High abundances of presolar grains and $^{15}$N-rich organic matter in CO3.0 chondrite Dominion Range 08006. *Geochimica et Cosmochimica Acta,* 226**,** 107-131.

Nittler, L. R., Alexander, C. M. O' D., Liu, N. & Wang, J. 2018b. Extremely $^{54}$Cr- and $^{50}$Ti-rich presolar oxide grains in a primitive meteorite: Formation in rare types of supernovae and implications for the astrophysical context of solar system birth. *The Astrophysical Journal Letters,* 856**,** L24.

Nittler, L. R., Stroud, R. M., Alexander, C. M. O' D. & Howell, K. 2020. Presolar grains in primitive ungrouped carbonaceous chondrite Northwest Africa 5958. *Meteoritics and Planetary Science,* 55**,** 1160-1175.

Nittler, L. R., Alexander, C. M. O' D., Patzer, A. & Verdier-Paoletti, M. J. 2021. Presolar stardust in highly pristine CM chondrites Asuka 12169 and Asuka 12236. *Meteoritics and Planetary Science,* 56**,** 260-276.

Nittler, L.R., Barosch, J., Alexander, C. M. O' D., Wang, J., Brearley, A. &Dobrică, E. 2023. Presoalr grain aggregates: a composite oxide-silicate from a low-metallicity AGB star and a possible circumstellar-interstellar grain aggregate. *86th Annual Meeting of the Meteoritical Society*, #6079.

Nollett, K. M., Busso, M. & Wasserburg, G. J. 2003. Cool bottom processes on the thermally pulsing asymptotic giant branch and the isotopic composition of circumstellar dust grains. *The Astrophysical Journal,* 582**,** 1036-1058.

Nomoto, K. i. 1987. Evolution of 8--10 $M_{sun}$ stars toward electron capture supernovae. II. Collapse of an O + NE + MG core. *The Astrophysical Journal,* 322**,** 206.

Nucci, M. C. & Busso, M. 2014. Magnetohydrodynamics and deep mixing in evolved stars. I. two- and three-dimensional analytical models for the asymptotic giant branch. *The Astrophysical Journal,* 787**,** 141.

Ott, U., Stephan, T., Hoppe, P. & Savina, M. R. 2019. Isotopes of barium as a chronometer for supernova dust formation. *The Astrophysical Journal,* 885**,** 128.

Pal, I., Jadhav, M., Savina, M. R., Shulaker, D. Z., Dory, C. J., Gyngard, F., Kita, N. & Amari, S. 2022. Heavy element isotopic measurements of high density presolar graphite grains with LION. *in 53rd Lunar and Planetary Science Conference*, The Woodlands, TX., #1262 (abstract).

Palmerini, S., Cristallo, S., Busso, M., Abia, C., Uttenthaler, S., Gialanella, L. & Maiorca, E. 2011. Deep mixing in evolved stars. II. interpreting Li abundances in red giant branch and asymptotic giant branch stars. *The Astrophysical Journal,* 741**,** 26.

Palmerini, S., Trippella, O. & Busso, M. 2017. A deep mixing solution to the aluminum and oxygen isotope puzzles in presolar grains. *Monthly Notices of the Royal Astronomical Society,* 467**,** 1193-1201.

Palmerini, S., Busso, M., Vescovi, D., Naselli, E., Pidatella, A., Mucciola, R., Cristallo, S., Mascali, D., Mengoni, A., Simonucci, S. & Taioli, S. 2021a. Presolar grain isotopic ratios as constraints to nuclear and stellar parameters of asymptotic giant branch star nucleosynthesis. *The Astrophysical Journal,* 921**,** 7.

Palmerini, S., Cristallo, S., Busso, M., La Cognata, M., Sergi, M. L. & Vescovi, D. 2021b. Low mass stars or intermediate mass stars? The stellar origin of presolar oxide grains revealed by their isotopic composition. *Frontiers in Astronomy and Space Sciences,* 7**,** 103.

Parikh, A., Wimmer, K., Faestermann, T., Hertenberger, R., José, J., Wirth, H.-F., Hinke, C., Krücken, R., Seiler, D., Steiger, K. & Straub, K. 2014. Isotopic $^{32}S/^{33}S$ ratio as a diagnostic of presolar grains from novae. *Physics Letters B,* 737**,** 314-319.

Pellin, M. J., Calaway, W. F., Davis, A. M., Lewis, R. S., Amari, S. & Clayton, R. N. 2000. Toward Complete Isotopic Analysis of Individual Presolar Silicon Carbide Grains: C, N, Si, Sr, Zr, Mo, and Ba in Single Grains of Type X. *in 31st Lunar and Planetary Science Conference*, Houston, TX., #1917 (abstract).

Piersanti, L., Cristallo, S. & Straniero, O. 2013. The effects of rotation on s-process nucleosynthesis in asymptotic giant branch stars. *The Astrophysical Journal,* 774**,** 98.

Pignatari, M., Hirschi, R., Wiescher, M., Gallino, R., Bennett, M., Beard, M., Fryer, C., Herwig, F., Rockefeller, G. & Timmes, F. X. 2013a. The $^{12}C$ + $^{12}C$ reaction and the impact on nucleosynthesis in massive stars. *The Astrophysical Journal,* 762**,** 31.

Pignatari, M., Wiescher, M., Timmes, F. X., de Boer, R. J., Thielemann, F.-K., Fryer, C., Heger, A., Herwig, F. & Hirschi, R. 2013b. Production of carbon-rich presolar grains from massive stars. *The Astrophysical Journal Letters,* 767**,** L22.

Pignatari, M., Zinner, E., Bertolli, M. G., Trappitsch, R., Hoppe, P., Rauscher, T., Fryer, C., Herwig, F., Hirschi, R., Timmes, F. X. & Thielemann, F.-K. 2013c. Silicon carbide grains of type C provide evidence for the production of the unstable isotope $^{32}Si$ in supernovae. *The Astrophysical Journal Letters,* 771**,** L7.

Pignatari, M., Zinner, E., Hoppe, P., Jordan, C. J., Gibson, B. K., Trappitsch, R., Herwig, F., Fryer, C., Hirschi, R. & Timmes, F. X. 2015. Carbon-rich presolar grains from massive stars: subsolar $^{12}C/^{13}C$ and $^{14}N/^{15}N$ Ratios and the Mystery of $^{15}N$. *The Astrophysical Journal Letters,* 808**,** L43.

Pignatari, M., Herwig, F., Hirschi, R., Bennett, M., Rockefeller, G., Fryer, C., Timmes, F. X., Ritter, C., Heger, A., Jones, S., Battino, U., Dotter, A., Trappitsch, R., Diehl, S., Frischknecht, U., Hungerford, A., Magkotsios, G., Travaglio, C. & Young, P. 2016. NuGrid stellar data set. I.stellar yields from H to Bi for stars with metallicities Z = 0.02 and Z = 0.01. *The Astrophysical Journal Supplement Series,* 225**,** 24.

Pignatari, M., Hoppe, P., Trappitsch, R., Fryer, C., Timmes, F. X., Herwig, F. & Hirschi, R. 2018. The neutron capture process in the He shell in core-collapse supernovae: Presolar silicon carbide grains as a diagnostic tool for nuclear astrophysics. *Geochimica et Cosmochimica Acta,* 221**,** 37-46.

Pozzo, M., Meikle, W. P. S., Fassia, A., Geballe, T., Lundqvist, P., Chugai, N. N. & Sollerman, J. 2004. On the source of the late-time infrared luminosity of SN 1998S and other Type II supernovae. *Monthly Notices of the Royal Astronomical Society,* 352**,** 457-477.

Pravdivtseva, O., Tissot, F. L. H., Dauphas, N. & Amari, S. 2020. Evidence of presolar SiC in the Allende Curious Marie calcium-aluminium-rich inclusion. *Nature Astronomy,* 4**,** 617-624.

Qin, L., Nittler, L. R., Alexander, C. M. O' D., Wang, J., Stadermann, F. J. & Carlson, R. W. 2011. Extreme $^{54}Cr$-rich nano-oxides in the CI chondrite Orgueil - Implication for a late supernova injection into the solar system. *Geochimica et Cosmochimica Acta,* 75**,** 629-644.

Qin, L., & Carlson, R. W. 2016. Nucleosynthetic isotope anomalies and their cosmochemical significance. *Geochemical Journal,* 50**,** 43-65.

Rauscher, T., Heger, A., Hoffman, R. D., & Woosley, S. E. 2002. Nucleosynthesis in massive stars with improved nuclear and stellar physics. *The Astrophysical Journal*, 576, 323-348.

Recio-Blanco, A. & de Laverny, P. 2007. The extra-mixing efficiency in very low metallicity RGB stars. *Astronomy and Astrophysics,* 461**,** L13-L16.

Richter, S., Ott, U. & Begemann, F. 1998. Tellurium in pre-solar diamonds as an indicator for rapid separation of supernova ejecta. *Nature*, 391, 261-263.

Roman-Duval, J., Jenkins, E. B., Tchernyshyov, K., Clark, C. J. R., Cia, A. D., Gordon, K. D., Hamanowicz, A., Lebouteiller, V., Rafelski, M., Sandstrom, K., Werk, J. & Merica-Jones, P. Y. 2022. METAL: the metal evolution, transport, and abundance in the Large Magellanic Cloud Hubble program. IV. Calibration of dust depletions versus abundance ratios in the Milky Way and Magellanic Clouds and application to damped Lyα systems. *The Astrophysical Journal,* 935**,** 105.

Russell, S. S. Arden, J. W. & Pillinger, C. T. 1991. Evidence for multiple sources of diamond from primitive chondrites. *Science*, 254, 1188-1191.

Russell, S. S., Lee, M. R., Arden, J. W. & Pillinger, C. T. 1995. The isotopic composition and origins of silicon nitride from ordinary and enstatite chondrites. *Meteoritics,* 30**,** 399.

Russell, S. S., Argen, J. W. & Pillinger, C. T. 1996. A carbon and nitrogen isotope study of diamond from primivitve chondrites. *Meteoritics & Planetary Science*, 31, 343-355.

Sarangi, A. & Cherchneff, I. 2015. Condensation of dust in the ejecta of Type II-P supernovae. *Astronomy and Astrophysics,* 575**,** A95.

Sarangi, A., Matsuura, M. & Micelotta, E. R. 2018. Dust in supernovae and supernova remnants I: Formation scenarios. *Space Science Reviews,* 214**,** 63.

Sanghani, M. N., Lajaunie, L., Marhas, K. K., Rickard, W. D. A., Hsiao, S. S.-Y., Peeters, Z., Shang, H., Lee, D.-C., Calvino, J. J. & Bizzarro, M. 2022. Microstructural and chemical investigations of presolar silicates from diverse stellar environments. *The Astrophysical Journal,* 925**,** 110.

Sasaki, H., Yamazaki, Y., Kajino, T., Kusakabe, M., Hayakawa, T., Cheoun, M.-K., Ko, H. & Mathews, G. J. 2022. Impact of hypernova νp-process nucleosynthesis on the Galactic chemical evolution of Mo and Ru. *The Astrophysical Journal,* 924**,** 29.

Savina, M. R., Davis, A. M., Tripa, C. E., Pellin, M. J., Clayton, R. N., Lewis, R. S., Amari, S., Gallino, R. & Lugaro, M. 2003. Barium isotopes in individual presolar silicon carbide grains from the Murchison meteorite. *Geochimica et Cosmochimica Acta,* 67**,** 3201-3214.

Savina, M.R., Pellin, M.J., Davis, A.M., Lewis, R.S., & Amari, S. 2007. p-Process signature in a unique presolar silicon carbide grain. *in 38th Lunar and Planetary Science Conference*, The Woodlands, TX., #2231 (abstract).

Schmidt, D. R., Woolf, N. J., Zega, T. J. & Ziurys, L. M. 2018. Extreme $^{13}$C,$^{15}$N and $^{17}$O isotopic enrichment in the young planetary nebula K4-47. *Nature,* 564**,** 378-381.

Schofield, J., Pignatari, M. Stancliffe, R. J., & Hoppe, P. 2022. Isotopic ratios for C, N, Si, Al, and Ti in C-rich presolar grains from massive stars. *Monthly Notices of the Royal Astronomical Society*, 517, 1803-1820.

Schulte, J., Bose, M., Young, P. A. & Vance, G. S. 2021. Three-dimensional supernova models provide new insights into the origins of stardust. *The Astrophysical Journal,* 908**,** 38.

Scott, E. R. D. & Krot, A. N. 2014. Chondrites and Their Components. *In:* DAVIS, A. M. (ed.) *Meteorites and Cosmochemical Processes.*

Shore, S. N., Kuin, N. P., Mason, E. & De Gennaro Aquino, I. 2018. Spectroscopic diagnostics of dust formation and evolution in classical nova ejecta. *Astronomy and Astrophysics,* 619**,** A104.

Silvia, D. W., Smith, B. D. & Shull, J. M. 2010. Numerical simulations of supernova dust destruction. I. cloud-crushing and post-processed grain sputtering. *The Astrophysical Journal,* 715**,** 1575-1590.

Singerling, S. A., Liu, N., Nittler, L. R., Alexander, C. M. O' D. & Stroud, R. M. 2021. TEM analyses of unusual presolar silicon carbide: insights into the range of circumstellar dust condensation conditions. *The Astrophysical Journal,* 913**,** 90.

Smartt, S. J. 2009. Progenitors of core-collapse supernovae. *Annual Review of Astronomy and Astrophysics,* 47**,** 63-106.

Speck, A. K., Thompson, G. D. & Hofmeister, A. M. 2005. The effect of stellar evolution on SiC dust grain sizes. *The Astrophysical Journal,* 634**,** 426-435.

Speck, A. K., Corman, A. B., Wakeman, K., Wheeler, C. H. & Thompson, G. 2009. Silicon carbide absorption features: Dust formation in the outflows of extreme carbon stars. *The Astrophysical Journal,* 691**,** 1202-1221.

Spitzer, F., Burkhardt, C., Budde, G., Kruijer, T. S., Morbidelli, A. & Kleine, T. 2020. Isotopic evolution of the inner solar system inferred from molybdenum isotopes in meteorites. *The Astrophysical Journal Letters,* 898**,** L2.

Šrámek, O., Milelli, L., Ricard, Y. & Labrosse, S. 2012. Thermal evolution and differentiation of planetesimals and planetary embryos. *Icarus,* 217**,** 339-354.

Starrfield, S., Truran, J. W., Sparks, W. M. & Kutter, G. S. 1972. CNO abundances and hydrodynamic models of the nova outburst. *The Astrophysical Journal,* 176**,** 169.

Starrfield, S., Bose, M., Iliadis, C., Hix, W. R., Woodward, C. E. & Wagner, R. M. 2020. Carbon-oxygen classical novae are Galactic $^{7}$Li producers as well as potential supernova Ia progenitors. *The Astrophysical Journal,* 895**,** 70.

Steele, R. C. J., Coath, C. D., Regelous, M., Russell, S. & Elliott, T. 2012. Neutron-poor nickel isotope anomalies in meteorites. *The Astrophysical Journal,* 758**,** 59.

Stephan, T., Trappitsch, R., Davis, A. M., Pellin, M. J., Rost, D., Savina, M. R., Yokochi, R. & Liu, N. 2016. CHILI - the Chicago Instrument for Laser Ionization - a new tool for isotope measurements in cosmochemistry. *International Journal of Mass Spectrometry,* 407**,** 1-15.

Stephan, T., Trappitsch, R., Davis, A. M., Pellin, M. J., Rost, D., Savina, M. R., Jadhav, M., Kelly, C. H., Gyngard, F., Hoppe, P. & Dauphas, N. 2018. Strontium and barium isotopes in presolar silicon carbide grains measured with CHILI-two types of X grains. *Geochimica et Cosmochimica Acta,* 221**,** 109-126.

Stephan, T., Trappitsch, R., Hoppe, P., Davis, A. M., Pellin, M. J. & Pardo, O. S. 2019. Molybdenum isotopes in presolar silicon carbide grains: details of s-process nucleosynthesis in parent stars and implications for r- and p-processes. *The Astrophysical Journal,* 877**,** 101.

Stephan, T., Trappitsch, R., Hoppe, P., Davis, A. M., Bose, M., Boujibar, A., Gyngard, F., Hynes, K. M., Liu, N., Nittler, L. R., and Ogliore, R. C. 2024. The presolar grain database. I. silicon carbide. *Astrophysical Journal Supplement Series*, 270, 27.

Stephan, T. & Trappitsch, R. 2024. Presolar Grain Database - Graphite (2024-05-13) [Data set]. *Zenodo*. https://doi.org/10.5281/zenodo.11188116.

Stephens, D., Herwig, F., Woodward, P., Denissenkov, P., Andrassy, R. & Mao, H. 2021. 3D1D hydro-nucleosynthesis simulations - I. Advective-reactive post-processing method and its application to H ingestion into He-shell flash convection in rapidly accreting white dwarfs. *Monthly Notices of the Royal Astronomical Society,* 504**,** 744-760.

Straniero, O., Gallino, R. & Cristallo, S. 2006. s process in low-mass asymptotic giant branch stars. *Nuclear Physics A,* 777**,** 311-339.

Stritzinger, M., Taddia, F., Fransson, C., Fox, O. D., Morrell, N., Phillips, M. M., Sollerman, J., Anderson, J. P., Boldt, L., Brown, P. J., Campillay, A., Castellon, S., Contreras, C., Folatelli, G., Habergham, S. M., Hamuy, M., Hjorth, J., James, P. A., Krzeminski, W., Mattila, S., Persson, S. E. & Roth, M. 2012. Multi-wavelength observations of the enduring Type IIn supernovae 2005ip and 2006jd. *The Astrophysical Journal,* 756**,** 173.

Stroud, R. M., Chisholm, M. F., Heck, P. R., Alexander, C. M. O'D. & Nittler, L. R. Supernova shock-wave-induced CO-formation of glassy carbon and nanodimaond. *The Astrophysical Journal Letters*, 738, L27 (5pp).

Taioli, S., Vescovi, D., Busso, M., Palmerini, S., Cristallo, S., Mengoni, A. & Simonucci, S. 2022. Theoretical estimate of the half-life for the radioactive $^{134}$Cs and $^{135}$Cs in astrophysical scenarios. *The Astrophysical Journal,* 933**,** 158.

Takigawa, A., Stroud, R. M., Nittler, L. R., Alexander C. M. O'D. & Miyake, A. 2018. High-temperature dust condensation around an AGB star: Evidence from a highly preistine presolar corundum. *The Astrophysical Journal Letters*, 862, L13.

Terzieva, R. & Herbst, E. 2000. The possibility of nitrogen isotopic fractionation in interstellar clouds. *Monthly Notices of the Royal Astronomical Society,* 317**,** 563-568.

Tielens, A. G. G. M. & Allamandola, L. J. 1986. Composition, structure, and chemistry of interstellar dust. *NASA Technical Memorandum*, 88350.

Timmes, F. X., Woosley, S. E. & Weaver, T. A. 1995. Galactic chemical evolution: hydrogen through zinc. *The Astrophysical Journal Supplement Series,* 98**,** 617.

Timmes, F. X. & Clayton, D. D. 1996. Galactic evolution of silicon isotopes: application to presolar SiC grains from meteorites. *The Astrophysical Journal,* 472**,** 723.

Trappitsch, R., Stephan, T., Savina, M. R., Davis, A. M., Pellin, M. J., Rost, D., Gyngard, F., Gallino, R., Bisterzo, S., Cristallo, S. & Dauphas, N. 2018. Simultaneous iron and nickel isotopic analyses of presolar silicon carbide grains. *Geochimica et Cosmochimica Acta,* 221**,** 87-108.

Trinquier, A., Birck, J.-L. & Allègre, C. J. 2007. Widespread $^{54}$Cr heterogeneity in the inner solar system. *The Astrophysical Journal,* 655**,** 1179-1185.

Trinquier, A., Elliott, T., Ulfbeck, D., Coath, C., Krot, A. N. & Bizzarro, M. 2009. Origin of nucleosynthetic isotope heterogeneity in the solar protoplanetary disk. *Science,* 324**,** 374.

Trippella, O., Busso, M., Palmerini, S., Maiorca, E. & Nucci, M. C. 2016. s-processing in AGB stars revisited. II. Enhanced $^{13}$C production through MHD-induced mixing. *The Astrophysical Journal,* 818**,** 125.

Troja, E., Segreto, A., La Parola, V., Hartmann, D., Baumgartner, W., Markwardt, C., Barthelmy, S., Cusumano, G. & Gehrels, N. 2014. Swift/BAT detection of hard X-rays from Tycho's supernova remnant: evidence for titanium-44. *The Astrophysical Journal Letters,* 797**,** L6.

Unterborn, C. T., Johnson, J. A. & Panero, W. R. 2015. Thorium abundances in solar twins and analogs: implications for the habitability of extrasolar planetary systems. *The Astrophysical Journal,* 806**,** 139.

Ventura, P. & D'Antona, F. 2005. Full computation of massive AGB evolution. I. The large impact of convection on nucleosynthesis. *Astronomy and Astrophysics,* 431**,** 279-288.

Verdier-Paoletti, M. J., Nittler, L. R. & Wang, J. 2019. High-resolution measurements of Mg, Si, Fe and Ni isotopes of O-rich presolar grains. *82nd Annual Meeting of The Meteoritical Society*, July 01, #6433 (abstract).

Vescovi, D., Cristallo, S., Busso, M. & Liu, N. 2020. Magnetic-buoyancy-induced mixing in AGB stars: presolar SiC grains. *The Astrophysical Journal Letters,* 897**,** L25.

Vescovi, D., Cristallo, S., Palmerini, S., Abia, C. & Busso, M. 2021. Magnetic-buoyancy-induced mixing in AGB stars: fluorine nucleosynthesis at different metallicities. *Astronomy and Astrophysics,* 652**,** A100.

Virag, A., Wopenka, B., Amari, S., Zinner, E., Anders, E. & Lewis, R. S. 1992. Isotopic, optical, and trace element properties of large single SiC grains from the Murchison meteorite. *Geochimica et Cosmochimica Acta,* 56**,** 1715-1733.

Wallis, M. K. 1980. Radiogenic melting of primordial comet interiors. *Nature,* 284**,** 431-433.

Wallner, A., Faestermann, T., Feige, J., Feldstein, C., Knie, K., Korschinek, G., Kutschera, W., Ofan, A., Paul, M., Quinto, F., Rugel, G. & Steier, P. 2015. Abundance of live $^{244}$Pu in deep-sea reservoirs on Earth points to rarity of actinide nucleosynthesis. *Nature Communications,* 6**,** 5956.

Wallner, A., Feige, J., Kinoshita, N., Paul, M., Fifield, L. K., Golser, R., Honda, M., Linnemann, U., Matsuzaki, H., Merchel, S., Rugel, G., Tims, S. G., Steier, P., Yamagata, T. & Winkler, S. R. 2016. Recent near-Earth supernovae probed by global deposition of interstellar radioactive $^{60}$Fe. *Nature,* 532**,** 69-72.

Warren, P. H. 2011. Stable-isotopic anomalies and the accretionary assemblage of the Earth and Mars: A subordinate role for carbonaceous chondrites. *Earth and Planetary Science Letters,* 311**,** 93-100.

Wasserburg, G. J., Boothroyd, A. I. & Sackmann, I.-J. 1995. Deep circulation in red giant stars: a solution to the carbon and oxygen isotope puzzles? *The Astrophysical Journal,* 447**,** L37.

Wesson, R., Barlow, M. J., Matsuura, M. & Ercolano, B. 2015. The timing and location of dust formation in the remnant of SN 1987A. *Monthly Notices of the Royal Astronomical Society,* 446**,** 2089-2101.

Wesson, R. & Bevan, A. 2021. Observational limits on the early-time dust mass in SN 1987A. *The Astrophysical Journal,* 923**,** 148.

Wongwathanarat, A., Janka, H.-Th., and Müller, E., Pllumbi, E., & Wanajo, S. 2017. Production and distribution of $^{44}$Ti and $^{56}$Ni in a three-dimensional supernova model resembling Cassiopeia A. *The Astrophysical Journal*, 842, 13 (20pp).

Woodward, P. R., Lin, P.-H., Mao, H., Andrassy, R. & Herwig, F. 2019. Simulating 3-D stellar hydrodynamics using PPM and PPB multifluid gas dynamics on CPU and CPU+GPU nodes. *Journal of Physics Conference Series*, May 01, 012020.

Woosley, S. E. & Weaver, T. A. 1995. The evolution and explosion of massive stars. II. Explosive hydrodynamics and nucleosynthesis. *The Astrophysical Journal Supplement Series,* 101**,** 181.

Woosley, S. E. & Heger, A. 2007. Nucleosynthesis and remnants in massive stars of solar metallicity. *Physics Reports,* 442**,** 269-283.

Xu, Y., Zinner, E., Gallino, R., Heger, A., Pignatari, M. & Lin, Y. 2015. Sulfur isotopic compositions of submicrometer SiC grains from the Murchison meteorite. *The Astrophysical Journal,* 799**,** 156.

Yada, T., Floss, C., Stadermann, F. J., Zinner, E., Nakamura, T., Noguchi, T. & Lea, A. S. 2008. Stardust in Antarctic micrometeorites. *Meteoritics and Planetary Science,* 43**,** 1287-1298.

Yamamoto, D., Kuroda, M., Tachibana, S., Sakamoto, N. & Yurimoto, H. 2018. Oxygen isotopic exchange between amorphous silicate and water vapor and its implications for oxygen isotopic evolution in the early solar system. *The Astrophysical Journal,* 865**,** 98.

Yokoyama, T., Nagai, Y., Fukai, R. & Hirata, T. 2019. Origin and evolution of distinct molybdenum isotopic variabilities within carbonaceous and noncarbonaceous reservoirs. *The Astrophysical Journal,* 883**,** 62.

Young, E. D. 2014. Inheritance of solar short- and long-lived radionuclides from molecular clouds and the unexceptional nature of the solar system. *Earth and Planetary Science Letters,* 392**,** 16-27.

Zhukovska, S., Gail, H.-P. & Trieloff, M. 2008. Evolution of interstellar dust and stardust in the solar neighbourhood. *Astronomy and Astrophysics,* 479**,** 453-480.

Zhukovska, S. & Henning, T. 2013. Dust input from AGB stars in the Large Magellanic Cloud. *Astronomy and Astrophysics,* 555**,** A99.

Zijlstra, A. A., Matsuura, M., Wood, P. R., Sloan, G. C., Lagadec, E., van Loon, J. T., Groenewegen, M. A. T., Feast, M. W., Menzies, J. W., Whitelock, P. A., Blommaert, J. A. D. L., Cioni, M.-R. L., Habing, H. J., Hony, S., Loup, C. & Waters, L. B. F. M. 2006. A Spitzer mid-infrared spectral survey of mass-losing carbon stars in the Large Magellanic Cloud. *Monthly Notices of the Royal Astronomical Society,* 370**,** 1961-1978.

Zinner, E. 2014. Presolar grains. *In:* DAVIS, A. M. (ed.) *Meteorites and Cosmochemical Processes.* Oxford: Elsevier.

Zinner, E., Nittler, L. R., Hoppe, P., Gallino, R., Straniero, O. & Alexander, C. M. O' D. 2005. Oxygen, magnesium and chromium isotopic ratios of presolar spinel grains. *Geochimica et Cosmochimica Acta,* 69**,** 4149-4165.

Zinner, E., Nittler, L. R., Gallino, R., Karakas, A. I., Lugaro, M., Straniero, O. & Lattanzio, J. C. 2006. Silicon and carbon isotopic ratios in AGB stars: SiC grain data, models, and the Galactic evolution of the Si isotopes. *The Astrophysical Journal,* 650**,** 350-373.

Zinner, E., Amari, S., Guinness, R., Jennings, C., Mertz, A. F., Nguyen, A. N., Gallino, R., Hoppe, P., Lugaro, M., Nittler, L. R. & Lewis, R. S. 2007. NanoSIMS isotopic analysis of small presolar grains: Search for $Si_3N_4$ grains from AGB stars and Al and Ti isotopic compositions of rare presolar SiC grains. *Geochimica et Cosmochimica Acta,* 71**,** 4786-4813.

Zoglauer, A., Reynolds, S. P., An, H., Boggs, S. E., Christensen, F. E., Craig, W. W., Fryer, C. L., Grefenstette, B. W., Harrison, F. A., Hailey, C. J., Krivonos, R. A., Madsen, K. K., Miyasaka, H., Stern, D. & Zhang, W. W. 2015. The hard X-ray view of the young supernova remnant G1.9+0.3. *The Astrophysical Journal,* 798**,** 98.